\documentclass[print,psfig]{aa}
\usepackage{float}
\usepackage{amsmath}
\usepackage{amssymb}
\usepackage[latin1]{inputenc}
\usepackage{graphicx}
\usepackage{epsfig}
\usepackage{natbib}
\usepackage{longtable,lscape,afterpage}

\def\int {\emph{INTEGRAL}}
\def\swi {\emph{Swift}}
\def\sax {Beppo\emph{SAX}}

\def\apjl{ApJ}                
\def\apjs{ApJS}               
\def\ao{Appl.~Opt.}           
\def\aap{A\&A}                

\begin{document}

\title{The updated spectral catalogue of \int\ gamma--ray bursts}

  \author{G.Vianello\inst{1,2}, D. G\"otz\inst{3} and S. Mereghetti\inst{1}}

   \offprints{G. Vianello, email: vianello@iasf-milano.inaf.it}
  \institute{INAF - Istituto di Astrofisica Spaziale e Fisica Cosmica Milano,
          Via Bassini 15, I-20133 Milano, Italy
          \and
              Dipartimento di Fisica e Matematica, Universit\`{a} dell'Insubria,
              via Valleggio 11, I-22100 Como, Italy
          \and
        CEA-Saclay, DSM/Irfu/Service d'Astrophysique, Orme des Merisiers,
        F-91191, Gif-sur-Yvette, France
   }

  \date{}

\abstract{ {We present a catalogue with the properties of all the bursts detected
and localized by the IBIS instrument onboard the \int\ satellite
from November 2002 to September 2008. The sample is composed of 56 bursts, corresponding to a rate of $\sim 0.8$ GRB per month. Thanks to the performances of the \int\ Burst Alert System, 50\% of the IBIS GRBs have detected afterglows, while 5\% have redshift measurements. A spectral analysis of the 43 bursts
in the \int\ public archive has been carried out using the most recent
software and calibration, deriving an updated, homogeneous and accurate catalogue with the spectral features of the sample. When possible a time-resolved spectral analysis also has been carried out. The GRBs in the sample have 20-200 keV fluences in the range $5 \times 10^{-8}$--$2.5 \times 10^{-4}$ erg cm$^{-2}$, and peak fluxes in the range $0.11$--$56$ ph cm$^{-2}$ s$^{-1}$.
While most of the spectra are well fitted by a power law with photon index $\sim 1.6$,
we found that 9 bursts are better described by a cut-off power law,
resulting in E$_{p}$ values in the range 35--190 keV. Altough these results are comparable to those obtained with BAT onboard \swi , there is marginal evidence that ISGRI detects dimmer bursts than \swi /BAT. 
Using the revised spectral parameters and an updated sky exposure map that also takes
into account the effects of the GRB trigger efficiency,
we strengthen the evidence for a spatial correlation with the super galactic
plane of the faint bursts with long spectral lag.}  }

\authorrunning{Vianello et al.}
\titlerunning{The updated spectral catalogue of \int\ GRBs}
\maketitle

\begin{figure*}[tb!]
\centering
\psfig{file=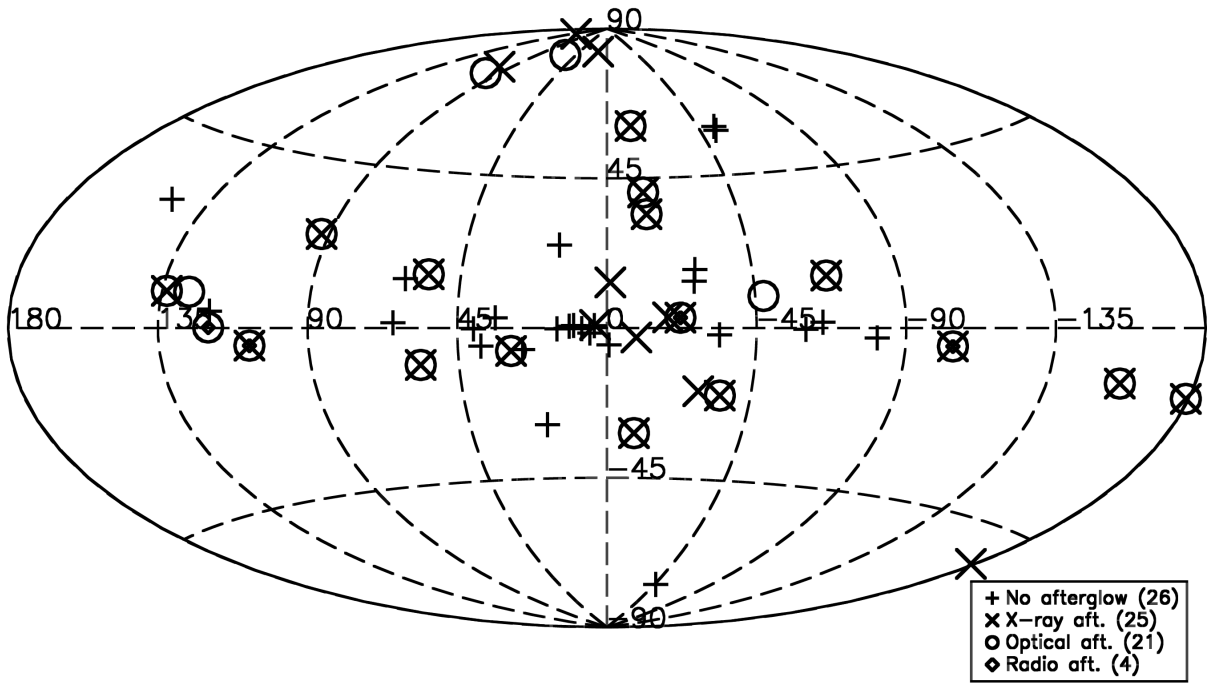,angle=90,width=0.9\linewidth}
\caption{Sky positions of the burst detected by IBAS in  Galactic coordinates. }
\label{fig:gal_positions}
\end{figure*}
\normalsize
\section{Introduction}

Enormous progress in the study of Gamma-Ray Bursts (GRBs) has been
achieved after the results of the \textit{Satellite per Astronomia X} (the Italian for ``Satellite for X-Ray Astronomy'') \sax\ , that enabled
the discovery of their counterparts at lower energy
\citep{Costa1997,Vanparadijs1997,Frail1997}. Our understanding of
GRBs has mainly profited from  dedicated instruments, or even
satellites, specifically designed for the study of these enigmatic
sources. For example, thanks to the very large GRB sample obtained
with the Burst And Transient Sources Experiment (BATSE) onboard the Compton Gamma-Ray Observatory (CGRO), we learned that GRBs are isotropically
distributed across the sky \citep{briggs96}, and that their
LogN-LogS is compatible with a cosmological origin
\citep{Meegan1992}. BATSE also provided accurate spectra
 \citep[e.g.][]{Band1993}, and confirmed that GRBs  can be divided in two
categories as a function of their duration and hardness, with
about 25\% of the GRBs being shorter than 2 s and harder compared
to bursts of longer duration \citep{Kouveliotou1993}. Quick
localizations with the High Energy Transient Explorer (HETE-2) satellite allowed investigators to study the GRB
emission in X-rays \citep[e.g.][]{vanderspeck04}, and to
detect the first afterglow of a short GRB \citep{fox05}. \swi\
\citep{swift}, launched in 2004, is currently localizing about 100
bursts per year, and is able to investigate the early phases of
the X-ray afterglow, which were unaccessible to former
instrumentation.
Many key parameters are derived from the afterglow observations.
Among them one of the most important is the distance, which is essential to infer
the intrinsic energy and luminosity. GRBs are the most powerful
explosions in the Universe, with isotropic equivalent energy
$E_{iso}$ in the range from 10$^{50}$ to 10$^{54}$ erg
\citep{Amati2006}, although such values can be reduced with the
hypothesis that GRBs are collimated sources \citep{rhoads97}, as supported by the detection of achromatic breaks in some afterglow light
curves. The jet opening angles
inferred from the afterglow break times 
\citep{Sari1999} indicate lower energy values, clustered around
10$^{51}$ erg with a reduced spread \citep{frail2001,Bloom2003}.
However, the observation of non-achromatic breaks in some
afterglows, and the apparent lack of any kind of break in some
others, casts doubts on this argument \citep[see][and reference
therein]{Curran2007}. In only a few years, the \swi\ satellite
has more than doubled   the number of GRBs with known redshift,
and discovered the most distant burst to date at $z \sim$6.7
\citep[GRB 080913,][]{Fynbo2008GCN}.

Recent progresses in GRB science concern also the identification
of their progenitors. There is general consensus in considering
the collapse of massive stars ($>$ 30 M$_{\odot}$) as the origin
of long GRBs \citep{Woosley1993,Vietri1998}. Indeed, in many cases
late-time bumps in the afterglow light curves of long GRBs have
been reported \citep[e.g.][]{Bloom1999, Zeh2004}, and are
interpreted as an emerging supernova component. In a handful of
cases the GRB/SN association has been spectroscopically confirmed
\citep[e.g.][]{Malesani2004}.
However, in some cases deep optical follow-up observations failed
to detect SN signatures from nearby long GRBs
\citep[e.g.][]{Dellavalle2006}, indicating the possible existence
of a different class of long GRBs. Models involving the merging of
two compact objects (black holes, neutron stars, or white dwarfs) are, on the other hand, preferred
for short GRBs \citep{Perna2002}.

Besides GRB-dedicated missions, also ``general purpose''
X--ray/gamma-ray satellites can significantly contribute to the
GRB quest. The most remarkable example is indeed the
aforementioned \sax\ mission, that thanks to its rapid and
accurate localizations led to the discovery of afterglows at
X-ray,  optical, and radio wavelengths, firmly proving the
cosmological origin of these sources by allowing a measurement of
their redshift.

The International Gamma-Ray Astrophysics Laboratory \citep[\int ,][]{integral}, launched in October 2002,
carries a set of coded mask instruments dedicated to fine imaging
and spectroscopy in the soft $\gamma$-ray energy domain (15
keV--10 MeV). Even if \int\ is not a  mission specifically
dedicated to GRBs, its imager instrument IBIS \citep[Imager on Board the Integral Satellite,][]{ibis}, thanks
to a large field of view (29$^{\circ} \times$29$^{\circ}$) and a
good point source location accuracy ($\sim$ 1--2$^{\prime}$), is
well suited for GRB studies, in particular with its low energy
detector plane ISGRI \citep[\int\ Soft Gamma-Ray Imager,][]{isgri} (15 keV--1 MeV) . Pre-launch
expectations of detecting about one GRB per month in the IBIS
field of view prompted the development of the \int\ Burst Alert
System (IBAS).

IBAS  \citep{ibas} is a software system, running on the ground at the
\int\ Science Data Centre \citep[ISDC;][] {isdc}, able to detect
and localize in real time the GRBs in the IBIS field of view.
Thanks to IBAS, the \int\ satellite  has been the first to
provide GRB positions with an uncertainty of only $\sim$2-3 arcmin
in near real time (less than a few tens of second after the burst
trigger). To date 56 bursts have been detected in IBIS/ISGRI
data\footnote{See {\tt http://ibas.iasf-milano.inaf.it}}, mostly
automatically by IBAS, and a few in off-line searches.

In this paper we present an updated catalog of all the GRBs
localized with \int\ , with a detailed spectral analysis of the
ones for which public data are available.
Some results have been published already by different authors on
individual bursts, especially the ones discovered in the first
months of the mission. More recently, a comprehensive analysis of a
sample of 46 \int\ bursts has been presented by
\citet{Foley2008}, who focused mainly on their timing properties. They also extract spectra of some GRBs, but using an old version of the software and of the calibration that now we know to be not adequate when dealing with off-axis bursts. The main
purpose of our work is to provide a homogeneous set of
results, obtained with the latest instrument calibration and
software, focusing mainly on the spectral properties. In several cases the spectral parameters we derived differ
from, and supersede, those reported in previous analyses, that
were based on earlier software versions and calibrations.

\section{Data analysis}
\label{sec:analysis method}
\subsection{The IBIS coded-mask telescope}
IBIS \citep{2003A&A...411L.131U} is an imaging gamma-ray telescope, based on the coded-mask
technique, with a square field of view  of $29^{\circ} \times
29^{\circ}$ (at zero sensitivity). The sensitivity is maximum and
nearly uniform within  the inner $8.5^{\circ} \times 8.5^{\circ}$,
corresponding to the so called fully coded field of view (FCFOV).
Our analysis is mainly based on data taken with the IBIS low energy
detector ISGRI \citep{isgri}, consisting of an array of 128$\times$128 CdTe
crystals sensitive in the nominal 15 keV--1 MeV energy range.
The CdTe detector thickness of 2 mm ensures an efficiency of 50\%
at 150 keV.  Above this energy the efficiency decreases rapidly
(10\% at 300 keV), and although the brightest GRBs are detected up
to $\sim$500 keV, the majority of the spectra presented here refer
only to the 18-300 keV band. The central part of the IBIS field of view is covered also by the Joint European X--ray Monitor \citep[JEM-X, ][]{2003A&A...411L.231L} and the Optical Monitoring Camera \citep[OMC, ][]{2003A&A...411L.261M}. When possible we have also analyzed data from these instruments. 
\subsection{GRB sample}

In Table~\ref{tab:observative} we list the 56 GRBs observed with
ISGRI until now (end of September 2008). Most of them were discovered and
localized with the \int\ Burst Alert System \citep[IBAS;][]{ibas}
and their coordinates were automatically distributed in near real
time within a few seconds. The remaining ones were detected during
the initial IBAS performance verification phase (when automated alert distribution was not yet enabled), in off-line searches, or
after interactive verification of low significance IBAS internal
alerts.

\begin{figure}[b!]
\psfig{file=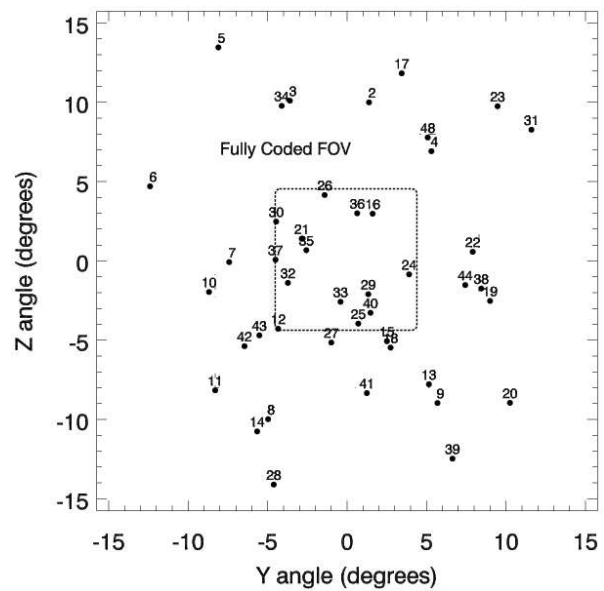,width=0.9\linewidth}
 \caption{GRB positions in the detector coordinates. The region inside the dotted rectangle is the fully coded field of view.}
 \label{fig:fov_positions}
\end{figure}

The sky distribution of the 56 bursts is
plotted in Fig.~\ref{fig:gal_positions}.
The concentration of events close to the galactic plane simply
reflects the greater exposure devoted to these regions by the
pointing program of \int\ .
The IBAS trigger efficiency is maximum in the fully coded field of view, where all of the detector is exposed to the flux of the source and the signal to noise ratio is maximized. Towards larger off-axis angle the efficiency decreases, so that we need a brighter burst to obtain the same signal-to-noise. On the other hand the sky area covered by the detector increases. Thus it is not surprising that the majority of GRBs have been detected at intermediate off-axis angles, as visible in Fig.~\ref{fig:fov_positions} where we show the instrumental positions
of the bursts. 23 \% of the sample have been detected in the FCFOV, 55 \% have been detected with an off-axis angle between 4.5$^{\circ}$ and 12$^{\circ}$, the remaining 22 \%  with an angle greater than 12$^{\circ}$. Some GRBs were detected at the very edge of the field of view, showing that the IBAS system is able to detect bursts in the whole field of view, provided the signal to noise ratio is large enough. 

\begin{figure}[tb!]
\psfig{file=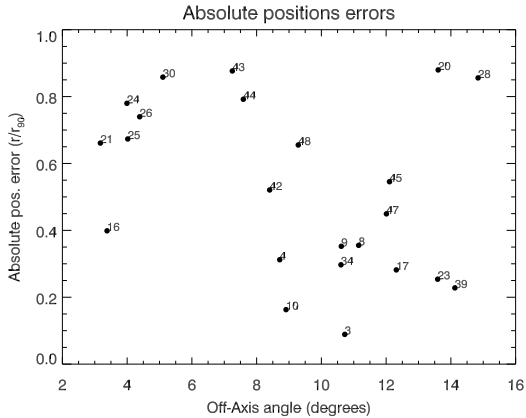,angle=90,width=0.9\linewidth}
\caption{Absolute position errors $r$ as a function of the off-axis angle, for bursts with a detected optical afterglow. $r$ is defined as the difference between the position obtained with ISGRI data and that of the optical afterglow. The error on the position of the optical afterglows is negligible. The absolute position error is given in units of $r_{90}$, that is, the 90 \% error radius estimated during the analysis of ISGRI data. All the afterglows were found inside the error circle provided by ISGRI.}
 \label{fig:poserr}
\end{figure}

The GRB positions reported in Table \ref{tab:observative} are
derived taking into account the most accurate information on the
satellite attitude. The errors on these coordinates are in most
cases between 1.5 and 3 arcmin, basically limited by the
statistical uncertainties.  We also report the position of the
afterglows (optical or X-ray) when available, as given in the
references indicated in the last column. For these bursts, the
difference between the ISGRI and the afterglow position is
compared with the quoted uncertainties in
Fig.~\ref{fig:poserr}, where we report the distance between the afterglow position and our position, normalized to the 90 \% error radius $R_{90}$ quoted in Table \ref{tab:observative}, as a function of the off-axis angle. All the positions of the afterglows are within the error regions provided by ISGRI, thus the localization of GRBs are accurate and the error regions are well sized. Moreover there is no evidence of systematic errors depending on the position of the burst on the detector.

Column 13 of Table \ref{tab:observative} gives the GRB peak flux in the 15-150 keV energy range
integrated over 1 s. This has been computed by using the
conversion factor from counts s$^{-1}$ to photons cm$^{-2}$
s$^{-1}$ derived from the time integrated spectrum of each GRB. This value is corrected for the instrument dead-time.

\subsection{Light curve extraction}
\label{sec:lc}
\begin{figure}[tb!]
\centering
\psfig{file=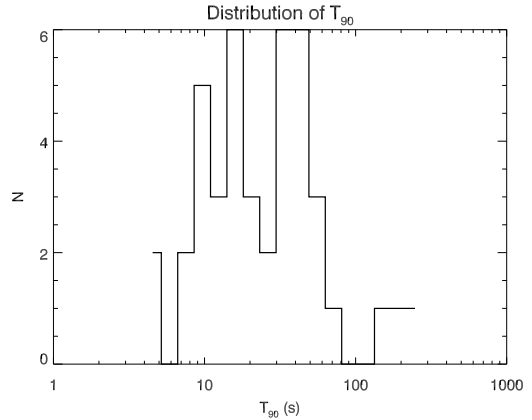,angle=90,width=0.9\linewidth}
\caption{Distribution of $t_{90}$ for \int\ GRBs we have analyzed in this work. No short bursts ($t_{90} < 2$ s) are present in the sample.}
\label{fig:t90_distr}
\end{figure}

\begin{figure*}[p]
\vspace{1.0cm}
\begin{tabular}{cc}
\psfig{figure=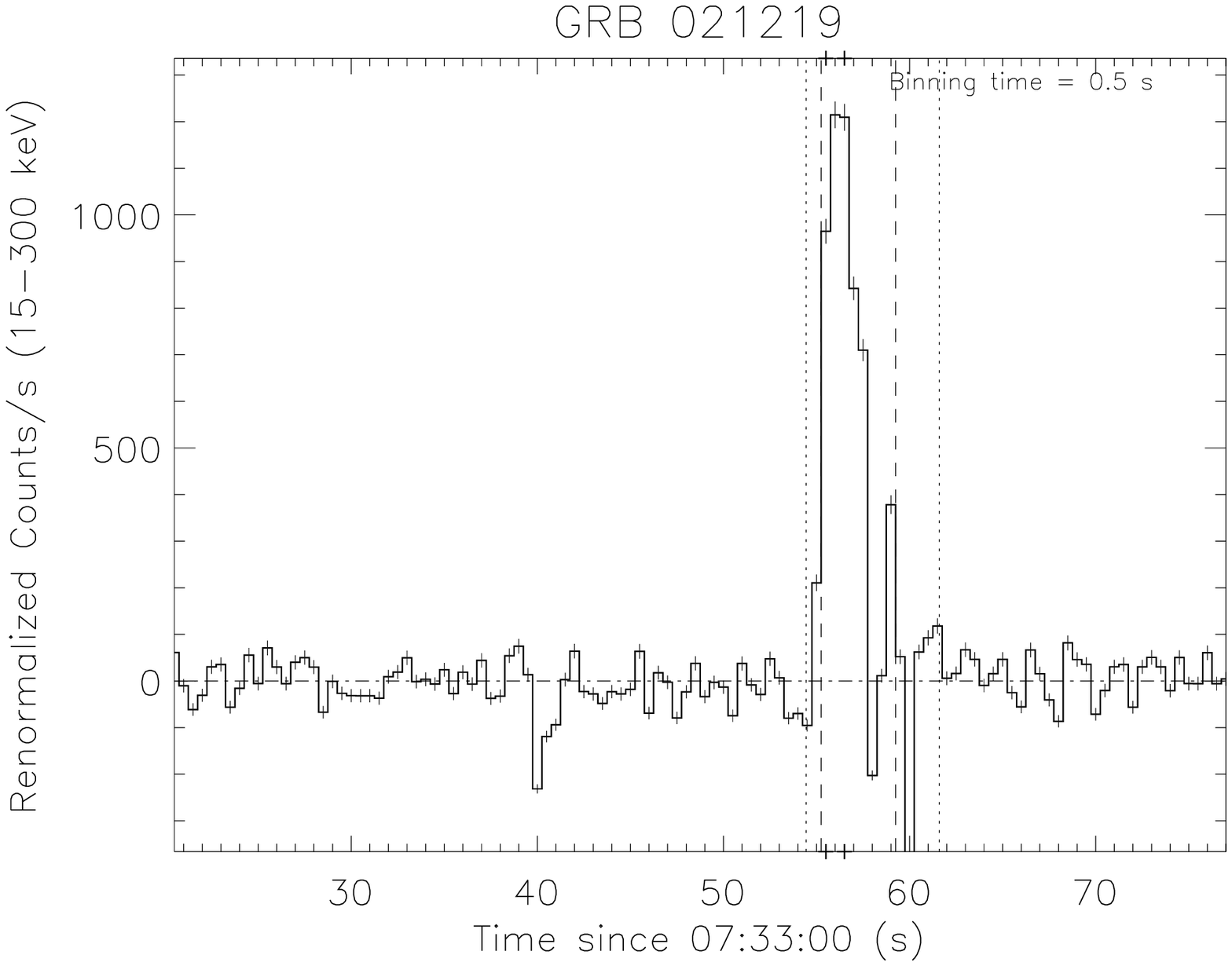,width=0.5\linewidth} &
 \psfig{figure=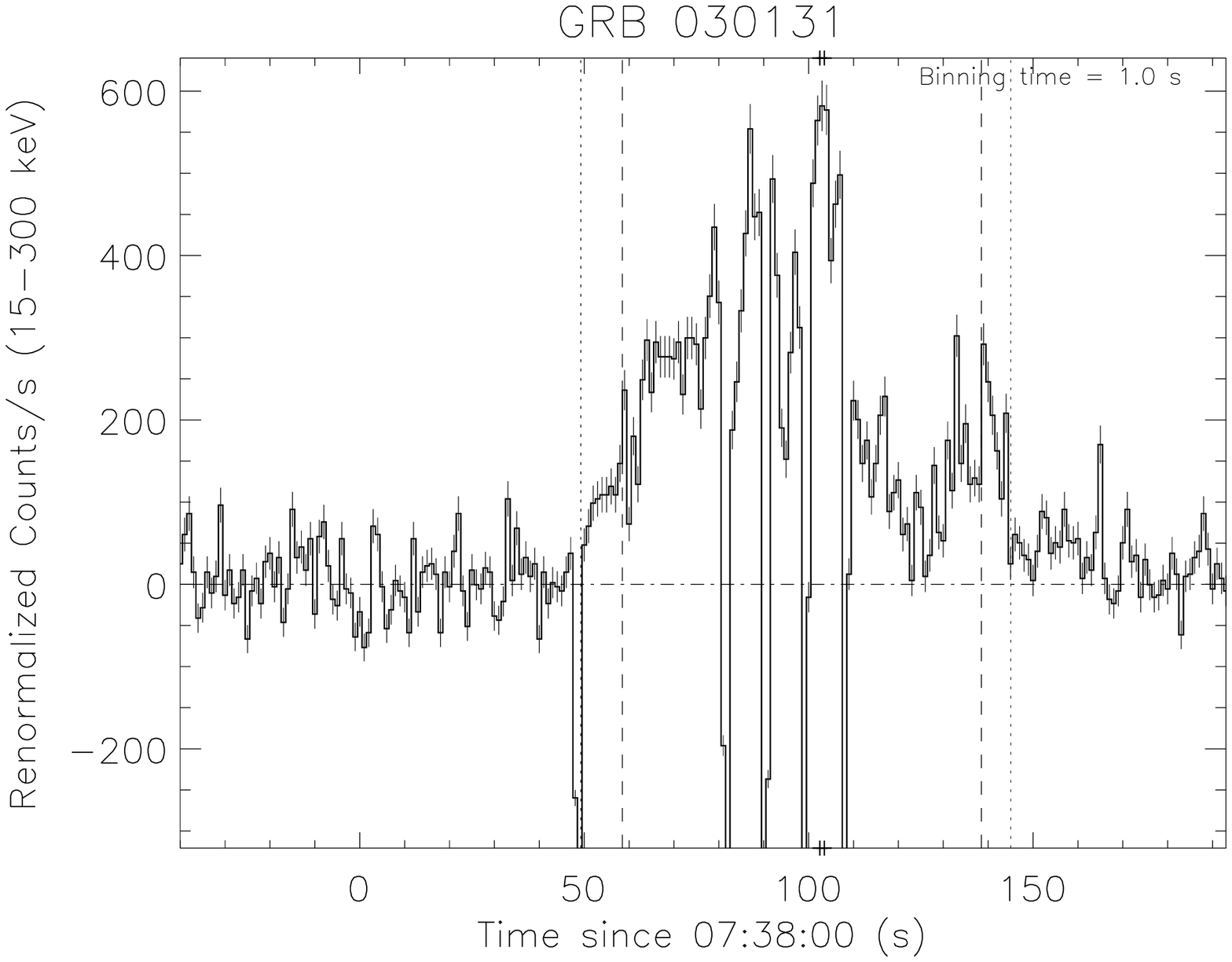,width=0.5\linewidth} \\
\psfig{figure=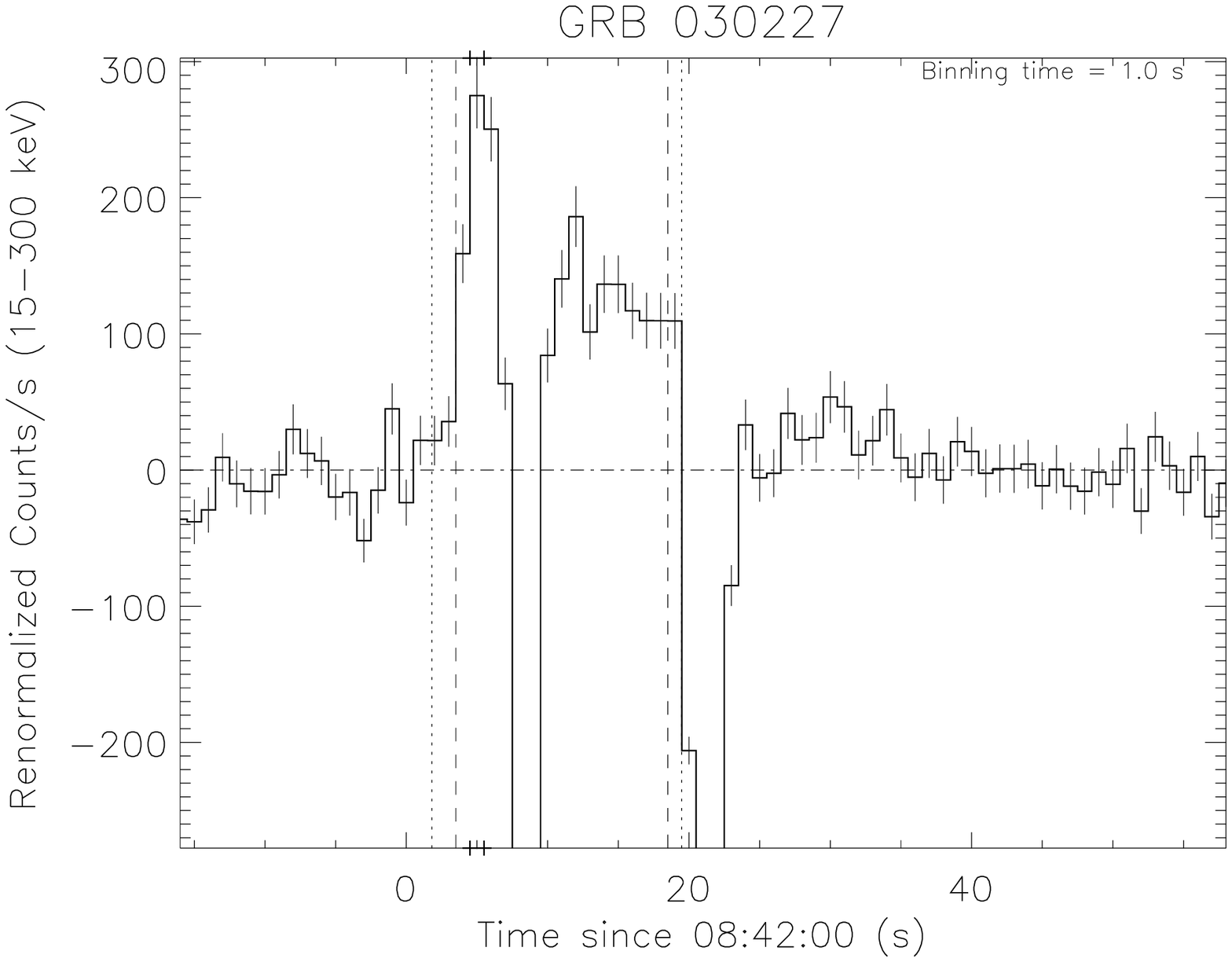,width=0.5\linewidth} &
 \psfig{figure=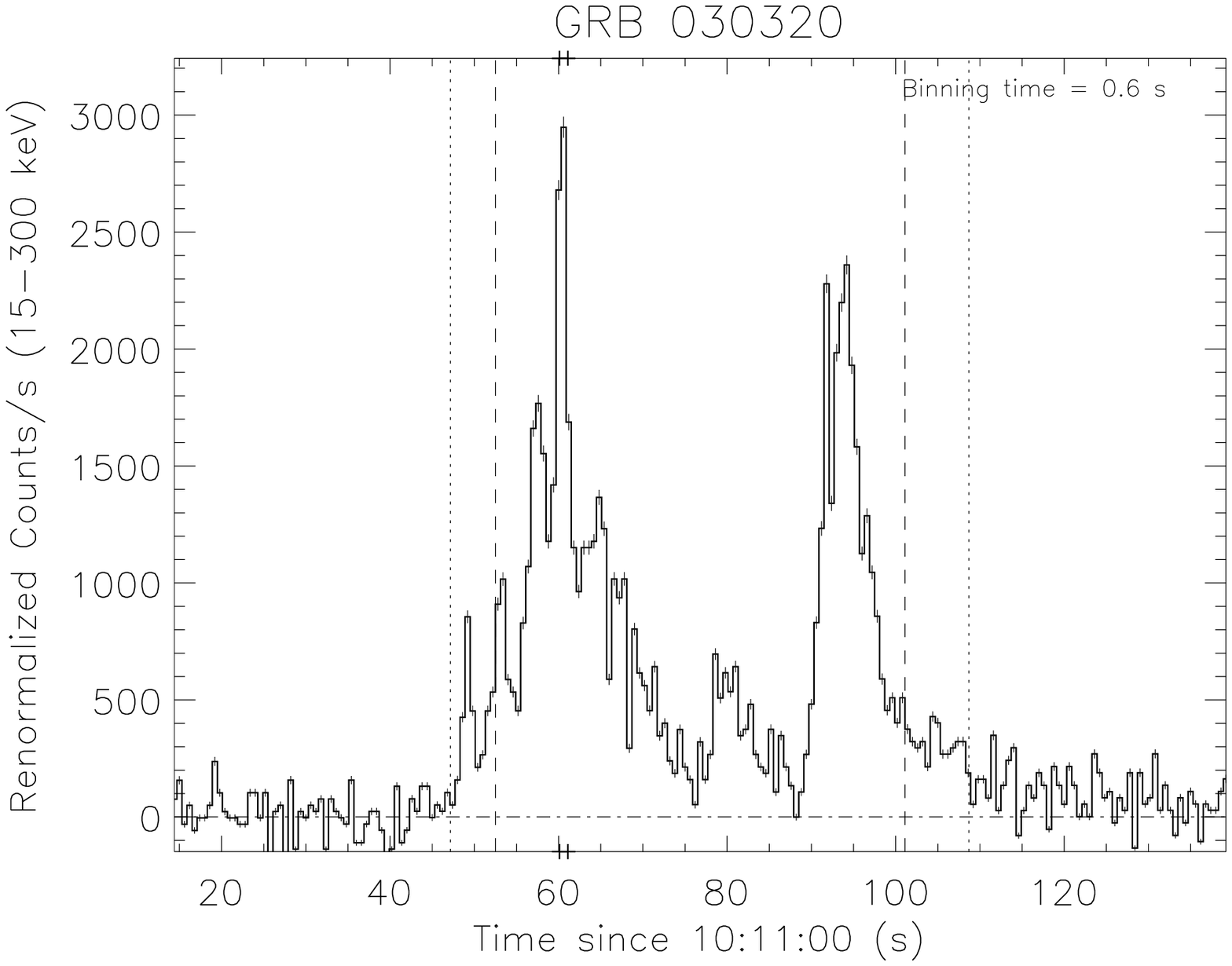,width=0.5\linewidth} \\
\psfig{figure=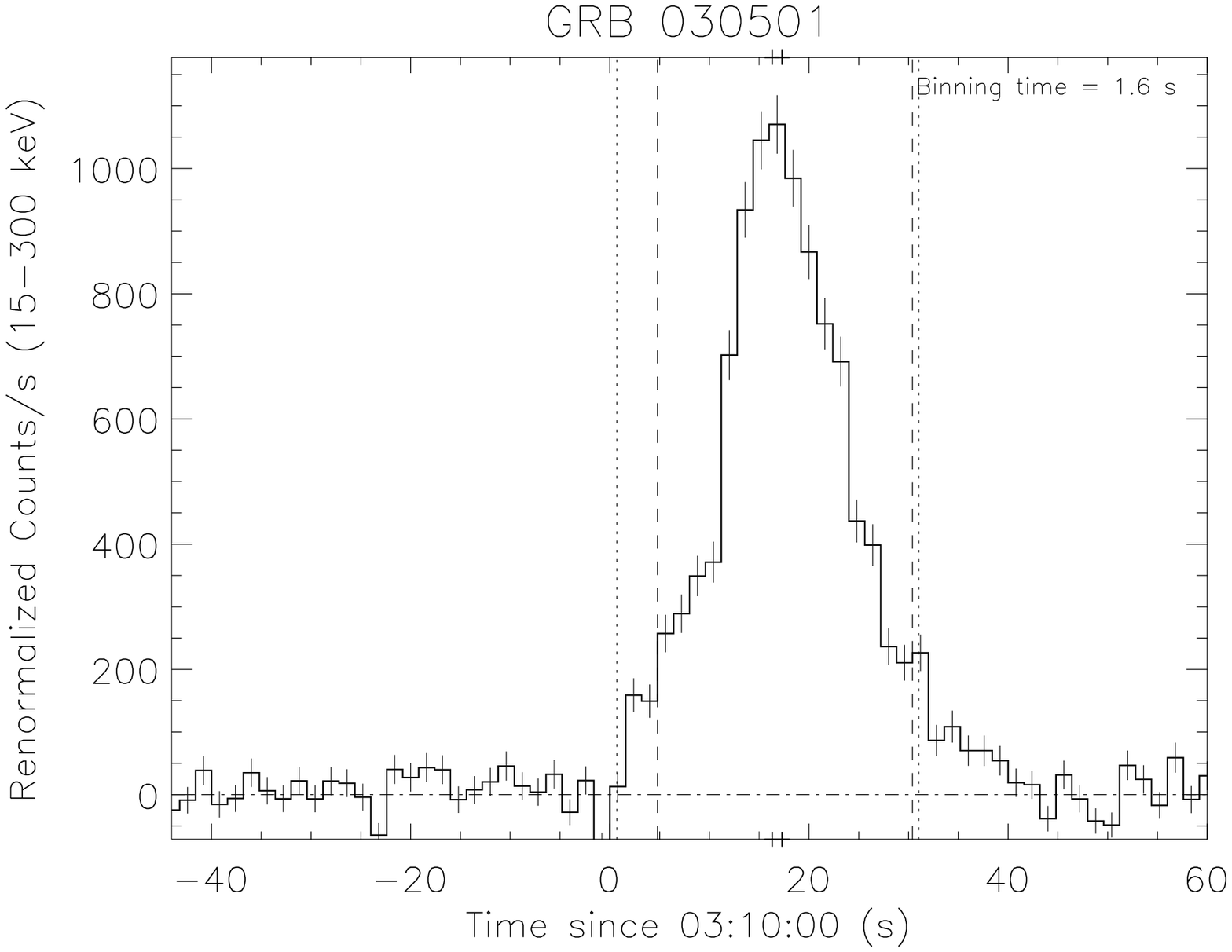,width=0.5\linewidth} &
 \psfig{figure=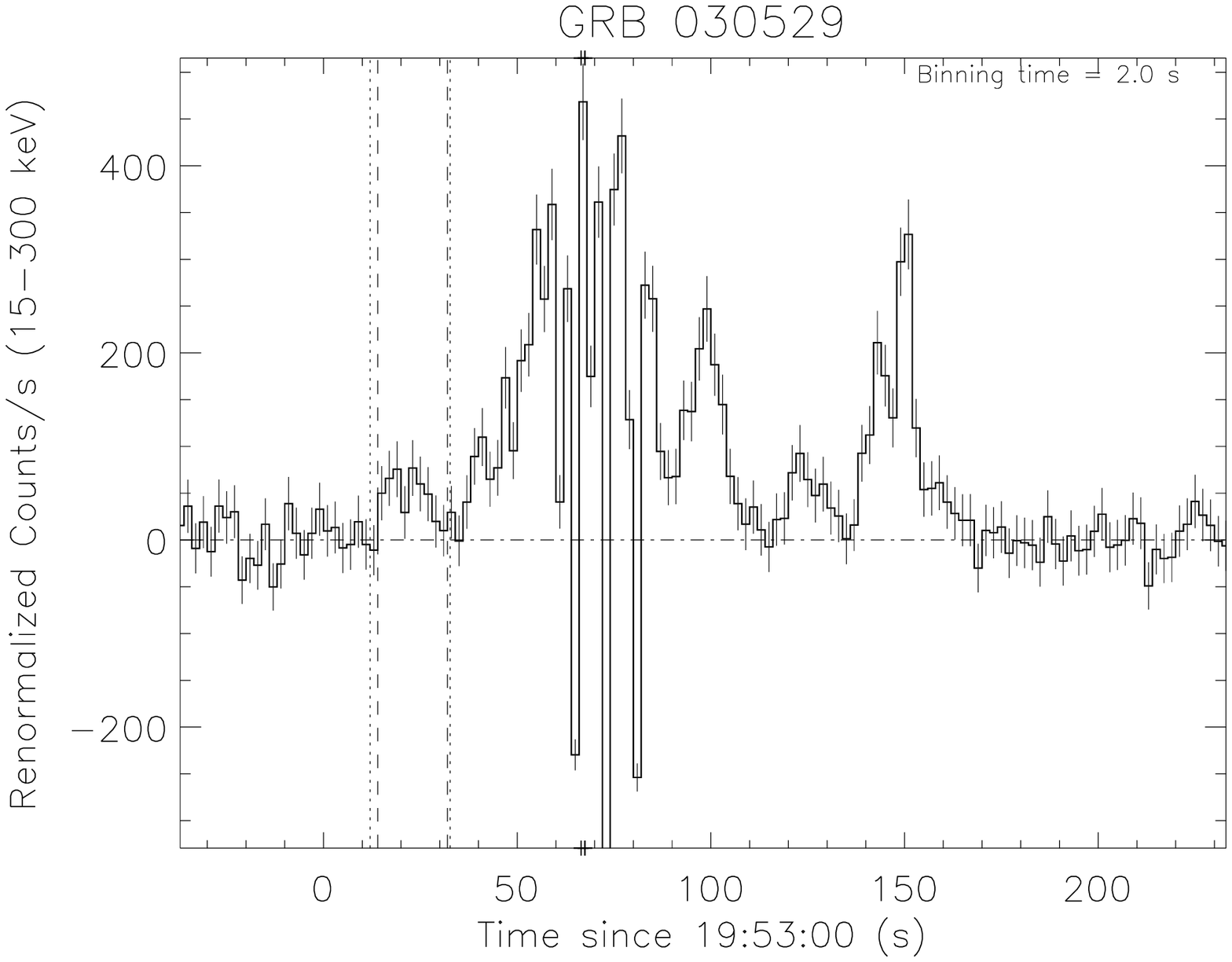,width=0.5\linewidth} \\
\end{tabular}
\caption{Light curves of \int\ GRBs}
\label{fig:light_curves}
\end{figure*}

To extract the GRB light curves, as well as for most of our timing
analysis, we used the software tools developed for the IBAS
interactive quick-look analysis. These tools are more suitable than the standard  software (the \textit{Off-line Science Analysis} software, OSA) for this kind of analysis.
The background subtracted light curves, in the 15-300 keV range,
are shown in Fig. \ref{fig:light_curves}. In order to obtain a better
signal to noise ratio, they were extracted  only from detector
elements where the Pixel Illuminated Fraction (PIF) was greater
than $0.5$. The background level is estimated with a linear fit to
time intervals of different durations appropriately chosen before
and after each GRB. The light curves have different bin sizes,
appropriate to the counts statistics, and the plotted count rates
are normalized taking into account the detector fraction
illuminated by the GRBs at  the different off-axis angles. All
these light curves are corrected for the instrumental dead time,
that increased from  $\sim$20\%  at the beginning of the mission
to the current value of $\sim$30\%. We have applied this
correction via a multiplicative factor computed bin-by-bin, in
order to avoid distortions due to dead time variations during the
GRBs (for example in the case of GRB 041219 the dead time varied
from 25\% to 40\% during the main peaks). Some gaps due to
telemetry saturation occurring during the brightest parts of the
bursts are visible in the light curves. Unfortunately there is no way
to correct for this effect, since no information on the
corresponding events is stored onboard or sent to earth. We have
used the light curve of each GRB to determine its $t_{90}$, the duration over
which 90\% of the fluence of the burst is observed. These
intervals are indicated by the dashed lines in Fig.~\ref{fig:light_curves}, while the dotted lines indicate the time interval from which we extracted the spectrum, when different from $t_{90}$. In Fig.\ref{fig:t90_distr} we report the distribution of the durations for the 43 GRBs we have analyzed in this work.
\subsection{Spectral extraction}

In this work we have considered for the spectral analysis only the
45 bursts observed until March of 2007, since the data of the remaining
ones are not yet publicly available. For two GRBs (GRB 021125 and GRB 030131) we were not able to perform the analysis (see Section \ref{sec:results}), so our sample is composed of 43 GRBs.
For the imaging and spectral analysis we used the latest release
of the standard software (OSA 7.0) distributed by the ISDC. The
time integrated spectra of the faintest bursts were extracted from
the $t_{90}$ time intervals. For brighter bursts we use longer time intervals, indicated by
the dotted lines in Fig.~\ref{fig:light_curves}. For the spectral extraction we
took into account the presence of all the other sources in the
field of view detected at more than 5 $\sigma$.  This is crucial
in the analysis of coded mask instruments with a wide field of
view, where many detector pixels are exposed to the flux of
several sources.
Spectra were usually extracted in 64 energy bins between 13 keV
and 1 MeV, and corrected for the instrumental dead time and for
the variation of the mask transparency as a function of the GRB
off-axis angle. When the GRB was too weak, or when other bright
sources were present in the field of view,  we  used only 30
energy bins between 13 and 500 keV.
The count spectra  were then fitted using XSPEC v. 11.3
\citep{xspec}, after excluding  the energy bins below 18 keV, for
which an accurate calibration is not available yet. We have also added a 3 \% systematic error to account for the calibration uncertainties. For the GRBs
affected by telemetry saturation, we assumed that they kept their average spectrum and flux during the time gaps.

We considered the following  spectral models:
\begin{itemize}
\item   power law (PL) with photon index $\alpha$:
\begin{equation}
f(E)=AE^{-\alpha}
\end{equation}

\item   power law with exponential cutoff (CPL):
\begin{equation}
f(E)=AE^{-\alpha}e^{-E/E_0}
\end{equation}

\item Quasi-thermal model (QT) \citep{Ryde2005,Thompson2007,Ghirlanda2007}, consisting of a power law and a blackbody:
\begin{equation}
f(E)=AE^{-\alpha}+ B \frac{E^{2}}{(kT)^{4}(e^{\frac{E}{kT}}-1)}
\end{equation}
\end{itemize}

We also performed some fits using the Band model, consisting of
two power laws joined with continuity \citep{Band1993}, which is
widely used to fit GRB spectra. However, we do not report the
results here since these fits to our ISGRI data could not constrain the
values of the high-energy photon index. This means in practice
that the Band model reduces to the CPL model when used in an energy band as small as ours.
\begin{figure}[t!]
\centering
 \psfig{file=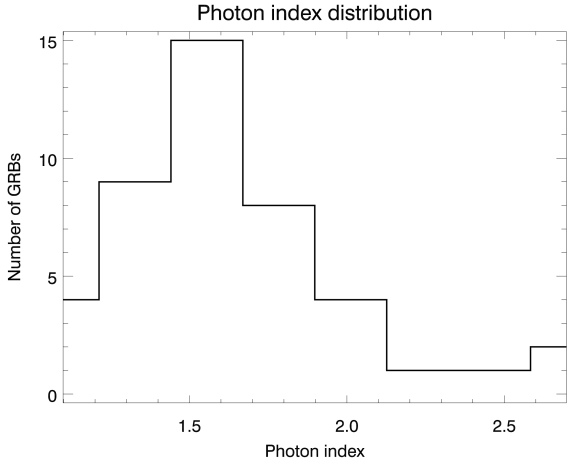,angle=90,width=0.9\linewidth}
 \caption{The distribution of the photon index of the single power law fits. All but two spectra are well described by a power law and are given in this plot, although for 9 of them a CPL or a QT model give a better fit.}
 \label{fig:photindex}
\end{figure}

\section{Results}
\label{sec:results}
\begin{figure*} [t!]
\centering
\begin{tabular}{cc}
\psfig{file=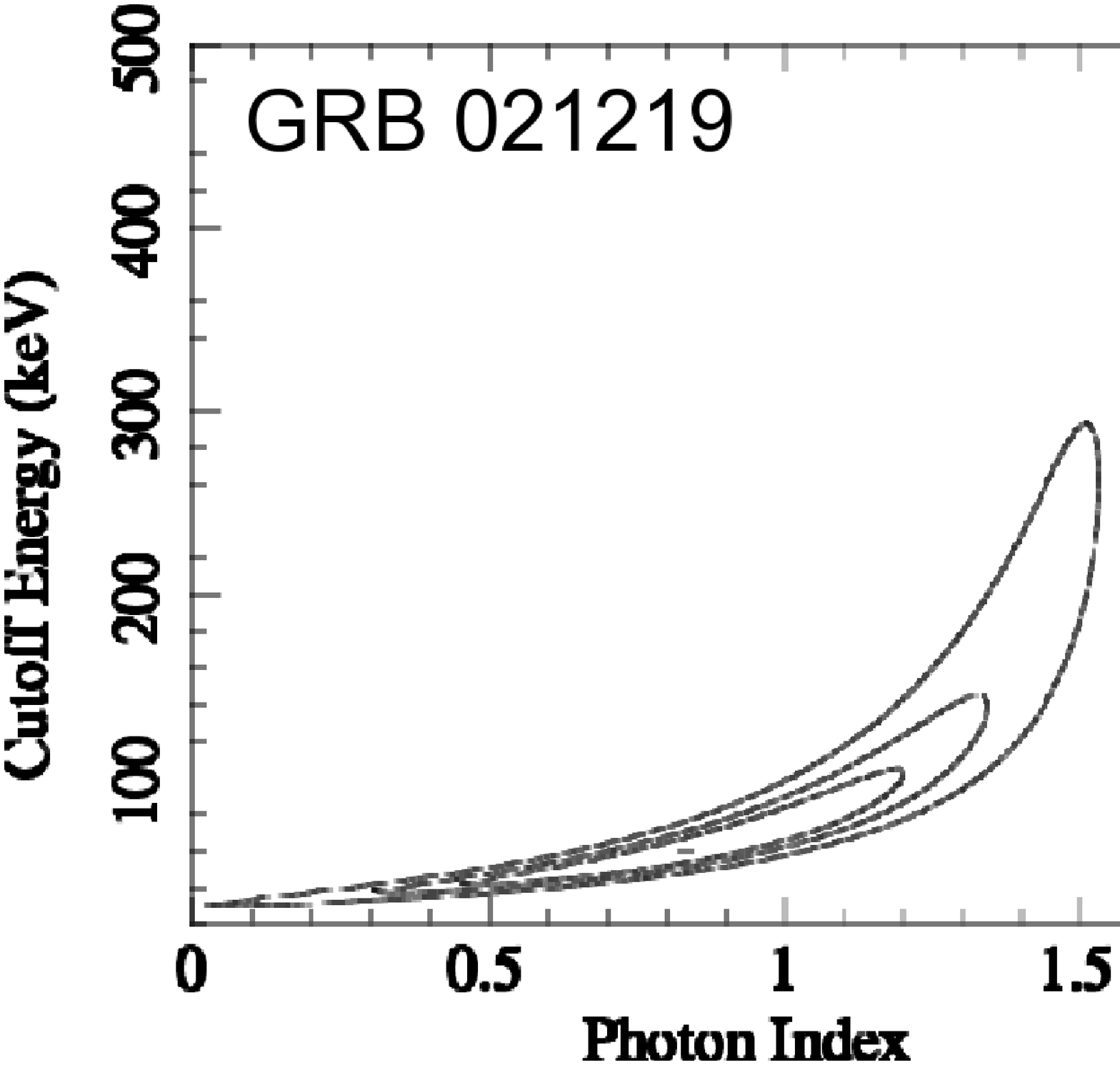,width=0.33\linewidth}
&
\psfig{file=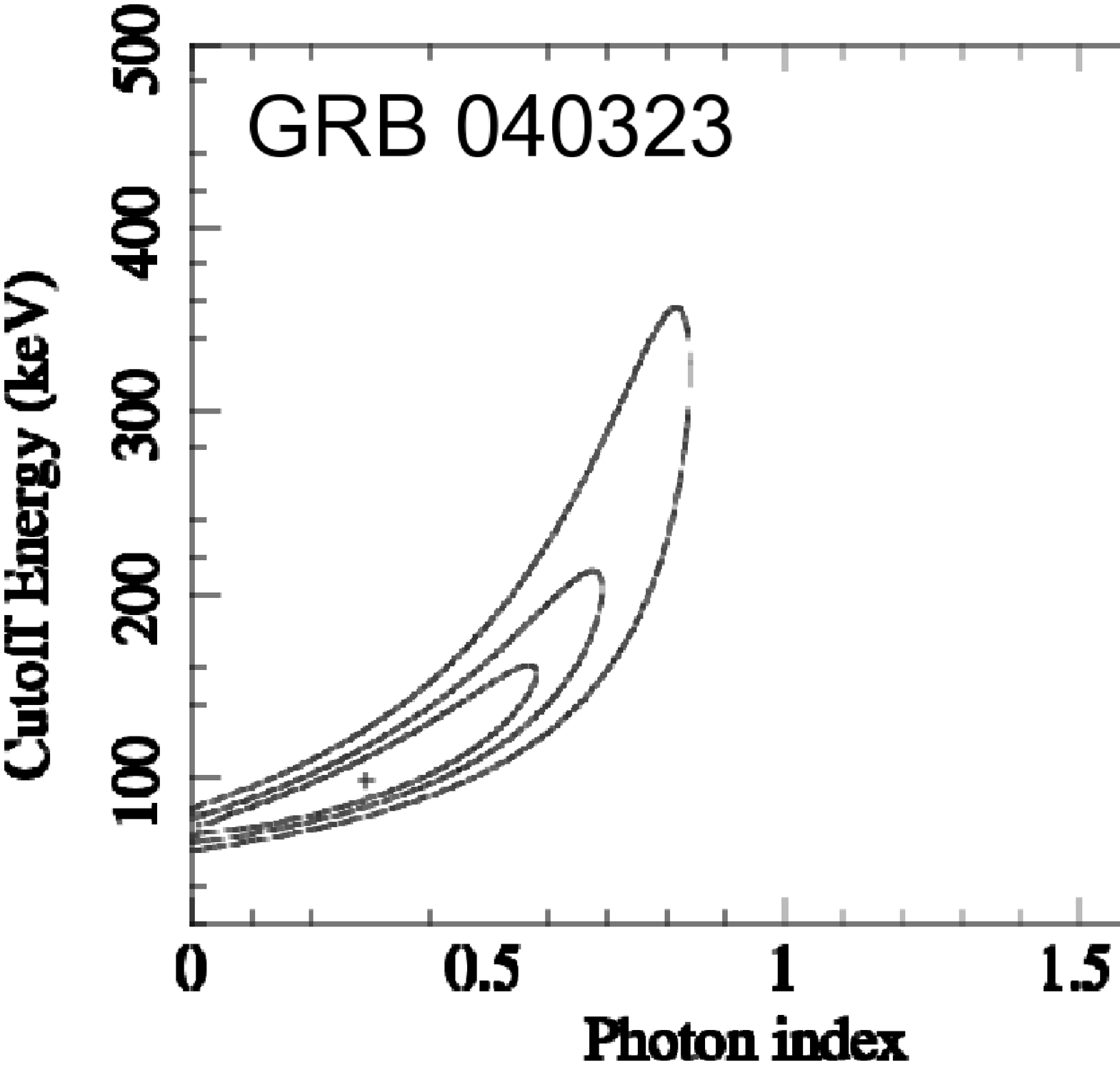,width=0.33\linewidth} \\
\psfig{file=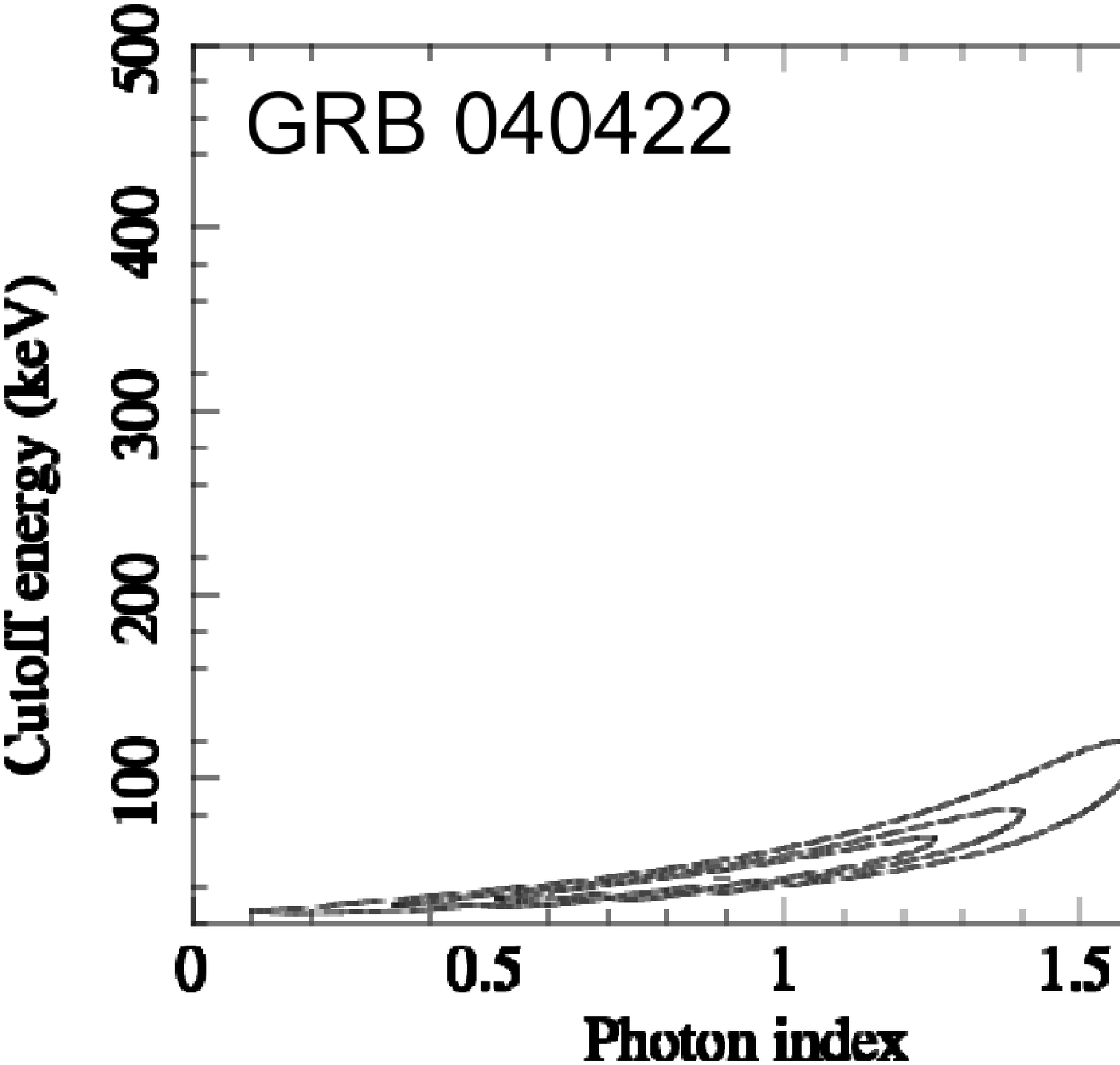,width=0.33\linewidth}
&
\psfig{file=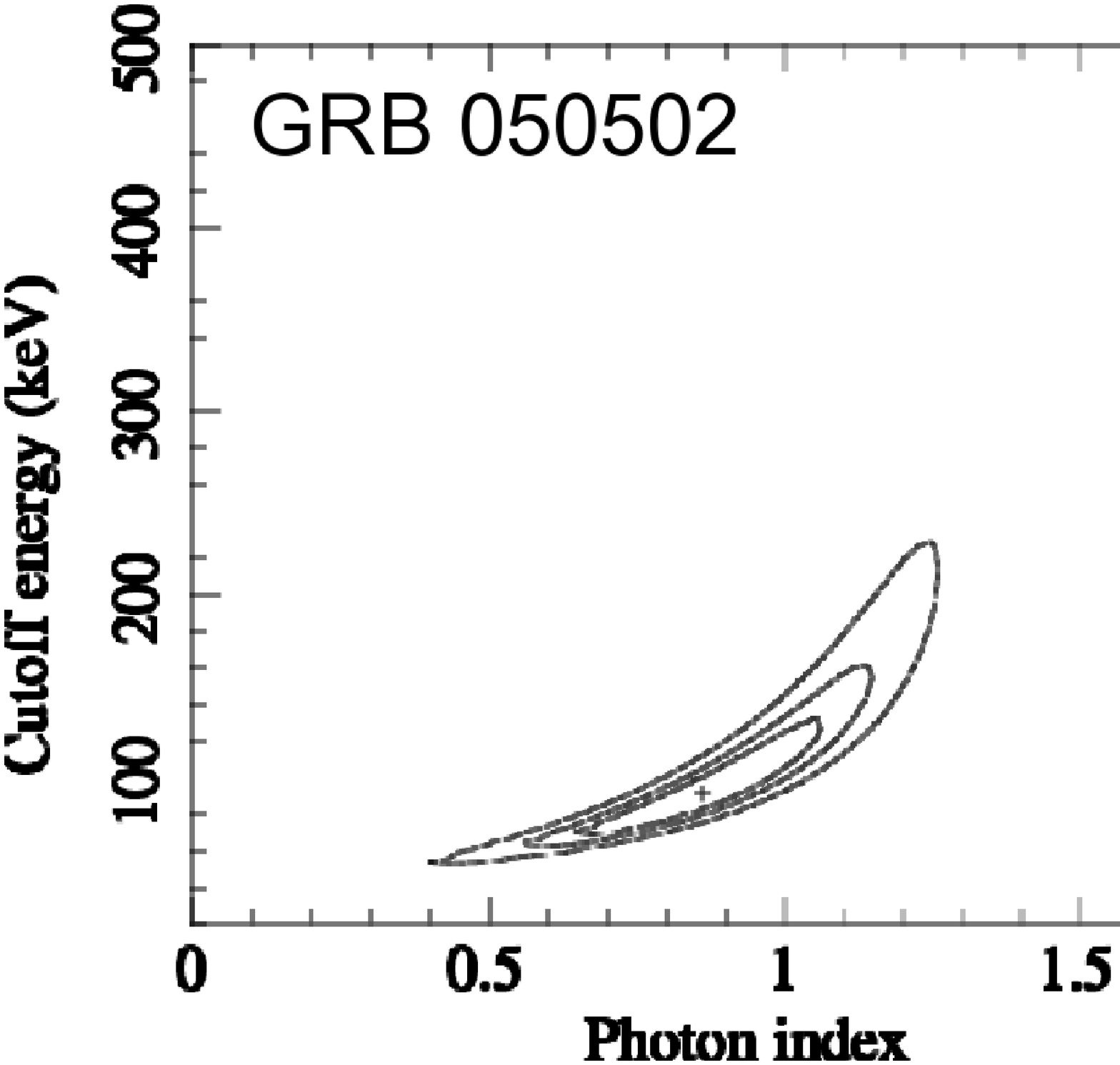,width=0.33\linewidth} \\
\psfig{file=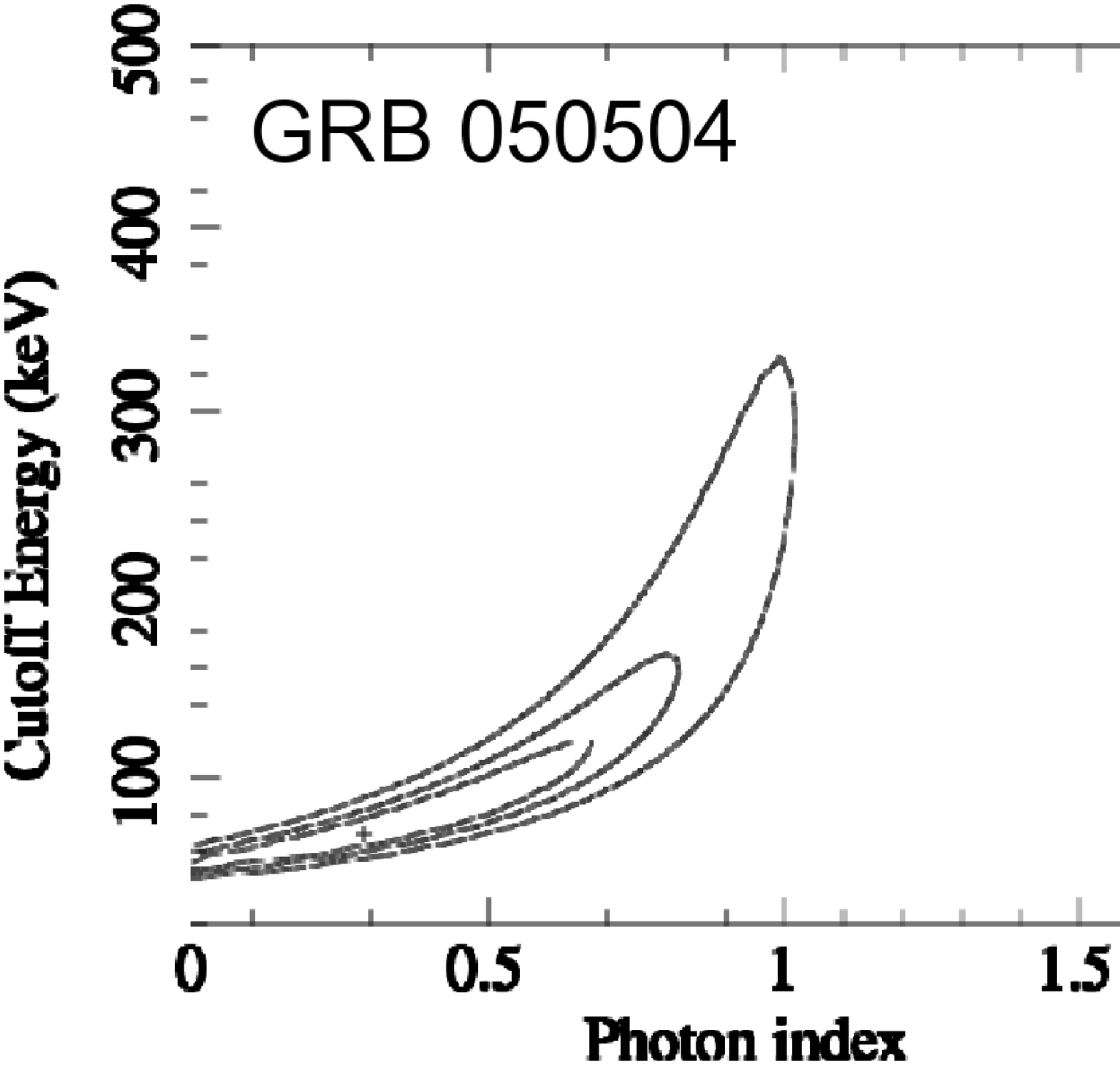,width=0.33\linewidth}
&
\psfig{file=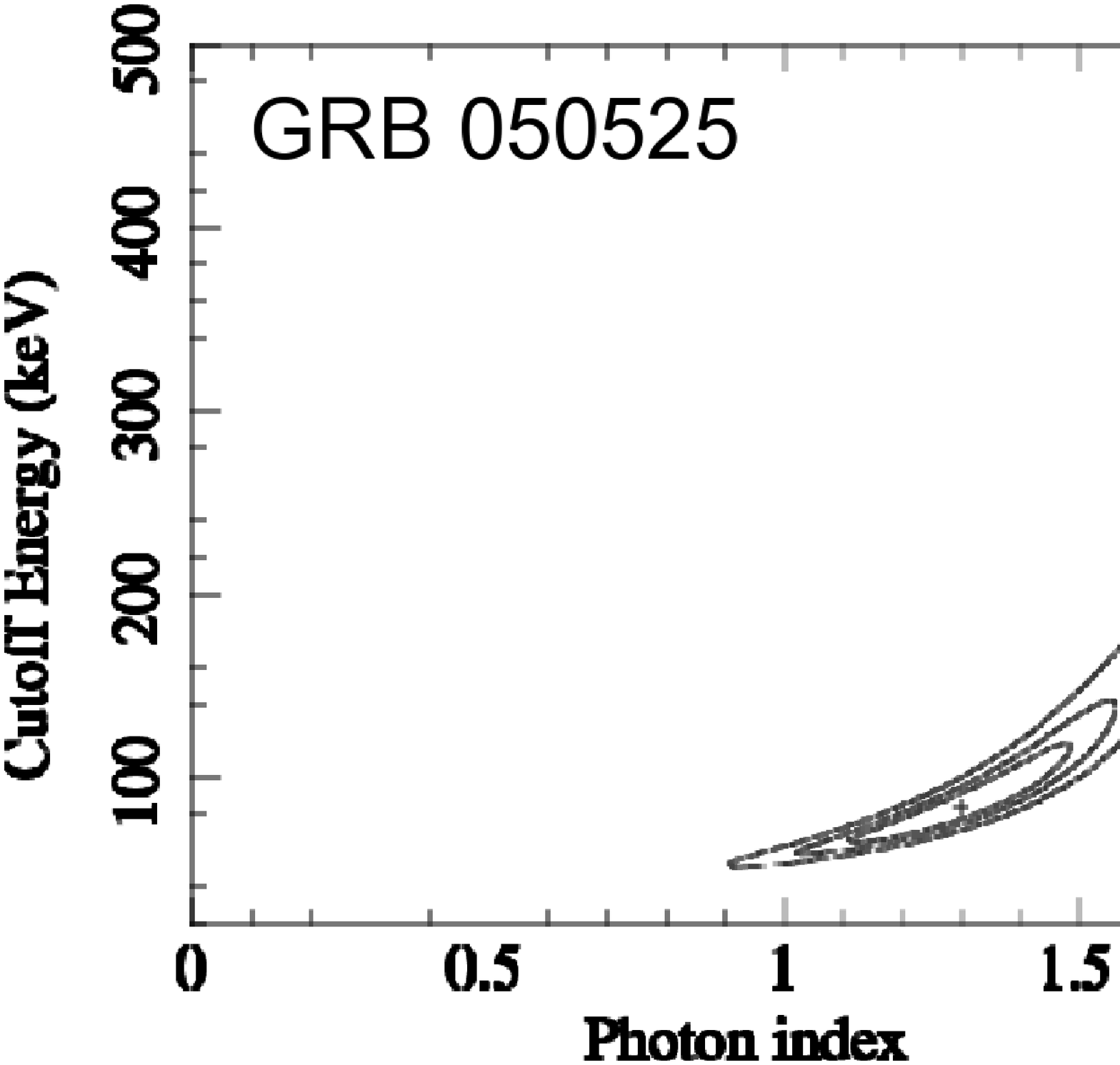,width=0.33\linewidth} \\
\psfig{file=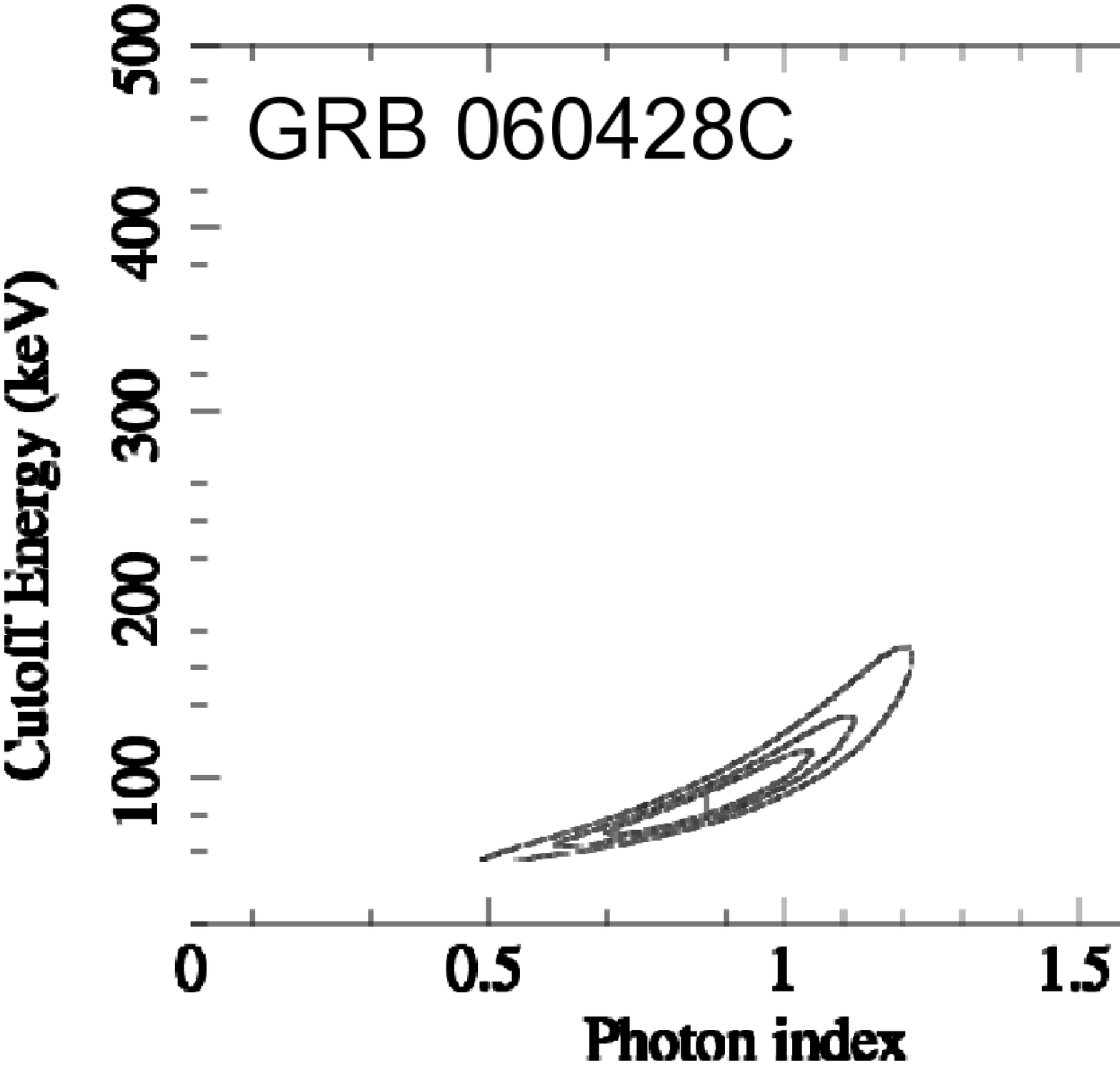,width=0.33\linewidth}
&
\psfig{file=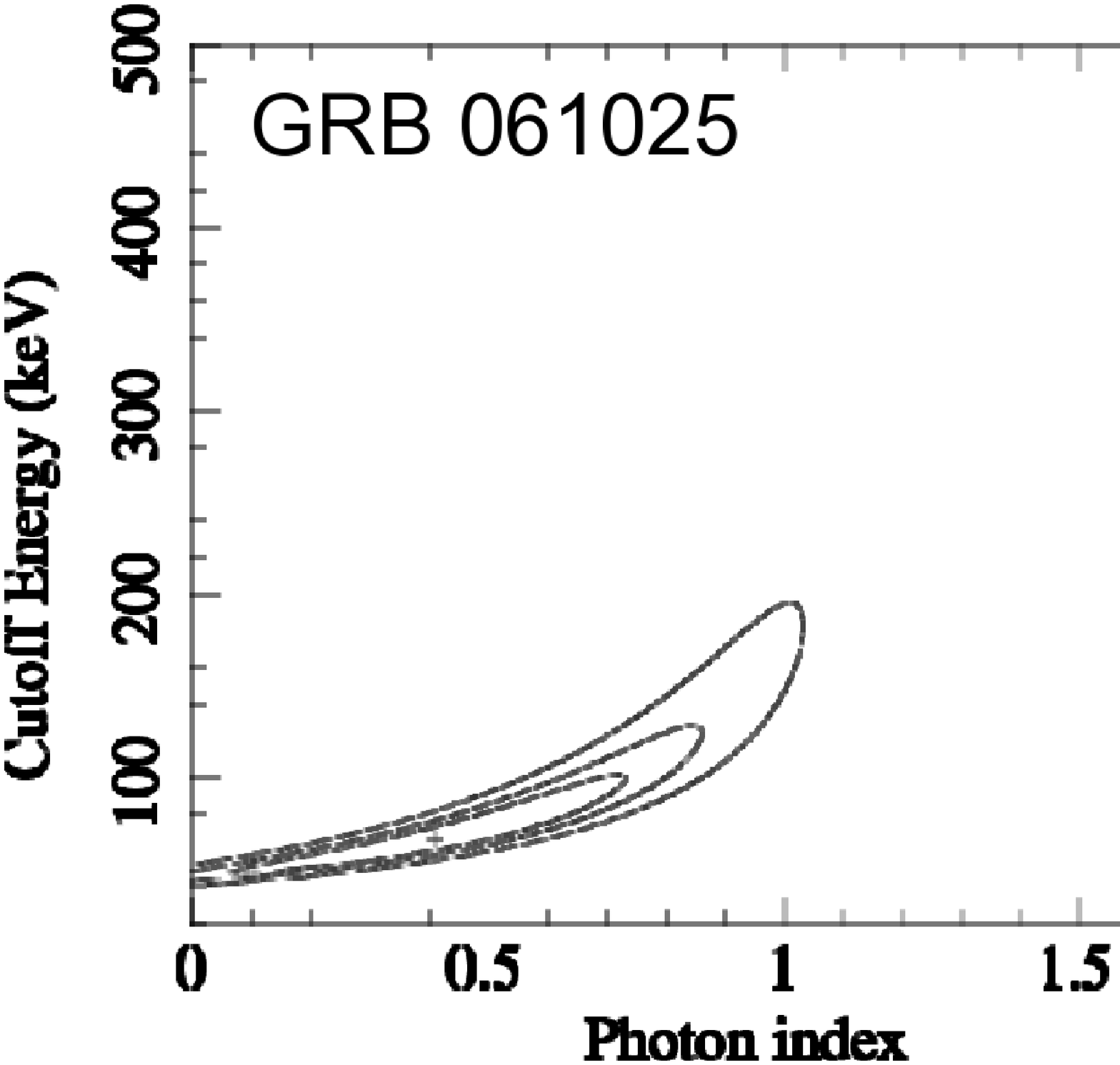,width=0.33\linewidth} \\
\psfig{file=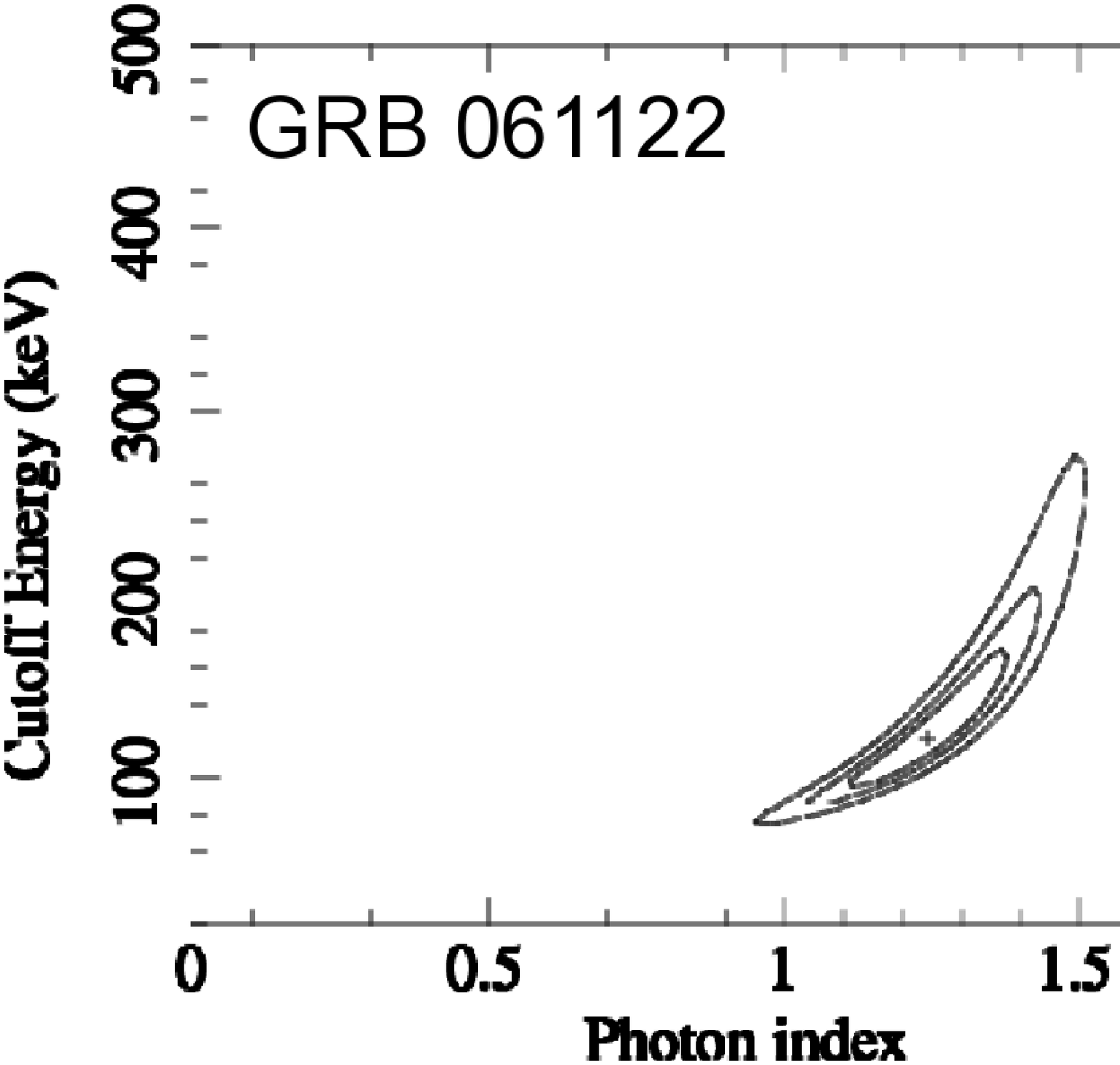,width=0.33\linewidth} & \\
\end{tabular}
\caption{Contour plot of the parameters of the CPL model. The lines
are respectively 68\%, 90\% and 99\% confidence levels. To make
comparisons easier the axes are the same in all plots.}
\label{fig:cut_contours}
\end{figure*}

The spectral results for the power law model are summarized in
Table \ref{tab:powerlaw}. These fits are in general acceptable and give photon
index values around $\sim$1.6 (see Fig.~\ref{fig:photindex}). 
Most of the burst spectra that result in unacceptable fits with this
model give better results with the introduction of an exponential
cut-off. The resulting best fit parameters are reported in Table
\ref{tab:cutoff}, where also the F-test results are given to show the
significance of the fit improvement. The confidence contours of
the photon index and cut-off energy are shown in
Fig.~\ref{fig:cut_contours}.
Almost all the bursts showing evidence for a curved spectrum in the ISGRI
energy range can also be fitted with the QT model, giving the best
fit parameters listed in Table~\ref{tab:qt}.

\subsection{GRB 021125}
This burst, the first one discovered by IBIS, was observed during
the \int\ Performance and Verification phase, while the
instruments were not yet fully operational. In particular, during
the observation of this burst, ISGRI could transmit to ground only
a limited amount of information because the majority of the
telemetry allocation was used for other data.
For this reason, the ISGRI data on GRB 021125
cannot be easily analyzed with the standard software and require
an ad hoc treatment, as reported in \citet{malaguti2003}. We have
not considered this burst in our work, but,  for completeness, we
give in Table \ref{tab:observative} the results obtained by
these authors.
\subsection{GRB 021219}

\begin{figure}[t]
\centering
\psfig{figure=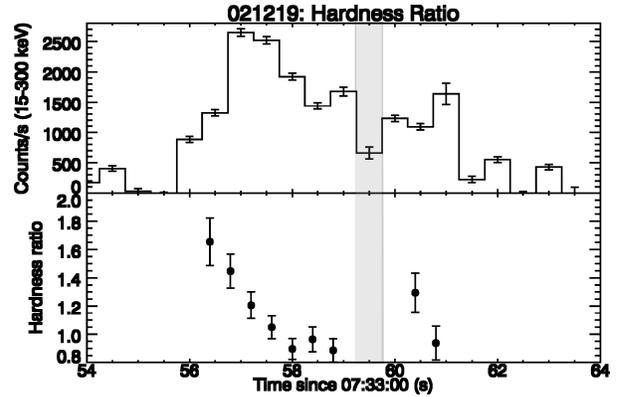,width=0.9\linewidth}
\caption{Spectral evolution of GRB 021219: the hard-to-soft
evolution is clearly visible. The hardness ratio is defined
as H/S, where H is the count rate in the 50-300 keV range and S is
the count rate in the 15-50 keV range. The gray region is affected by a telemetry gap, so no meaningful hardness ratio could be computed.}
\label{fig:021219_hr}
\end{figure}

The results of this GRB, the first to be detected and localized
by the IBAS software, were first reported by
\citet{mereghetti2003}.  By comparison with the Crab, due to the
lack at that time of a proper calibration, they found a power law
photon index $\alpha =2$ for the average spectrum, with evidence
for a hard-to-soft evolution.
We obtain a consistent value for the fluence, but a slightly
harder spectrum (see Table \ref{tab:powerlaw}). Moreover, thanks
to the improved spectral calibration now available, we have found
convincing evidence of a spectral break at $E_0=50^{-11}_{+66}$
keV. Indeed the fit with the CP model results in a significant
improvement, with an F-test chance probability of only $4\times
10^{-4}$ (see Table \ref{tab:cutoff}).
We confirm the spectral evolution, as shown by the hardness ratio
reported in Fig.~\ref{fig:021219_hr}.

\subsection{GRB 030131}
This burst, lasting about two minutes, was partly observed during
a satellite slew. The slew started at 07:39:09 UT, 20 s after the
beginning of the burst, and resulted in a variation of
$\sim2^{\circ}$ of the GRB position in the ISGRI field of view
during the source activity period. The OSA software cannot be used to analyze data collected during slews, so we could not analize this GRB. We refer to the work of \cite{Gotz2003} for a proper analysis using the IBAS software. 

\subsection{GRB 030227}
This is the first \int\ GRB for which  X--ray and optical
afterglow searches were successful
\citep{Mereghetti2003b,castrotirado2003}. Evidence for intrinsic
absorption, as well as possible emission lines, were reported in
the spectrum of its X--ray afterglow
\citep{Mereghetti2003b,watson2003}. The ISGRI spectrum reported
here is slightly harder, but still consistent with that derived by
 \cite{Mereghetti2003b} using preliminary calibrations.

\subsection{GRB 030320}
This burst shows three different emission episodes, with several
peaks. The average spectrum is poorly fitted by a power law
($\chi^{2}_{red} = 1.8$ with 24 d.o.f), and also the CP (or Band) model
and the QT model do not give better fits. On the contrary, separately fitting the three emission episodes, we found that each of them
is well described by a  power law. The corresponding photon
indexes are $1.27 \pm 0.08$ ($\chi^{2}_{red}=1.39$ with 39 d.o.f.), $2.01
\pm 0.3$ ($\chi^{2}_{red}=1.41$ with 34 d.o.f.) and $1.49 \pm 0.1$
($\chi^{2}_{red}=1.09$ with 24 d.o.f.). The first peak is the hardest, then there is a second softer peak, and a third hard peak again.
A similarly complex spectral evolution has been seen in a few
other GRBs, for example  GRB 060124 \citep{Romano2006}.
\cite{Foley2008} reported a
time dependent spectral lag: $0.33 \pm 0.03$ s for the first peak,
and a value close to zero ($0.05 \pm 0.03$ s) for the third one.
No spectral lag could be measured for the second peak, because it
is observed only in the softer band. We can speculate, in the
context of the fireball model, that the second peak could be the
onset of the afterglow, i.e. it could be due to the external shock
caused by the contact between the circumstellar medium and the expanding fireball, resulting from the merging of the two shells that
produced the first peak. The time
between the onset of the first peak and the onset of the second
one ($\Delta t\simeq 30$ s) is compatible with this hypothesis, provided we are in the thin shell case \citep{SariPiran1999}, as expected when the prompt and the afterglow phases are distinct: using a typical Lorentz factor for the expanding shell ($\Gamma \sim 100$), the start of the emission from the external shock would have taken place at the deceleration radius $R = 2\Gamma^{2} c \Delta t \simeq 1.8 \times 10^{16}$ cm from the center, that is a typical value in this scenario \citep[see for example ][]{1999PhR...314..575P}.\\
Our spectral results on GRB 030320 differ from those reported by
\citet{vonkienlin2003}, who found systematically softer spectra.
This is probably due to the inadequacy of the preliminary
calibrations they used for this burst at a large off-axis angle
(15.7$^{\circ}$).

\subsection{GRB 030501}
This  single-peaked GRB  was seen  in the partially coded field of
view, at $\sim 13^{\circ}$  from the instrument axis. Due
to the very large off-axis angle, only 10\% of the detector plane
was   illuminated by the source. The spectrum of GRB 030501 is
well fitted by a power law, with no evidence for spectral cutoff
or spectral evolution. While these could be intrinsic properties
of this burst, it is also likely that they are simply due to the
poor statistics of the data. Our results are fully compatible with
those reported by \cite{Beckmann2003}.

\subsection{GRB 030529}
GRB 030529 was discovered during an off-line re-analysis of the
\int\ data,  performed with an updated version of the IBAS
tools, more sensitive than those used in real time during the first months of the
mission. GRB 030529 was a 20 s long, faint
burst, that occurred during a period of highly variable
instrumental background (see  Fig. \ref{fig:light_curves}).
This explains why it was missed by the first version of the IBAS  system.
The spectrum we extracted is well fitted by a power law with a
very soft photon index $\alpha=3.5^{-0.43}_{+0.51}$. Thus GRB
030529 belongs to the class of X-Ray Flashes (XRF).
\subsection{GRB 031203}
GRB 031203 is among the most interesting GRBs discovered by
\int\ . Its afterglow  was observed at  X-ray
\citep{2004ApJ...605L.101W}, IR/Optical \citep{2004ApJ...608L..93C},
and radio \citep{Frail2003GCN} wavelengths, and spectroscopic
evidence of an associated Type Ic Supernova was found
\citep{2004ApJ...609L...5M}. The discovery of its
host galaxy led to a redshift determination of z=0.106
\citep{2004ApJ...611..200P},  implying a surprisingly small
isotropic-equivalent  energy E$_{iso}$=(6--14)$\times10^{49}$ erg. This value and the measure of E$_{peak}=151 \pm 50$ keV,
derived from the average spectrum as observed by Konus-Wind
\citep{Ulanov2005}, make GRB 031203 an outlier in the
E$_{peak}$-E$_{iso}$ relation \citep{Amati2006}.\\
The X--ray images obtained with \textit{XMM-Newton} led to the
discovery of  expanding  rings due to the scattering of the GRB
X-ray emission by dust grains in our Galaxy \citep{Vaughan2004}.
The modelling of this GRB ``echo'' provides an indirect means to
estimate the intensity of the prompt GRB  emission at  X--ray
energies. This gives some evidence for an X-ray flux component in
excess of the low-energy extrapolation of the \int\ spectrum
\citep{Vaughan2004,TM2006}. Taking into account this phenomenon
and other effects, GRB 031203 could be made consistent with the Amati correlation \citep{Ghisellini2006} . \\
The spectrum we extracted is well described by a power law. No
improvements are achieved by using more complicated models, and we
found no evidence of a spectral break. Our spectral parameters
basically confirm the results given by \cite{Sazonov2004}: their values for the fluence ($2.0 \times
10^{-6}$ erg cm$^{-2}$) and for the photon index ($\alpha = 1.63 \pm
0.06$) are slightly different from ours, but this differences
could easily be due to the much better calibration we used. Using
Konus-Wind data, \cite{Ulanov2005} found a cutoff around 150 keV.
We can obtain only a lower limit on the cutoff ($E_{0} > 100$ keV, 90 \% c.l.), because the fluence at high energy is too small.

\subsection{GRB 040106}
This GRB is composed of two peaks separated by  a quiescent period
of ~30 s. To optimize the signal to noise ratio, we  extracted the
average spectrum  removing the quiescent period.
The best fit model is a power law: no cutoff is detected. Our
photon index is a little harder (but compatible) with that
reported by \cite{Moran2005}.\\
We found evidence of spectral evolution between the two peaks.
Their spectra are well described by a power law, with a photon index respectively of $\alpha=1.80 \pm 0.14$
($\chi^{2}_{red}=1.05$ with 46 d.o.f) and $\alpha=1.41 \pm 0.17$
($\chi^{2}_{red}=1.35$ with 46 d.o.f). This soft-to-hard evolution, also
noted by \cite{Moran2005}, is not usual. We note however that the
quiescent time between the two peaks is quite long compared to the
duration of the peaks, so that the two emission episodes are 
distinct and in principle could be uncorrelated.
This
is not the only peculiarity of this GRB: the X--ray afterglow,
promptly observed with \textit{XMM-Newton}, decays as expected like a
power law with an index of 1.46$\pm$0.04, but is spectrally harder
than usual. This behavior is difficult to describe with a normal
fireball model, and requires ad-hoc adjustments \citep{Moran2005}.

\subsection{GRB 040223}
This faint GRB occurred in the Galactic plane (b=3.2$^{\circ}$) at only
19$^{\circ}$ from the Galactic center, in a region where there are
several bright sources. Three low-mass X-ray binary systems were detected by ISGRI with
a significance above $5\sigma$ in the time interval of the GRB, namely 4U 1636--536, H 1705-440, and GX 340+0. The presence of these sources has been properly taken into account
in the spectral analysis. GRB 040223  is composed by
three peaks, with a soft integrated spectrum well fitted by a
power law.
The lack of adequate statistics prevents us from testing more
complex models or detecting any kind of spectral evolution. Our
results are fully in agreement with those reported by
\cite{Filliatre2006}.\\
This burst was observed by \textit{XMM-Newton} 5 hours after the
trigger, and a fading afterglow was detected \citep{Tiengo2004GCN}.

\subsection{GRB 040323}
This GRB was single-peaked and quite intense. Nevertheless it is
apparently a dark burst, as no afterglow was detected in the
following hours and days. A tentative detection of an optical
counterpart \citep{Gal-Yam2004GCN} has not been
confirmed. The spectrum of the burst is well described by a CPL
model, with a cutoff energy  $E_{0}=120^{-39}_{+84}$ keV (see
Table \ref{tab:cutoff}). The F-test gives a probability of only $2
\times 10^{-4}$ that the improvement obtained using the CPL model instead of the power law is obtained by chance. We report a
contour plot of the CPL fit parameters in Fig. \ref{fig:cut_contours}. \\
To our knowledge, there is no other published work about this burst made
using ISGRI data. \cite{Foley2008} used SPI data, fitting the spectrum of GRB 040323 with a power law. They found a best-fit photon index of $1.44 \pm 0.18$. They could not detect the cutoff, due to the poor statistics of their spectrum. As a consequence they found a fluence slightly higher than ours.

\subsection{GRB 040403}
This burst is quite faint and soft, with an integrated spectrum
well fitted by a power law with a photon index
$1.84^{-0.15}_{+0.16}$. Our results   agree  with those
reported in \cite{Mereghetti2005}, who found a  slightly higher
fluence, probably due to the  non optimal calibration used at that
time. We did not detect  significant spectral evolution.
Unfortunately this GRB happened during a full-Moon night, so
optical follow-ups were difficult and no optical afterglow was
detected. Nevertheless, a quite deep limit in magnitude was
obtained 17 hours after the trigger ($R > 24.2$,
\cite{Mereghetti2005}), indicating a rather faint afterglow, as
seen in other soft and faint bursts. No X-ray or radio follow-up
observations were performed.

\subsection{GRB 040422}
This GRB was seen 9.4$^{\circ}$   off-axis, but it was bright
enough to obtain a good quality spectrum that showed  evidence of
a cutoff at $E_{0}=45^{-13}_{+25}$ keV (see Table
\ref{tab:cutoff}). The F-test gives a probability of only $10^{-5}$
that the improvement of the fit with respect to the power law is obtained by chance. Our
results are compatible with those reported by
\cite{Filliatre2005}: they found a similar cutoff value, but they
used the Band model fixing the high-energy photon index to $-4$.
We have also obtained a good fit using the QT model, with parameters fully compatible with those reported by \cite{Foley2008} but with smaller errors. The F-test is not applicable here to
compare this fit with that obtained with the power law \citep{2002ApJ...571..545P}. We obtain
the same $\chi^{2}_{red}$ using both the CPL model and the QT
model, but the latter model requires one more parameter. \\
The afterglow was detected by \cite{Filliatre2005} in the near infrared 2 hours after the burst. They also identified the host galaxy, but it is too faint to measure the redshift. 

\subsection{GRB 040624}
This burst was observed 12$^{\circ}$ off-axis, so that, despite
its intrinsic brightness, it resulted in  quite a low signal to
noise level. Its spectrum is well described by a power law, with a
photon index fully compatible with that   reported by
\cite{Filliatre2006}. These authors derived a duration for the
burst of $T_{90}=46$ s, while we found $T_{90}=27$ s. Moreover
they reported a fluence nearly two times higher than ours. We do not understand the  reason for these differences, but we think they could be due to an inadequate calibration. \\
This GRB was promptly observed from the ground, but no optical
afterglow was detected down to the magnitude $R > 23.8$  only half
a day after the trigger, despite the very low Galactic absorption
present in the burst direction. Thus this GRB could be
intrinsically dark \citep{Filliatre2006}.

\subsection{GRB 040730}
This GRB was located very close to the Galactic plane, and only
$\sim$34$^{\circ}$     from the Galactic center direction.
In this direction the extinction  is very intense, so it is not
surprising that no afterglows were detected at any wavelength. The
spectrum of this burst is well fitted by a power law, with results
in full agreement with those reported by \cite{Foley2008}.

\subsection{GRB 040812}
This is the first burst that occurred in the field of view of the
JEM-X instrument. It was weak, but quite soft so that
it was well detected in the JEM-X energy band (3-35 keV).  We have analyzed
the JEM-X data  finding  the burst best position at coordinates
R.A.=246.463, Dec.=--44.714 (J2000), with a 2$'$  error radius
(90\% c.l.). This position is fully in agreement with the ISGRI
one. The spectrum we have extracted from ISGRI data is well
described by a power law.
Adding the JEM-X data does not change significantly the best fit
parameters, owing to the large statistical errors. An extensive
multi wavelength campaign was performed to search for an afterglow
candidate. Only the Chandra/ACIS X-ray telescope found convincing evidence of a decaying source inside the IBIS error
circle  \citep{Patel2004GCN,Campana2004GCN}. Its coordinates,
R.A.=246.5093, Dec.=--44.7304,  are  1.6$'$ away from the IBIS
position, and at 2.1$'$ from the JEM-X position. Following this
detection, \cite{2006astro.ph.12644D} found a decaying optical
source and a candidate host galaxy. Due to the lack of emission
features from the galaxy they could only derive a tentative
redshift range of $0.3 \le z \le 0.7$.

\subsection{GRB 040827}
This burst was rather faint and observed at an off-axis angle of
12$^{\circ}$, with only $\sim$10\% of the ISGRI detector surface
exposed to the source flux. Thus the trigger significance was
below the threshold for the automatic alert delivery. The burst
was confirmed by an interactive analysis and announced to the
scientific community about one hour later \citep{2004GCN..2670....1M}. Several
follow-up observations were carried out, leading to the detection
of the afterglow  in the X--ray band with XMM-Newton and in the
NIR with the VLT. The accurate NIR position (R.A.=229.25558,
Dec.=--16.14142, \cite{DeLuca2005}) is
only 0.6$'$  away from the refined ISGRI position.\\
The analysis of this burst is made impossible by the presence in the
IBIS field of view of the very bright and variable source Sco X-1.
This low-mass X-ray binary is detected at 14 $\sigma$ above the
background in the GRB time interval, and is only $7^{\circ}$ away
from the IBIS optical axis. The GRB is detected instead only 10
$\sigma$ above the background and is $12^{\circ}$ off-axis, so that the field is dominated by the flux of Sco X-1. The spectra of both the GRB and Sco X-1  extracted at the same time, as requested by the nature of the coded-mask telescope, are distorted and cannot be fitted with reasonably simple models. To better understand the problem we have extracted two spectra of Sco X-1 in two 
intervals before and after the GRB, lasting $\sim$30 s, like the GRB. Due to the short exposure, Sco X-1 is well
detected only in energy bins below 70 keV, while its contribution
is negligible above that energy. Both spectra are well
described by the best fit (averaged) model proposed by
\cite{DiSalvo2006}.  The spectrum extracted during the GRB is compatible with these ones only below 50 keV, while it shows a lack of counts in the 50-70 keV energy range. The GRB spectrum presents an excess in the same range. We conclude that the software is not able to properly disentangle the contributions of
the two sources around 50 keV, so that both spectra are
distorted. We have tried to reduce the number of energy
bins, and even to extract the spectra only around the peak of the
GRB (maximizing the signal to noise ratio), but without results.

\subsection{GRB 040903}
This relatively faint and spectrally soft burst came from a
direction close to the Galactic center (l=5.2$^{\circ}$,
b=--1.5$^{\circ}$). Due to the presence of a faint ROSAT X-ray
source in its error region,  the possibility that it was due to a
Type I X-ray burst from an unidentified low mass X-ray binary was
considered in the initial reports \citep{Gotz2004GCN}. However,
subsequent analysis \citep{Kuulkers2004GCN}, confirmed here,
indicated that a Type I X-ray burst origin is rather unlikely.
We therefore include this event in our sample of GRBs.
Indeed this burst is detected only below 100 keV and, although its
poorly constrained spectrum is consistent also with a blackbody,
there are several factors at variance with a type I X-ray burst
origin. The blackbody temperature of 6.7 $\pm$ 1.1 keV is much
higher than that typically observed in X-ray bursts and there is
no evidence for a spectral softening. The  ISGRI spectrum is also
well fit with a power law with photon index $2.9^{-0.39}_{+0.46}$.
Finally, no persistent emission from the
\textit{ROSAT} source has been detected by \int\ before or
after the burst. Thus we conclude that the nature of this burst is
that of an X-ray flash.

\subsection{GRB 041015}
This faint GRB was detected $\sim 10^{\circ}$ off-axis, and its significance was below the threshold for the
automatic delivery of the alert message. A
manual alert was distribuited about 1 hour
later \citep{2004GCN..2805....1M}. Despite the follow up observations carried out by optical
and infrared observatories, no afterglow candidates were detected.

\subsection{GRB 041218}
The  spectrum of GRB 041218 is  well described by a power law.
\cite{Foley2008} obtained peak flux and fluence values very similar to
those reported here, but they were unable to fit the spectrum with
a power law and obtained only marginally acceptable fits with the
CPL and the QT model. This is probably due to the use of old
spectral extraction software and calibration responses not
optimized  at large off-axis angles. In fact this bright burst was
at an off-axis angle of $\sim$13$^{\circ}$.
The light curve of GRB 041218 shows three main peaks, for which we
extracted individual spectra: they are well described by power
laws, with photon index respectively $\alpha_{1}=1.52 \pm 0.08$,
$\alpha_{2}=1.49 \pm 0.09$ and $\alpha_{3}=1.72 \pm 0.10$. While
the first and the second peaks are almost identical, the third
one is slightly softer. \\
An optical afterglow has been detected, and two breaks in the
temporal decay have been reported by \cite{Torii2005}.

\subsection{GRB 041219A}
Thanks to the long duration of this GRB, the IBAS alert was issued when the burst was still on going \citep{Gotz2004GCN2}. This allowed 
robotic telescopes to detect a prompt optical and IR flash 
\citep{2005Natur.435..181B,2005Natur.435..178V} whose position was consistent with the IBAS one. GRB 041291A is the longest and brightest GRB in the IBIS sample,
and could be studied in detail. For example, using SPI data a high degree of polarization of the prompt emission was reported for this burst, but with low significance \citep{2007A&A...466..895M}.
The burst is composed of three distinct emission episodes: two close precursor peaks, and then after a quiescent interval the main GRB emission. The spectra of the two precursors are well described by power laws, with a photon index respectively of $1.38 \pm 0.04$ ($\chi^{2}_{red}=1.46$ with 46 d.o.f) and $2.16 \pm 0.08$ ($\chi^{2}_{red}=0.5$ with 36 d.o.f). The spectral evolution is evident. Also the spectrum of the GRB is well described by a power law, with a photon index of $1.89 \pm 0.01$ ($\chi^{2}_{red}=1.12$ with 46 d.o.f). In order to study the spectral evolution, we have divided the GRB in two parts: the first one from 240 s to 356 s and the second one from 356 s to the end of the burst (see Fig.\ref{fig:light_curves}).
The corresponding spectra can be described with power laws, with a photon index of $1.62 \pm 0.02$ ($\chi^{2}_{red}=1.36$ with 46 d.o.f) and $2.17 \pm 0.02$ ($\chi^{2}_{red}=1.4$ for 46 d.o.f) respectively. Again the softening is evident. For a more detailed analysis of this peculiar event, and a comparison with the analysis of SPI data, see 
G\"otz et al. (2008, in preparation).

\subsection{GRB 050129}
This very faint GRB was below the significance level requested for
the automatic delivery of the IBAS alert. Its spectrum is quite soft, so it is revealed only below $~120$
keV. No afterglow candidates have been detected.

\subsection{GRB 050223}
This GRB was discovered by \swi  \citep{Mitani2005GCN}.
Just by chance \int\ was observing in the direction of this
burst, that was thus detected by the IBIS instrument at an
off-axis angle of 14$^{\circ}$. The burst triggered the IBAS
programs, however the alert was not automatically distributed,
because the burst occurred during a short time interval (200 s) in
which  one of the 8 modules composing the ISGRI detector was
switched off (this is done autonomously by the on board software
handling the noisy pixels). Under these circumstances there is the
possibility that IBAS incorrectly assigns to the GRB the
coordinates of one of its ghost images, so automatic alert distributions are disabled. In the off-line analysis
it was possible to take these effects into account.
Our results are fully compatible with those obtained by \cite{Page2005} with \swi\ data.\\
Following the \swi /BAT trigger and the \swi /XRT detection
\citep{Giommi2005GCN}, several follow-up observations were carried
out. An afterglow was detected with XMM-Newton in the X--ray band
\citep{DeLuca2005GCN},  but not at optical/NIR wavelengths. A
galaxy at redshift  $z=0.584$ is contained in  the 1.5 arcsec
radius error region provided by XMM-Newton \citep{Pellizza2006}.
Assuming that this is the GRB host galaxy,  the isotropic energy
output can be estimated as $~10^{51}$ erg, a quite low value.
However, the probability of a chance superposition between the
putative host galaxy and the X-ray error circle is $\sim$19\%
\citep{Cobb2008}.

\subsection{GRB 050502}
\begin{figure*}[tb]
\centering
\begin{tabular}{cc}
\psfig{file=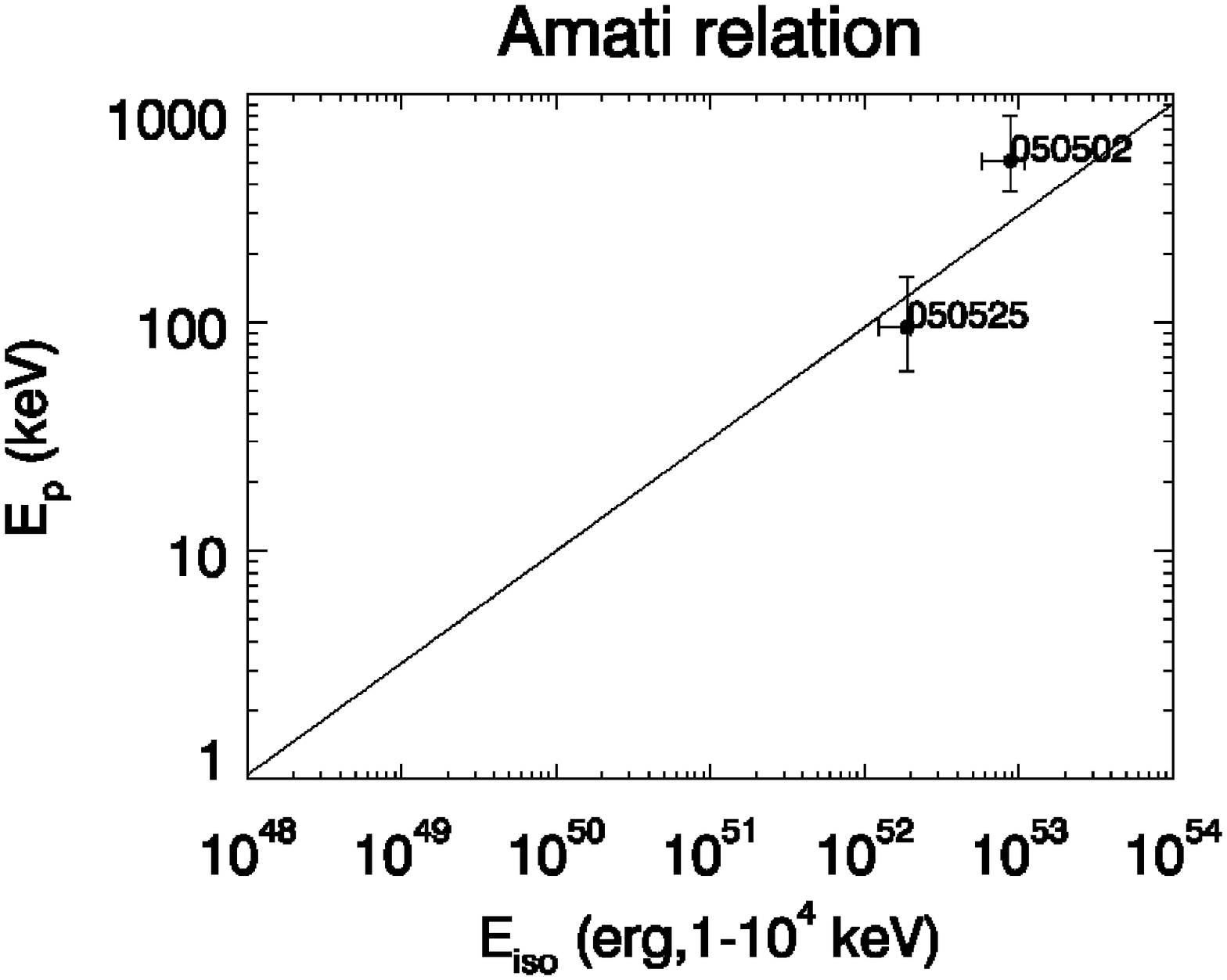,angle=90,width=0.47\linewidth}
&
\psfig{file=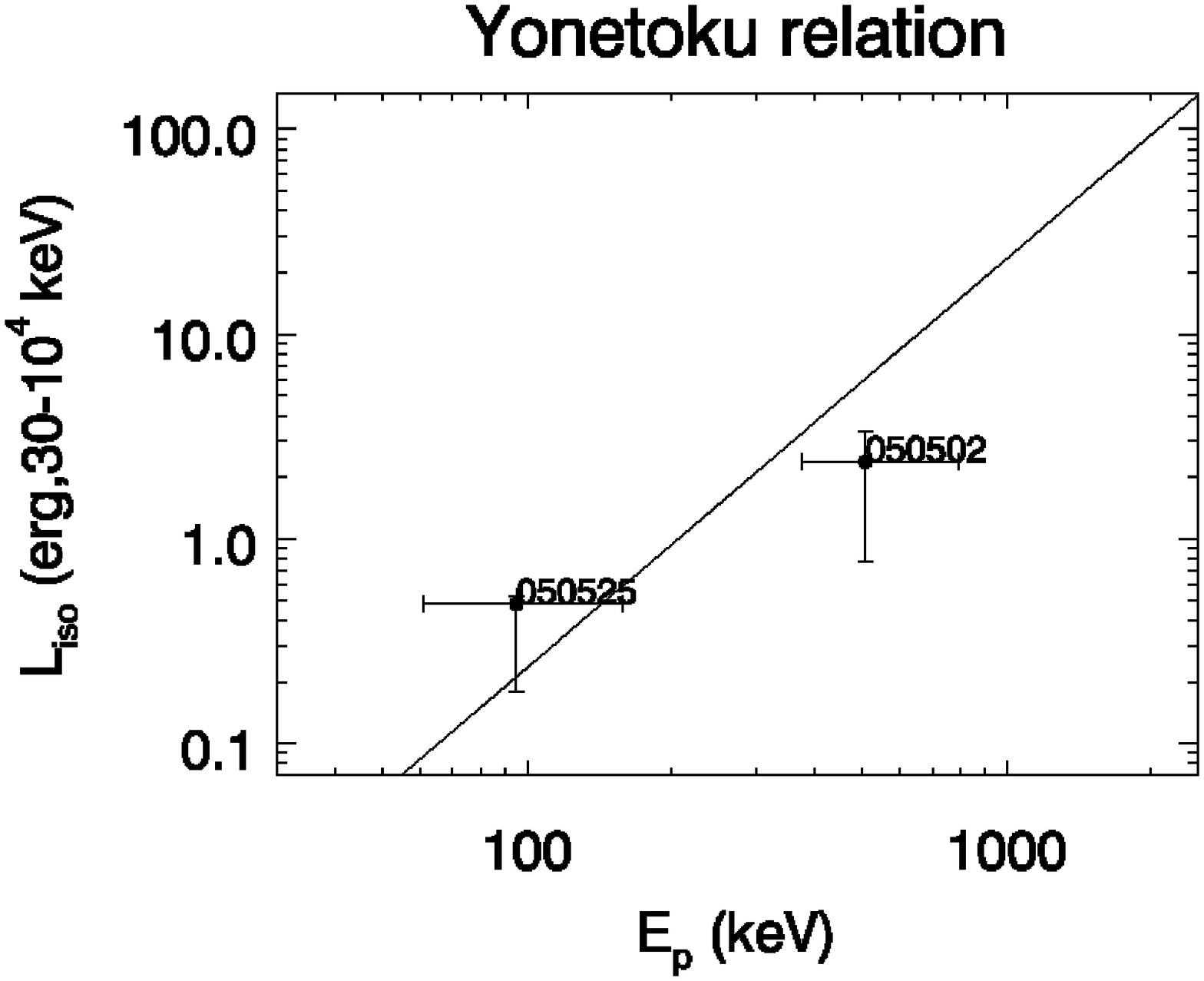,angle=90,width=0.47\linewidth} \\
\end{tabular}
\caption{Position of two \int\ GRBs on the E$_{p}$ Vs E$_{\mbox{iso}}$ \citep[left panel,][]{Amati2006} and L$_{\mbox{iso}}$ Vs $E_{p}$ \citep[right panel,][]{Yonetoku2004} relations.}
\label{fig:correlations}
\end{figure*}

This burst was in the fully coded field of view of IBIS, but it
occurred   just after a satellite slew. This is the reason why
time bins before 02:13:56 UT are lacking in the light curve.
Moreover there was a telemetry saturation causing data loss at
02:14:10 UT. Thus our measures of $t_{90}$ and fluence are only
lower limits. The spectrum has a very good signal-to-noise ratio
and is poorly described by a power law ($\chi^{2}_{red}=1.6$ with 46
d.o.f). A CPL model gives a better fit ($\chi^{2}_{red}=1.0$ with 45
d.o.f). The F-test gives a chance probability of observing such a 
$\chi^2$ improvement of only $3\times 10^{-6}$. The measured
cutoff energy is $E_{0}=91^{-25}_{+47}$ keV and the CPL  photon
index   is $\alpha=0.86^{-0.24}_{+0.22}$. These values correspond
to a peak energy   $E_{p}=104^{-38}_{+55}$ keV. Our measure is
fully compatible with that reported by \cite{Schaefer2007}. In
Fig.~\ref{fig:cut_contours} we report the confidence contour for
the two parameters of the CPL model. The spectrum of this GRB can
also be well described with a QT model
($\chi^{2}_{red}=1.0$ with 44 d.o.f), with the parameters reported in Table \ref{tab:qt}. \\
Only $\sim$23 seconds after the IBAS trigger, the ROTSE-IIIb
robotic telescope started to observe the field of GRB 050502 and
discovered its optical afterglow. Follow-up observations allowed investigators
to detect a break in the light curve that seemed to be achromatic
\citep{Yost2006,Guidorzi2005}. Moreover, the Keck-I optical
telescope acquired a high-resolution spectrum of the afterglow
only 3 hours after the burst, measuring a redshift of $z=3.793$
\citep{Prochaska2005GCN}. This is the first \int\ GRB with both
a measured cutoff and a known redshift. With this information we
can derive the GRB parameters in the cosmological rest frame, using the procedure
described by \cite{Amati2002}. Using the total duration of the
burst (20 s), we computed the total emitted energy (in the
isotropic hypothesis) $E_{iso}=8.8 \times 10^{52}$ erg. The
rest-frame peak energy resulted in $E_{p}^{iso}=508^{-135}_{+289}$
keV, while the isotropic (peak) luminosity resulted in $L_{iso}=2.38
\times 10^{52}$ erg s$^{-1}$. We used the standard values for the
cosmological parameters ($H_{0}=70$ km~s$^{-1}$~Mpc,
$\Omega_{\lambda}=0.7$ and $\Omega_{m}=0.3$). In
Fig.~\ref{fig:correlations} we report our measured values for
$E_{p}$, $E_{iso}$ and $L_{iso}$: we can see that this GRB lies
quite well on the Amati and Yonetoky correlations. We note, however, that our
value for $E_{iso}$ could be affected by the unknown duration of
the GRB, as mentioned above.

\subsection{GRB 050504}
This burst was detected  in the fully coded field of view, with
the whole detector area exposed to the flux of the source.
The resulting good statistics allowed us to measure a spectral
cutoff. In fact the CPL model gave a better fit than a power law,
with an F-test chance probability of $6 \times 10^{-4}$. We also tried the QT model, but the best fit value for the blackbody
normalization is compatible with zero and without any improvement
in the fit quality.\\
Despite the location of this burst at high Galactic latitude
(b=$+75^{\circ}$), the presence of a 10$^{th}$ magnitude K type
star prevented the observation of part of the error region (90$''$
radius). The precise position derived for the X-ray afterglow,
later discovered with \swi \citep{Kennea2005GCN}, is
inconsistent with that of the bright star. The upper limits for
the optical afterglow are thus rather constraining:  independent
groups reported values of R$\sim$19 and R$\sim$21  a few minutes
after the burst \citep[see for example][]{Cenko2005}.

\subsection{GRB 050520}
The integrated spectrum of this multi-peaked GRB is well described
by a power law.
In order to investigate the spectral evolution of the burst, we
have divided the emission in three main episodes: from $T_{0}+55$
s to $T_{0}+85$ s, from $T_{0}+85$ s to $T_{0}+100$ s, and
$T_{0}+100$ s to $T_{0}+120$ s (see Fig.~\ref{fig:light_curves}). The three corresponding spectra
are well fitted by power laws, with photon index $1.52 \pm 0.09$
($\chi^{2}_{red}=1.13$ with 33 d.o.f), $1.58 \pm 0.14$ ($\chi^{2}_{red}=1.04$
with 33 d.o.f) and $1.72 \pm 0.1$ ($\chi^{2}_{red}=1.15$ with 33 d.o.f),
respectively. Thus, there is no evidence of spectral
evolution. Spectral differences could be present in the several
peaks contained in each of these intervals, but the burst is not
bright enough to allow a spectral analysis for each
peak.\\
\swi /XRT observed the field of GRB 050520 about 2 hours after the
trigger, detecting a fading X-ray counterpart
\citep{Kennea2005GCN2}.

\subsection{GRB 050522}
This is one of the faintest bursts detected with \int\ . No
emission above $\sim$40 keV was detected with IBIS/ISGRI for this
burst that can thus be classified as an X--ray flash.
\swi\ reported an X--ray source outside the IBAS error
region, but subsequent observations showed it to be unrelated to
the burst. The same observations led to the discovery of a fading
X--ray source inside the error box, tentatively identified with
the afterglow of this burst \citep{Capalbi2005GCN}. However, the
variability of this source has not been confirmed, so the
identification remains uncertain.\\

\subsection{GRB 050525A}
This very bright burst, discovered by \swi , occurred by
chance within the  IBIS field of view, although at a very large
off-axis angle  (14.7$^{\circ}$).  A telemetry saturation caused a
gap in its light curve. The spectrum of this burst cannot be well
fitted by a power law, and requires a CPL or a QT model. Our best-fit
parameters of the CPL model are compatible with those obtained
with \swi  \citep{Cummings2005GCN} and with Konus-WIND
\citep{Golenetskii2005GCN}. The fit with the QT model is good, but
gave a slightly worse $\chi^{2}$  with one degree of freedom less
than the CPL model.\\
The high level of coding noise in the deconvolved image, due to the
very large off-axis position of the burst, and the simultaneous
presence of Cyg X-1 in the field of view, conspired to make GRB
050525A undetectable in real time by the IBAS programs. This was
unfortunate, since the \swi\ localization had a delay of
six minutes and the initial part of the optical afterglow was
missed by ground based telescopes. The ROTSE-III and TAROT robot
telescopes detected the optical afterglow at R$\sim$15 ten minutes
after the burst \citep{Rykoff2005GCN,Klotz2005GCN}, while earlier
optical data were obtained with UVOT on board \swi
\citep{Holland2005GCN}. Further observations led to the
determination of the redshift z=0.6 \citep{Foley2005GCN} and to
the detection of the radio afterglow \citep{Cameron2005}. The
rest frame cutoff energy is $E_{0}^{rest}=133^{-36}_{+66}$ keV, the
isotropic emitted energy of $E_{iso}=2.1 \times 10^{52}$ erg and the
peak luminosity $L_{iso}=3.4 \times 10^{52}$ erg~s$^{-1}$. All these
values are compatible with those found with \swi
\citep{Nava2006}. In Fig.~\ref{fig:correlations} we report our measured values for
$E_{p}$, $E_{iso}$ and $L_{iso}$: we can see that GRB 050525 lies
on both the Amati and the Yonetoku correlations.

\subsection{GRB 050626}
This GRB was nearly on axis, lying also within the field of view
of the OMC optical telescope. IBAS correctly localized the burst
in the IBIS image and a telecommand to place a CCD window at the
burst coordinates was automatically sent to the OMC.
Unfortunately, the GRB was at only 2$'$ from one of the brightest
stars in the sky, $\alpha$ Crucis, that has a magnitude of 0.8. Thus the
region where the GRB took place is severely affected by saturation
by this very bright star, making the OMC data for this burst useless. This burst was also detected by JEM-X at coordinates
R.A.=186.74, Dec.=--63.13, 25 arcsec away from the ISGRI position.
The ISGRI spectrum is well fitted by a power law. Adding the
JEM-X spectrum does not change significantly the
best fit parameter. No afterglow has been detected \citep{Gorosabel2005GCN,Mangano2005GCN}.
\subsection{GRB 050714}
The spectrum of this faint burst was well fitted by a power law.
The light curve is composed of three peaks, but the burst is too
weak to perform a time-resolved spectral analysis. A few minutes
after the onset of the burst two sources were proposed as optical
counterparts \citep{Klose2005GCN}, but these detections were not
confirmed. About 14 hours later an X-ray candidate was detected
using \swi /XRT  \citep{Racusin2005GCN}, and soon after an optical
source was discovered in the XRT error circle
\citep{Klose2005GCN2}. The optical variability of the source has
not been confirmed, so its identification  with the afterglow has
to be taken with care.

\subsection{GRB 050918}
This burst was  at 14$^{\circ}$ off-axis, with only a few percent
of the detector surface exposed. No prompt IBAS Alert had been
issued for this burst because its significance was below the
threshold for automatic alert delivery. The light curve shows two
distinct emission episodes, separated by a quiescent period of $\sim180$
s. The missing data visible in the light curve are due to telemetry
saturation. We have extracted the spectrum excluding the quiescent
period. It is well fitted by a power law.
Contrary to the results of \cite{Foley2008}, we could not find
improvements in the fit using the CPL or QT models. This difference is probably due to the better calibration for
large off-axis angles now available.
We have also extracted the spectra of the two emission episodes
separately. Although they are separated by a long quiescent
period, they are both fitted with practically the same power law
photon index: $\alpha=1.72 \pm 0.14$ for the first episode
($\chi^{2}_{red}=1.04$ with 30 d.o.f), and $\alpha=1.76 \pm 0.12$ for
the second episode ($\chi^{2}_{red}=0.6$ with
30 d.o.f).\\
A follow-up observation of this burst was carried out with \swi,
revealing a previously uncatalogued faint X--ray  source inside
the 2.8 arcmin radius \int\ error circle
\citep{Kennea2005GCN3}. There was no evidence of fading in this
source, however as this observation occurred more than 2 days
after the burst, this result is expected.

\subsection{GRB 050922}
The significance of this faint burst was below the threshold for automatic alert delivery.
The burst is detected only up to 100 keV, with a very low signal-to-noise
ratio, so its spectrum is affected by large
uncertainties. No afterglow was detected
\citep{Sonoda2005GCN}.

\subsection{GRB 051105B}
This faint burst was observed only 2.5$^{\circ}$  off-axis. This
position is within the field of view of JEM-X, but just outside that of
the OMC. The position derived with JEM-X,
R.A.=9.442, Dec=--40.495, is 1.5 arcmin away from the ISGRI
position.  Due to the low significance of the detection at soft
X-ray energies, is not possible to extract a spectrum with JEM-X
data. The ISGRI spectrum is well fitted by a power law. \\
Despite the position of the burst, it was observed with optical
instruments only a few hundred seconds after its onset, no
afterglow was detected
\citep{Kinugasa2005GCN,Torii2005GCN,Distefano2005GCN}. The field
was observed two days later with XRT and UVOT onboard \swi.
Again, no afterglow was detected
\citep{Mineo2005GCN,Blustin2005GCN}.

\subsection{GRB 051211B}
This  GRB was observed about 10$^{\circ}$  off-axis. Its spectrum
is well described by a power law.
This burst was promptly observed by optical telescopes and by
\swi /XRT. An afterglow candidate was detected in X-rays
\citep{LaParola2005GCN}, then confirmed in the optical
\citep{Jelinek2005GCN}. There has been also a detection at radio
wavelength, but it was not confirmed \citep{Frail2005GCN}.

\subsection{GRB 060114}
This faint GRB was only 2.6$^{\circ}$ from the IBIS optical axis. Thus it was in the JEM-X field of view, but too faint to be detected. Its spectrum is well fitted by a simple power law, with a quite hard photon index. No afterglow was detected.

\subsection{GRB 060130}
This is another faint GRB, seen near the IBIS optical axis (3$^{\circ}$ off-axis). Despite it being inside the JEM-X field of view, it is not detected due to its faintness. Its spectrum is well fitted with a power law. No afterglow was detected.

\subsection{GRB 060204}
This faint GRB was observed at the edge of the fully coded field of view of IBIS, and slightly outside of the field of view of JEM-X. Its spectrum is well described by a power law. No afterglow was detected.

\subsection{GRB 060428C}
This bright GRB occurred during a short period in which IBAS was not running. It was detected off-line by \cite{Grebenev2007}. The
light curve shows several peaks, with no (or very short) periods of
quiescence. The gaps visible after the highest peak is due to
telemetry saturation. The integrated spectrum cannot be
described by a power law ($\chi^{2}_{red}=1.9$ with 46 d.o.f),
while is well fitted by a CPL model with photon index $\alpha
=0.86 \pm 0.2$ and cutoff energy $E_{0}=85^{-19}_{+34}$ keV
($\chi^{2}_{red}=0.9$ with 45 d.o.f). The F-test chance probability of
obtaining such an improvement is only $2 \times 10^{-8}$. The
spectrum can be described also by a QT model with blackbody
temperature kT=$19.47^{-2.56}_{+2.34}$ keV and photon index
$\alpha=1.70^{-0.15}_{+0.20}$. We have also perfomed a time-resolved spectral analysis, dividing the burst in three parts, and fitting with a power law, a CPL and a QT model. The results, reported in Table \ref{tab:060428C}, are fully in
agreement with those of \cite{Grebenev2007}. Note, however, that those authors gave errors at a $1\sigma$ level. The GRB shows a clear spectral evolution, going from quite a soft spectrum to a harder one, then coming back to soft emission.\\
Owing to the lack of notification by the IBAS system no follow-up observations were done.
\begin{table*}[tb!]
\caption{Results of the time-resolved spectral analysis of GRB 060428C. Errors are at 90\% c.l.} 
\label{tab:060428C}
\centering
\begin{tabular}{cccccccc}
\hline  \hline
Name & T$_{Start}$ & T$_{Stop}$ & Spectral model & Phot. index & $E_{0}$ (CPL) or $kT$ (QT)& $\chi^{2}_{red}$&d.o.f \\ 
& UTC & UTC & & & keV & & \\ 
\hline
Interval 1 & 02:30:35 & 02:30:40 & PO & $1.9 \pm 0.1$ & & 1.4 & 28 \\
           &          &          & CPL& $1.3 \pm 0.5$ & $74^{-32}_{+155}$ & 1.2 & 27 \\
           &          &          & QT & $2.3 \pm 0.5$ & $14^{-5}_{+5}$    & 1.2 & 26 \\
Interval 2 & 02:30:40 & 02:30:44 & PO & $1.46 \pm 0.07$ &                 & 1.95 & 31 \\
           &          &          & CPL& $0.5 \pm 0.3$ & $64^{-16}_{+30}$  & 1.25 & 30 \\
           &          &          & QT & $1.4 \pm 0.3$ & $17^{-3}_{+3}$    & 1.15 & 29 \\
Interval 3 & 02:30:45 & 02:30:48 & PO & $2.2 \pm 0.3$ &                   & 1.29 & 21 \\
           &          &          & CPL& \multicolumn{4}{c}{Not constrained} \\
           &          &          & QT & \multicolumn{4}{c}{Not constrained} \\
\hline 
\end{tabular}
\end{table*}

\subsection{GRB 060901}
\begin{figure}[tb]
\centering
\psfig{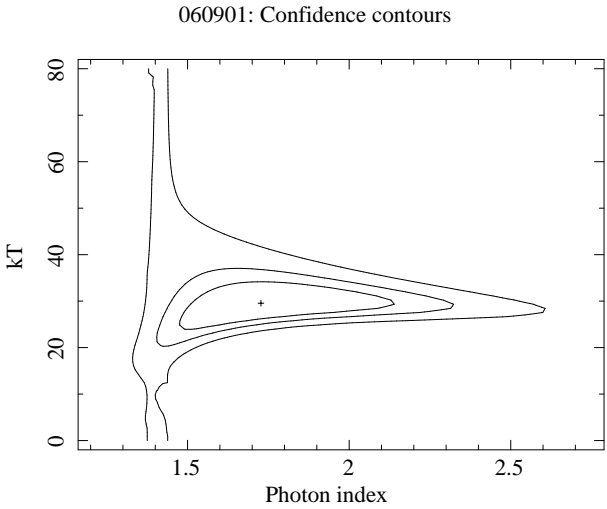}
\caption{Contour plot of the QT model parameters applied to GRB 060901. The contours are at 68\%, 90\% and 99\% confidence levels for two parameters of interest.}
\label{fig:060901_qtcontour}
\end{figure}

This burst was observed 14$^{\circ}$ off-axis, with only a few percent of the detector surface exposed to the flux from the source. In the light curve there is a gap due to telemetry saturation. The spectrum is well described by a power law ($\chi^{2}_{red}=1.1$ with 46 d.o.f). While the CPL model does not improve the fit, the QT model describes the spectrum slightly better ($\chi^{2}_{red}=0.9$ with 44 d.o.f) than the  power law. However, the contour plot in Fig.~\ref{fig:060901_qtcontour} shows that the best fit parameters are not well constrained. The limited fit improvement and the poor constraints of the parameters indicate that probably this result is not significant. \cite{Foley2008} did not succeed in fitting the spectrum of GRB 060901 with a power law, and use a QT model with parameters compatible with ours. However, they obtain larger uncertainties and a worse $\chi^{2}_{red}$. This again could be due to the different calibration they used, that is not optimal in analyzing bursts with such large off-axis angles. GRB 060901 was also detected by Konus-Wind \citep{Golenetskii2006GCN}. They found that its integrated spectrum is well fitted by a Band model in the range 20 keV-2 MeV, with $\alpha=0.77^{-0.26}_{+0.23}$, $\beta=2.31^{-0.18}_{+0.36}$ and $E_{0}=156^{-71}_{+46}$ keV. \\
An afterglow candidate has been discovered by \swi /XRT
\citep{Racusin2006GCN}. A tentative optical counterpart has been
also proposed \citep{Wiersema2007GCN}, but not confirmed.

\subsection{GRB 060912B}
This is a very long and faint GRB, with a slow rise and and a slow
decay. Its spectrum is well fitted by a power law, while
there are not enough statistics to constrain both the CPL and the QT
models. No afterglow was detected.

\subsection{GRB 060930}
This is another quite faint GRB, detected 8.5$^{\circ}$ off-axis. Its spectrum
is well described by a power law, with no improvement in the
fit using the CPL model or the QT model. No afterglow
was detected.

\subsection{GRB 061025}
This burst was detected 8.4$^{\circ}$ away from the IBIS optical axis. Its spectrum is not described very well by a power law ($\chi^{2}_{red}=1.6$ with 46 d.o.f), while a CPL model describes it much better ($\chi^{2}_{red}=1.1$ with 45 d.o.f, F-test chance probability of $1 \times 10^{-5}$). We can obtain a good fit with the QT model too, but with one more parameter ($\chi^{2}_{red}=1.1$ with 44 d.o.f). We obtain best fit parameters for the QT model compatible with that of \cite{Foley2008}, but with smaller errors. On the contrary, they could not fit the spectrum with a CPL model, probably because of their software version: OSA 5.1 is not optimal in extracting spectra from a source with an off-axis angle so large.\\
Following the localization provided by the IBAS system, the
\swi /XRT detected the X--ray counterpart \citep{Mineo2006GCN} and
the ROTSE III optical telescope confirmed it \citep{Yuan2006GCN}.

\subsection{GRB 061122}
This is a bright GRB, observed at 8.2$^{\circ}$ from the IBIS optical axis. This burst also triggered Konus-Wind \citep{Golenetskii2006GCN2}. Two gaps due to telemetry saturation are visible in the light curve, so that the peak flux reported in Table \ref{tab:observative} should be regarded as a lower limit. The spectrum is poorly described by a power law model ($\chi^{2}_{red}=1.6$ with 39 d.o.f), while we obtained a good fit with the CPL model ($\chi^{2}_{red}=0.8$ with 45 d.o.f.). The F-test probability of obtaining a similar improvement by chance is $2 \times 10^{-7}$. We measured a photon index $\alpha=1.24 \pm 0.16$ and a cutoff energy  $E_{0}=122^{-31}_{+60}$ keV. Also the spectrum measured by Konus-Wind is well fitted by a CPL model, with $\alpha=1.03^{-0.07}_{+0.06}$ and $E_{0}=160^{-7}_{+8}$ keV \citep{Golenetskii2006GCN2}. These values are compatible with ours. Our spectrum also could be fitted with the QT model, with a similar quality ($\chi^{2}_{red}=0.8$ with 43 d.o.f.). \cite{Foley2008} obtain the best description of the spectrum of GRB 061122 using a QT model, while they were not able to constrain the cutoff energy in the CPL model. Moreover, their best fit value for the blackbody temperature in the QT model ($12 \pm 3$ keV) is only marginally compatible with ours ($19 \pm 3$ keV). This could be due again to the off-axis angle of this burst. \\
An afterglow has been
detected at X-ray wavelengths \citep{Halpern2006GCN} and in the
optical band \citep{Halpern2006GCN2}.

\subsection{GRB 070903}
This long and faint GRB was observed with an off-axis angle of 7.6$^{\circ}$. Its integrated spectrum is affected by large statistical errors due to the low signal to noise ratio. It is well described by a power law ($\chi^{2}_{red}=1.3$ with 16 d.o.f), while neither the CPL model nor the QT model give a better fit.\\
The X--ray afterglow of this GRB has been
observed by \swi /XRT \citep{Vergani2007GCN}.

\subsection{GRB 070311}
This burst was observed at a position 12$^{\circ}$ away from the IBIS optical axis. Its spectrum is well described by a power law, with no improvement adopting other models. Our results are compatible with that of \cite{Sazonov2007} and \cite{Guidorzi2007}.\\
Thanks to the prompt localization provided by the IBAS system, the REM
telescope detected the afterglow of GRB 070311 only 51 s after the
onset of the GRB \citep{Covino2007GCN}. The afterglow
was later observed by other observatories at optical wavelengths, and by
the \swi /XRT at soft X--ray energies \citep{Guidorzi2007GCN}. The
light curve of the afterglow showed a major rebrightnening and
other interesting features, described in detail in
\cite{Guidorzi2007}.

\begin{figure}[tb]
\centering
\psfig{figure=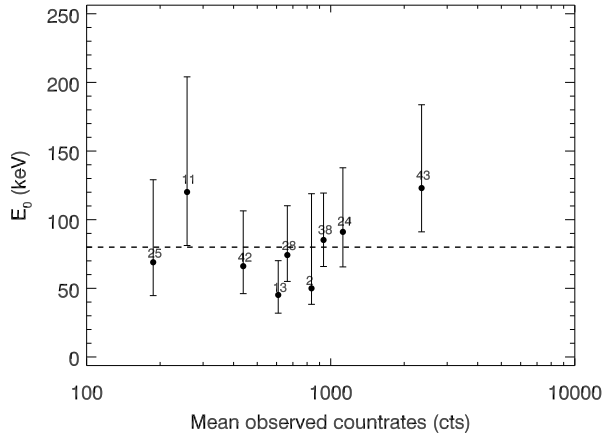,width=0.75\linewidth,angle=90}
\caption{Cut-off energy $E_{0}$ as a function of the mean countrate (background subtracted, not renormalized) measured during each GRB. The error bars are at the 90\% confidence level for a single parameter of interest. The dashed line is the mean cutoff energy $<E_0>=80$ keV.}
\label{fig:e0_vs_rates}
\end{figure}

\section{Discussion}
\label{sec:discussion}
\subsection{Spectral results}

We have measured the cutoff energy E$_{0}$ in 9 out of 43 GRBs,
corresponding to  $\sim 20\%$ of our sample. The values of E$_{0}$
are in the range 50-150 keV, and are clustered around 80 keV (see
Fig.~\ref{fig:e0_vs_rates}). This is not surprising, because most
of the spectra have a significant signal from 18 keV to 300 keV,
but the spectral bins above 200 keV  usually have large
statistical errors. In order to constrain the cutoff energy, a
significant number of energy bins with  good statistics are needed
both below and above the cutoff value. Thus, the limited bandpass of
ISGRI introduces a bias in the cutoff measurement, as also seen in
similar instruments, like Swift/BAT \citep[see, i.e., ][]{Ghirlanda2008b}. \cite{Sakamoto2008} studied a
large sample of BAT bursts detecting a cutoff energy in 32 long
GRBs out of 216, a percentage ($\sim 15 \%$) even smaller than
that found in our sample.

The largest GRB sample with good spectral information currently
available is that obtained with the BATSE/CGRO experiment. It
demonstrates that the majority of GRBs are well fitted by the Band
model \citep{Band1993,Preece2000}. Analyzing the subsample of very
bright GRBs, with peak flux $>$10 ph cm$^{-2}$ s$^{-1}$ and 
fluence $>$ $10^{-5}$ erg cm$^{-2}$ (20-2000 keV),
\cite{Kaneko2006} (hereafter K06) found that the average values
for the low- and high-energy photon index are $\alpha \simeq 1.1$
and  $\beta \simeq 2.3$, respectively. The two photon index
distributions are clustered around these values and are well
separated. Our photon index distribution
(Fig.~\ref{fig:photindex}) peaks around 1.6, a value between the
average values of $\alpha$ and $\beta$. A similar result was found
for  Swift/BAT GRBs  \citep{Sakamoto2008}. If the results found by
K06 using only  the brightest bursts apply to the whole GRB
population, we can conclude that we are unable to see some cutoff
energies either because they are within the ISGRI energy range but
the spectra do not have high enough statistics, or because the cut-off
lies outside the ISGRI energy band. In both cases the
observed spectrum would have some curvature, so that in both
cases the best fit with a single power law would give a photon
index intermediate between $\alpha$ and $\beta$. \\
To estimate the number of undetected cutoffs in our ISGRI spectra,
we should know the real distribution of the cutoff values.  From
the existence of correlations between the energetics of a burst and
its spectral parameters \citep[for example the Amati
relations,][]{Amati2002}, we know that the distribution of $E_{0}$
is not independent of the fluence distribution of the sample.
Only four  GRBs of our sample satisfy  the brightness criteria of
K06, so the cutoff energy distribution derived by those authors is
not useful for our purposes. \cite{Nava2008Astro-ph} extended the
spectral analysis of BATSE bursts to lower fluences, down to  $F
\geq 10^{-6}$ erg cm$^{-2}$ (20-2000 keV), and found that 35\% of
them have a cutoff below 150 keV.  In our sample there are 30
bursts above this fluence\footnote{For a typical spectrum with
$\alpha=1.1$, $\beta=2.3$, and $E_{0}=200$ keV, the fluence
threshold used by \cite{Nava2008Astro-ph} translates to $F \geq F_{0}=4.5 \times 10^{-7}$ erg
cm$^{-2}$ in our energy band (20-200 keV)}, and the 9 bursts for
which we could measure the cutoff are in this sample
(corresponding to the 30 \%). The photon index distribution of the
remaining 21 bursts is again centered on $\alpha\sim1.6$. Thus we
conclude that we are probably not missing more than a few cutoffs
in the subsample with $F \geq F_{0}$, so we have not
measured a cutoff energy for most of the remaining 21
bursts because it lies above 150 keV or below 20 keV. We can also
say that the population of bursts observed by ISGRI with $F \geq
F_{0}$ has spectral properties compatible with those observed by
BATSE. For the burst with a fluence lower than $F_{0}$ we cannot
say if we are missing the measurement or if the cutoff is
outside our band.

\subsection{Spectral evolution}

When the fluence of the burst was high enough, we performed
a time-resolved spectral analysis that showed  different kinds of
spectral evolution. In particular, we can divide the spectrally
evolving bursts in two classes: single-peaked bursts with
spectral evolution, and multi-peaked bursts with
spectral differences   \textit{between the peaks}. In the first
class we detected only one case of the classical hard-to-soft
evolution (GRB 021219). In the second class we detected
hard-to-soft evolution (GRB 041218, GRB 041219), soft-to-hard evolution (GRB
040106), more complex evolution (GRB 030320, GRB 060428C) and also one case of
no evolution at all (GRB 050918). All these
cases are described individually in section \ref{sec:results}.

\subsection{Peak fluxes distribution}

\begin{figure*}[tb!]
\centering
\begin{tabular}{cc}
\psfig{file=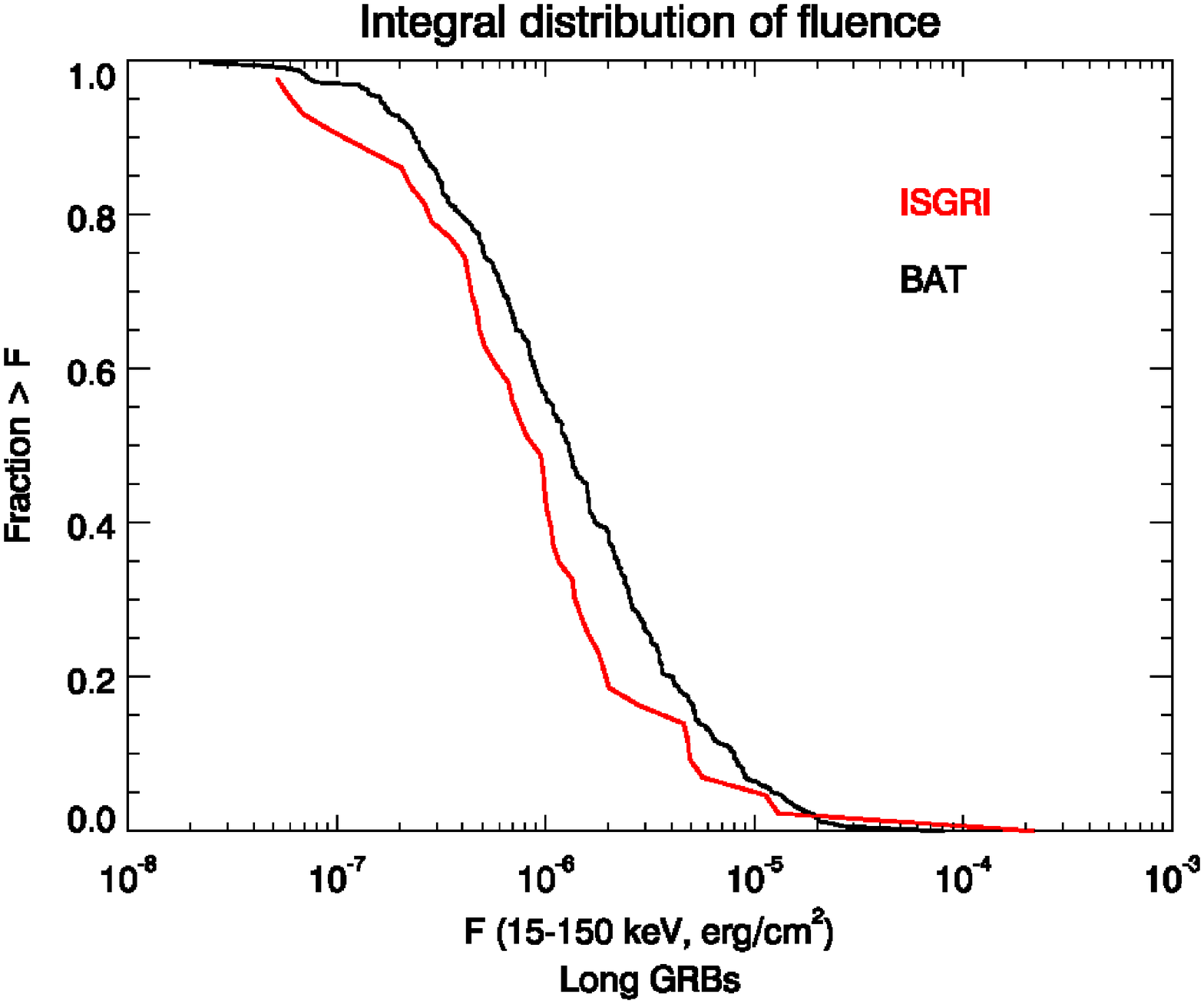,width=0.45\linewidth} &
\psfig{file=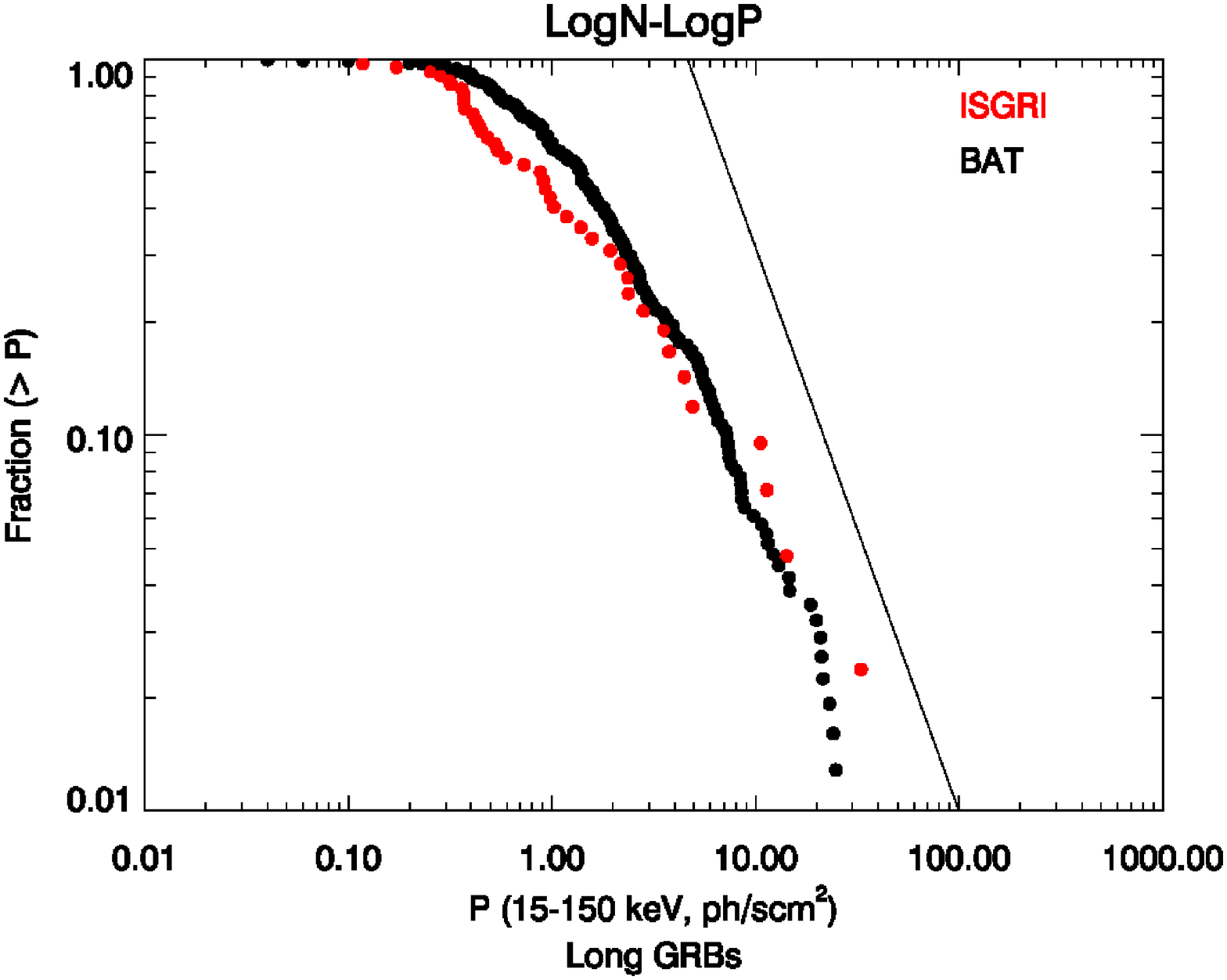,width=0.45\linewidth} \\
\end{tabular}
\caption{Normalized integral distributions of fluences (left) and
peak fluxes (right) for both ISGRI and BAT. } \label{fig:logN-log}
\end{figure*}

In Fig.~\ref{fig:logN-log} we show the normalized integral
distributions of fluences and peak fluxes for both ISGRI and BAT.
The two instruments are very similar from several points of view,
with the most important difference being the larger field of view
covered by BAT. Thus we expect that the burst populations seen by
the two instruments should be very similar. As discussed above,  this
seems to be true from the spectral point of view. To check  if
the distributions of fluence and peak flux of the two samples also are similar, we applied a Kolmogorov-Smirnoff (KS) test to the
data of Fig.~\ref{fig:logN-log}.  This gave a probability of 12\%
that the ISGRI and BAT fluences are drawn from the same
distribution, while for the  peak fluxes the  probability is 3\%.
The latter result suggests that the two populations are somewhat
different, with  ISGRI  detecting fainter GRBs than BAT.
However, larger samples are required to eventually confirm this
possible difference.

\subsection{The rate of short bursts}
The IBAS system detected 2 out of 56 short hard bursts, corresponding to $\sim$ 3.5 \%, while Swift/BAT found $\sim$ 8\%. This small difference is consistent with the small statistics of the samples. However, the BATSE experiment onboard CGRO found that 1 burst out of 4 was a short hard GRB (25 \%). This discrepancy is more compelling. We have to take into account, however, that BATSE was better suited to detect hard bursts: it was based on a different technology than both ISGRI and BAT, and it had a broader bandpass and a larger effective area at high energy. This fact alone could explain the difference in the measured rate of short hard bursts. We will soon have an independent measurement of this rate by the Gamma-ray Burst Monitor (GBM) onboard \textit{Fermi}.

\subsection{On the relation between spectral lags and peak fluxes}
 \label{subsec:lags-peaks}
Peaks of GRBs occur at slightly different times in different
energy bands \citep{Cheng1995}. In general they  migrate to later times and
become wider going to lower energy, although some GRBs with the
opposite trend have been observed \citep{Chen2005}. This
so-called ``spectral lag'' correlates with the peak flux of the
GRBs \citep{Norris2000,Norris2002,Tsutsui2008}: long spectral lags
have been measured mainly in GRBs with low peak fluxes, while
short spectral lags have been observed in bursts with both high
and low peak fluxes. Based on these findings, one can divide GRBs
in three groups: bright with
short lags, faint with short lags, and faint with long lags. \\
\cite{Foley2008} measured the  spectral lags  of \int\ GRBs, and
divided the sample in two classes: long-lag GRBs ($\Delta t
\gtrsim 0.75$ s) and short-lag GRBs ($\Delta t \lesssim 0.75$ s).
Based on peak fluxes taken from literature or,  when unavailable,
analyzing the data themselves, they could verify that the three
groups mentioned above are also visible in the \int\ GRB sample.
In some cases we found GRB peak fluxes quite different from those used by these authors. We therefore repeated their analysis, using their time-lag values, but the peak fluxes computed by us in a consistent way for all the GRBs. We found that
the conclusions on the spectral lag-peak flux correlation in \int\
GRBs found by \cite{Foley2008} are confirmed.

\subsection{Spatial distribution}
\label{sec:spatialdistrib}
While the whole \int\ GRB sample
is isotropically distributed, the
sub-sample of bursts with  long lag and low luminosity  is
spatially correlated with the supergalactic plane \citep{Foley2008}. This result,
which supports  the existence of a local GRB population, relies on
an accurate estimate of the \int\ sky coverage. In computing
the sky coverage, these authors used the \int\ exposure maps,
but neglected other factors affecting the sensitivity of GRB
triggers, thus potentially biasing the observed sky distribution of
the faintest bursts (which are indeed those for which an anisotropy
was found).

Therefore, to better investigate this interesting point, we
computed an IBIS exposure map by also taking into account the
sensitivity variations of the IBAS triggers. These are dominated
by changes in the background level resulting from the presence of
bright sources in the field of view,  particle induced events such
as solar flares or radiation belt passages, as well as the long
term background increase during the mission due to the solar
cycle. Based on the actual  ISGRI   background level measured in
each pointing, we rescaled the exposure time with respect to that
of an ideal reference pointing  (i.e. a high galactic latitude
pointing performed at the beginning of the mission and without
strong sources in the field of view).
The effective exposure time, $t_{eff}$, has been computed as
$t_{eff}=t_{0}\cdot(\sqrt{bkg_{i}}/\sqrt{bkg_{ref}})$, where
$bkg_{i}$ is the background level for the considered pointing,
$bkg_{ref}$ is the reference background, and $t_{0}$ is the raw
exposure time. As an example, the effective exposure time with
this improved calculation gives $\sim$12 Ms for the Galactic
Center region, compared to   $\sim$17 Ms obtained with the simple
time exposure map.

After having computed the improved exposure map, we simulated a
population of 50000 GRBs uniformly distributed over the entire sky
following our new map. For this burst population we derived the
quadrupole moment with respect to the supergalactic plane, in
order to derive our bias, which is $Q$=0.046. Then we derived the
GRB quadrupole moment for the 11 long-lag GRBs which results in $Q$=--0.225$\pm$0.089, while for all \int\ GRBs we obtain
$Q$=0.039$\pm$0.042. Once one subtracts the bias, one obtains
$Q$=--0.271$\pm$0.089 for long lag GRBs and $Q$=--0.007$\pm$0.042
for the whole sample. These results confirm the findings of
\cite{Foley2008}, but with our improved method we enhanced the
statistical significance of the quadrupole moment for long lag
bursts from $\sim$2.5 to $\sim$ 3 $\sigma$.

\section{Conclusions}
We have provided a complete spectral catalogue of the publicly available \int\ GRBs, based on
IBIS data. By using the latest available software and calibration, we were able to derive
accurate spectral parameters for all the bursts in a coherent way,
superseding and harmonising all the previous results that could be found in literature.  
In addition, the spectral evolution of the bursts could be studied in detail for most of the sources.

By computing an accurate exposure map, which takes into account the variation of the instrumental background
over our dataset, we were able to confirm, with a higher significance, the clustering of faint, long-lag bursts around the 
supergalactic plane, supporting the existence of a local GRB population.

We have shown that, despite the good ISGRI sensitivity, which is slightly better then the BAT one, 
only for a limited number of bursts could the spectral description go beyond a simple power law.  
Indeed, our analysis and other works in the literature have shown that, despite the increasing number of GRBs 
with measured redshift, provided mainly by \swi, the number of bursts for which 
both good spectral and redshift information are available is still limited. 
Although IBIS has a slightly broader band pass, and better chances to determine $E_{peak}$ with respect to BAT,
it is limited by a smaller field of view, and by its pointing strategy biased towards the low  Galactic latitudes, where optical absorption 
is severe. This explains the low redshift determination rate for \int\ GRBs ($\sim$5\%).

The \int\ and \swi\ experience has shown that wide field coded mask instruments are a powerful tool for
GRB detection and localization, but the lack of sensitivity over a broad bandpass hampers the progress in GRB science. 
Even in the presence of a redshift determination, with the simple power law description, a key parameter like 
the overall energy budget of the source cannot be determined, and accurate attempts to model the prompt GRB emission cannot be performed.

The combined use of these satellites with the non-imaging Gamma-ray Burst Monitor (GBM) on board {\it Fermi}, or the 
Wide-band All-sky Monitor (WAM) on board {\it Suzaku} can improve the quality of GRB prompt data, but the rate of simultaneous detections
is small, due to different sensitivities, pointing strategies or earth occultation constraints.
Future GRB missions should carry adequate instruments in order to simultaneously obtain prompt localisations and broad band spectroscopy.

\begin{acknowledgements}
Based on observations with \int , an ESA project with instruments and
science data centre funded by ESA member states (especially the PI
countries: Denmark, France, Germany, Italy, Switzerland, Spain), Czech
Republic and Poland, and with the participation of Russia and the USA.
ISGRI has been realized and maintained in flight by CEA-Saclay/Irfu with
the support of CNES.
This research has been supported by the Italian Space
Agency through contract ASI-INAF I/023/05/0.
G.V. thanks A. Paizis for helping in the \int\ data reduction and for her effort in solving related problems; M.Turler of the \int\ helpdesk; L.Nava for sharing her knowledge and for interesting discussions.
D.G. acknowledges the French Space Agency (CNES) for financial support.

\end{acknowledgements}
\bibliographystyle{aa.bst}
\bibliography{pap_IntegralGRBs_v16}
\newpage
\onecolumn
\begin{landscape}
\begin{longtable}{llccccccc@{ }c@{ }cccccc}
\caption[]{Observed properties of \int\ GRBs. GRBs detected after March 2007 are given here for completeness, but we have only extracted their coordinates and their $t_{90}$. \label{tab:observative}}\\
\hline\hline
\# & Name & Time & R.A. & Dec. & L & B & Pos.Error & \multicolumn{3}{c}{Afterglow} & z & T$_{90}$ & T$_{50}$ & Peak Fl. & Position\\
& & (UTC) & (deg.) & (deg.) & (deg.) & (deg.) & (arcmin) & X & O & R & & (s) & (s) & (ph cm$^{-2}s^{-1}$) & Reference \\
\hline\hline
\endfirsthead
\caption[]{(continued)}\\
\hline\hline

\# & Name & Time & R.A. & Dec. & L & B & Pos.Error & \multicolumn{3}{c}{Afterglow} & z & T$_{90}$ & T$_{50}$ & Peak Fl. & Position\\
& & (UTC) & (deg.) & (deg.) & (deg.) & (deg.) & (arcmin) & X & O & R & & (s) & (s) & (ph cm$^{-2}s^{-1}$) & Reference \\
\hline\hline
\endhead
\hline
\multicolumn{16}{c}{(Continued on the next page)}\\
\endfoot
\hline
\hline
\endlastfoot
1  &  021125\footnote{Observed during commissioning, Malaguti et al. (2003)}   & 17:58:30 & 296.983 &  28.391 & 64.347 &   1.476 &     2.0   &     - &     - &     - &	- &  24   &  n.a.  &	   $>$ 22  &        -\\
2  &  021219 & 07:33:54 & 282.613 &  31.950 & 61.88 &   14.14 &     1.14   &     - &     - &     - &	- &  5	  &  1.0   &           $>$4.5 &	    - \\
3  &  030131\footnote{Observed during a slew, Gotz et al. (2003)} & 07:38:49 & 202.086 &  30.673 & 58.809 &   81.183 &     2.1   &     - &     Y &     - &	- &  124  &  97   &	 $>$ 2.1 & - \\
4  &  030227 & 08:42:02 &  74.394    &  20.492     & 180.90 &  -13.76 & 	1.9     &     Y &       Y &     - &     - &  15 &  9.0   &           1.4 &	           \\
   &         &          & 74.387711  & 20.484694   &        &         & 0.0005   &       &	  &	  &	  &     &        & 	         &  Opt, GCN 1907  \\
5  &  030320 & 10:11:52 & 267.892    & -25.330     & 3.79   &  0.71   &      1.4 &     - &	- &	- &     - &  48	&  33.0  &	    10.6 &	         - \\
6  &  030501 & 03:10:02 & 286.386    & 6.277       & 40.12  & -0.26   &   1.01   &     - &	- &	- &     - &  25	&  8.0   &	     3.6 &	         - \\
7  &  030529 & 19:53:18 & 145.107    & -56.3425    & 278.69 &  -2.74  &   3.0    &     - &	- &	- &     - &  16	&  8.0   &	     0.5 &	         - \\
8  &  031203 & 22:01:27 & 120.634    & -39.851     & 255.74 &  -4.80  &     1.53 &     Y &	Y &	Y & 0.105 &  19 &  8.0   & 	     2.2 &                 \\
   &         &          & 120.62579  & -39.85112   &        &         &    0.012 &       &	  &	  &       &     &        & 	         &  X, GCN 2490    \\ 
9  &  040106 & 17:55:10 & 178.042    & -46.797     & 292.50 &   14.88 &     2.23 &     Y &      Y &     - &	- &  48 &  43.0  &	     0.9 &	           \\
   &         &          & 178.05179  & -46.78775   &        &         &    0.012 &       &	  &	  &       &     &        & 	         &  X, GCN 2510\\ 
10 &  040223 & 13:24:51 & 249.877    & -41.927     & 341.61 &   3.19  &   0.025  &     Y &      - &     - &     - & 198 &  150.0 &	    0.45 &	           \\
   &         &          & 249.87571  & -41.93325   &        &         &    0.025 &       &	  &	  &       &     &        & 	         &  X, GCN 2547\\ 
11 &  040323 & 13:02:58 & 208.469    & -52.354     & 312.52 &   9.36  &    1.26  &     - &      ? &     - &     - &  14 &  6.0   &	     1.9 &	         - \\
12 &  040403 & 05:08:03 & 115.226    & 68.215      & 147.59 &   29.62 &    1.16  &     - &      - &     - &	- &  15 &  6.0   &           0.6 &	         - \\
13 &  040422 & 06:57:59 & 280.505    & 1.981       & 33.62  &   3.00  &    0.95  &     - &      - &     - &	- &  4  &  1.8   &           2.8 &	         - \\
14 &  040624 & 08:21:35 & 195.030    & -3.583      & 307.18 &   59.21 &    2.8   &     - &      - &     - &	- & 27  &  10.0  &           0.9 &	         - \\
15 &  040730 & 02:12:06 & 238.302    & -56.470     & 326.13 &  -2.02  &    1.35  &     - &      - &     - &	- & 42  &  18.0  &           0.5 &	         - \\
16 &  040812 & 06:01:52 & 246.511    & -44.703     & 337.89 &   3.09  &    1.65  &     Y &      ? &     ? & (\footnote{\cite{2006astro.ph.12644D} derived a photometric redshift $0.3 \le z \le 0.7$}) &  8  &  3.0   &           0.7 &	           \\
   &         &          & 246.52083  & -44.70889   &        &         &    0.017 &       &	  &	  &       &     &        & 	         &  X, GCN 2649    \\ 
17 &  040827 & 11:50:50 & 229.255    & -16.152     & 346.46 &   34.15 &    2.36  &     Y &      Y &     - &	- & 32  &  12.0  &           1.0 &	           \\
   &         &          & 229.25558  & -16.14142   &        &         &    0.003 &       &	  &	  &       &     &        & 	         &  NIR \footnote{\cite{DeLuca2005}}\\ 
18 &  040903 & 18:17:58 & 270.824    & -25.258     & 5.14   &  -1.52  &    3.20  &     - &      - &     - &     - &  7  &  3.0   &	    0.42 &	         - \\
19 &  041015 & 11:11:33 & 4.690      & 66.853      & 119.71 &   4.19  &    2.81  &     - &      - &     - &     - &  30 &  12.0  &	     0.4 &	         - \\
20 &  041218 & 15:45:44 & 24.794     &  71.345     & 126.77 &   8.82  &    0.86  &     - &      Y &     - &     - &  38 &  27.3  &	     3.8 &	           \\
   &         &          & 24.78167   &  71.34167   &        &         &    0.025 &       &	  &	  &       &     &        & 	         &  X, GCN 2861    \\ 
21 &  041219 & 01:42:13 & 6.1143     &  62.839     & 119.86 &   0.13  &    0.34  &     - &      Y &     Y &     - & 239 &  129   &	$>$ 17.2 &	           \\
   &         &          & 6.1153330  &  62.842640  &        &         &    0.003 &       &	  &	  &       & (460) \footnote{if we include precursors, see text}&     (152)$^{6}$  &          	 &  Opt, GCN 2893\\
22 &  050129 & 20:03:05 & 252.789    &   -3.097    & 15.23  &   24.88 &    2.32  &     - &      - &     - &     - &  30 &  12.0  &	     0.4 &	          - \\
23 &  050223 & 03:09:00 & 271.398    & -62.459     & 331.52 &  -18.82 &    4.46  &     Y &      - &     - & 0.591 &  30 &  12.0  &	     0.9 &	            \\
   &         &          & 271.38538  & -62.472519  &        &         &    0.025 &       &	  &	  &       &     &        &          	 &  X, GCN 3109     \\ 
24 &  050502 & 02:14:01 & 202.436752 &  42.671132  & 98.76  &   72.61 &    0.62  &     - &      Y &     - & 3.793 &$>$11&$>$6.0  &	     2.4 &	            \\
   &         &          & 202.44292  & 42.674362   &        &         &    n.d.  &       &	  &	  &       &     &        &          	 &  Opt, GCN 3322   \\ 
25 &  050504 & 08:00:50 & 200.998    &  40.697     & 98.66  &   74.85 &    0.80  &     Y &      - &     - &     - &  44 &  22.0  &           0.5 &	            \\
   &         &          & 201.0054   & 40.70333    &        &         &    0.11  &       &	  &	  &       &     &        &          	 &  X, GCN 3359     \\ 
26 &  050520 & 00:05:57 & 192.511    & 30.450      & 127.88 &   86.66 &    1.16  &     Y &      - &     - &     - &  52 &  30.8  &           1.2 &	            \\
   &         &          & 192.52583  & 30.450556   &        &         &    0.08  &       &	  &	  &       &     &        &          	 &  X, GCN 3343     \\ 
27 &  050522 & 06:00:21 & 200.137    & 24.793      & 14.90  &   83.04 &    3.0   &     ? &      - &     - &     - &  11 &  2.7   &	     0.3 &	          - \\
28 &  050525 & 00:02:53 & 278.117    & 26.332      & 54.95  &   15.54 &    1.4   &     Y &      Y &     - & 0.606 &  9  &  6.0   &	$>$ 56.2 &	            \\
   &         &          & 278.13571  &  26.339582  &        &         &   0.002  &       &	  &	  &       &     &        &          	 &  Opt, GCN 3493   \\ 
29 &  050626 & 03:46:07 & 186.741    & -63.137     & 300.18 & -0.38   &    1.35  &     - &      - &     - &     - &  56 &  24.0  &	     0.4 &	          - \\
30 &  050714 & 00:05:53 & 43.612     &  69.120     & 133.56 & 8.82    &    1.46  &     Y &      ? &     - &     - &  34 &  24.0  &	     0.4 &	            \\
   &         &          & 43.592499  &  69.112778  &        &         &   0.115  &       &	  &	  &       &     &        &          	 &  X, GCN 3649     \\ 
31 &  050918 & 15:36:38 & 267.650    & -25.406     & 3.57   &  0.91   &    3.33  &     ? &      - &     - &     - & 280 &  226.0 &	     2.4 &	          - \\
32 &  050922 & 13:43:20 & 271.094    & -32.041     & 359.39 &  -5.09  &    2.9   &     - &      - &     - &     - &  10 &  10.0  &	    0.17 &	          - \\
33 & 051105B & 11:05:41 & 9.468      & -40.479     & 313.95 &  -76.35 &    1.1   &     - &      - &     - &     - &  14 &  6.0   &	     0.4 &	          - \\
34 & 051211B & 22:06:07 & 345.667    &  55.081     & 107.72 &  -4.55  &    1.34  &     Y &      Y &     ? &     - &  47 &  22.0  &	     1.0 &	            \\
   &         &          & 345.67322  &  55.080971  &        &         &   0.006  &       &	  &	  &       &     &        &          	 &  Opt, GCN 4358   \\ 
35 &  060114 & 12:39:31 & 195.276    & -4.735      & 307.54 &   58.04 &    1.47  &     - &      - &     - &     - &  80 &  32.0  &	     0.3 &	          - \\
36 &  060130 & 04:56:29 & 229.248    & -36.921     & 332.72 &   17.41 &    2.0   &     - &      - &     - &     - &  19 &  7.0   &	    0.25 &	          - \\
37 &  060204 & 13:19:39 & 232.254    & -39.467     & 333.25 &   13.96 &    1.7   &     - &      - &     - &     - &  52 &  30.0  &	     0.3 &	          - \\
38 & 060428C & 02:30:35 & 285.227    & -9.556      & 25.46  &  -6.43  &    0.72  &     - &      - &     - &     - &  10 &  4.8   &	     4.9 &	          - \\
39 &  060901 & 18:43:55 & 287.164    & -6.639      & 28.95  &  -6.83  &    1.5   &     Y &      ? &     - &     - &  16 &  3.2   &	    11.4 &	            \\
   &         &          & 287.15805  & -6.6394440  &        &         &   0.07   &       &	  &	  &       &     &        &          	 &  X, GCN 5496     \\ 
40 & 060912B & 17:32:11 & 271.203    & -19.872     & 10.01  &  0.76   &    2.6   &     - &      - &     - &     - & 140 &  60.0  &	     0.1 &	          - \\
41 &  060930 & 09:04:12 & 304.511    & -23.635     & 19.56  &  -28.94 &    2.3   &     - &      - &     - &     - &  9  &  4.0   &	     0.4 &	          - \\
42 &  061025 & 18:35:53 & 300.904    & -48.244     & 350.95 &  -31.58 &    0.87  &     Y &      Y &     - &     - &  11 &  4.0   &	     1.6 &	            \\
   &         &          & 300.91177  & -48.242981  &        &         &   0.008  &       &	  &	  &       &     &        &          	 &  Opt, GCN 5754   \\ 
43 &  061122 & 07:56:50 & 303.836    & 15.511      & 56.65  &  -10.66 &   0.50   &     Y &      Y &     - &     - &  12 &  4.9   &	   $>$33.0 &	            \\
   &         &          & 303.83267  &  15.517361  &        &         &   0.008  &       &	  &	  &       &     &        &          	 &  Opt, GCN 5849   \\ 
44 &  070309 & 10:00:39 & 263.680    & -37.9525    & 351.16 &  -2.83  &    2.1   &     Y &      - &     - &     - &  22 &  12.0  &	     0.4 &	            \\
   &         &          & 263.6629   & -37.93006   &        &         &   0.08   &       &	  &	  &       &     &        &          	 &  X, GCN 6187     \\ 
45 &  070311 & 01:52:34 & 87.543158  &  3.39136    & 202.76 &  -12.00 &   2.04   &     Y &      Y &     - &     - &  32 &  16.0  &	     1.1 &	            \\
   &         &          & 87.534210  &  3.3750800  &        &         &   0.02   &       &	  &	  &       &     &        &          	 &  X, GCN 6190     \\ 
\hline
46 &  070615 & 02:20:37 & 44.316     & -4.410      & 181.41 &  -52.36 &   1.92   &     ? &      - &     - &     - &  15 &  7.0   &	    n.a. &	          - \\
47 &  070707 & 16:08:38 & 267.738    & -68.924     & 324.61 &  -20.00 &   1.85   &     Y &      Y &     - &     - & 0.7 &  0.3   &	    n.a. &	            \\
   &         &          & 267.74396  & -68.924225  &        &         &   0.008  &       &	  &	  &       &     &        &          	 &  X, GCN 6612     \\ 
48 &  070925 & 15:52:32 & 253.214    & -22.036     & 358.96 &   13.71 &    0.1   &     Y &      - &     - &     - &  19 &  7.0   &	    n.a. &	            \\
   &         &          & 253.21699  & -22.028561  &        &         &   0.1    &       &	  &	  &       &     &        &          	 &  X, GCN 6826     \\ 
49 &  071017 & 00:58:08 & 274.706    & -15.968     & 15.03  & -0.27   &    2.43  &     - &      - &     - &     - & 0.5 &  0.2   &	    n.a. &	          - \\
50 &  071109 & 20:35:55 & 289.913    & 2.048       & 37.99  & -5.33   &   2.5    &     - &      - &     - &     - & 30  & n.a.   &          n.a. &                - \\
51 &  080120 & 17:28:28 & 225.254    & -10.885     & 346.86 & 40.68   &   2.0    &     Y &      Y &     - &     - & 15  & n.a.   &          n.a. &                  \\
   &         &          & 225.2636   & -10.8749    &        &         &   0.01   &       &        &       &       &     &        &               & Opt, GCN 7198   \\
52 &  080414 & 22:33:30 & 272.133    & -18.829     & 11.36  & 0.55    &   2.0    &     - &      - &     - &     - &   8 & n.a.   &          n.a. &                - \\
53 &  080603 & 11:18:15 & 279.408    & 62.735      & 92.50  & 25.46   &   2.0    &     Y &      Y &     - &     - & 150 & n.a.   &          n.a. &                  \\
   &         &          & 279.4082   & 62.7441     &        &         &   0.01   &       &        &       &       &     &        &               & Opt, GCN 7976   \\
54 &  080613 & 09:35:21 & 213.275    & 5.169       & 348.09 & 60.66   &   1.5    &     Y &      Y &     - &     - & 30  & n.a.   &          n.a. &                 \\
   &         &          & 213.2709   & 5.1732      &        &         &   0.01   &       &        &       &       &     &        &               & Opt, GCN 7872   \\
55 & 080723B & 13:22:15 & 176.833    & -60.245     & 295.05 & 1.64    &   1.5    &     - &      - &     - &     - &  95 & n.a.   &          n.a. &                - \\
56 &  080922 & 11:03:36 & 270.837    & -22.509     & 7.56   & -0.19   &   1.5    &     - &      - &     - &     - &  60 & n.a.   &          n.a. &                - \\

\end{longtable}
\end{landscape}

\begin{longtable}{lccccccc}
\caption[]{Results of fitting spectra of \int\ GRBs with a  power law model. \label{tab:powerlaw}}\\
\hline\hline
\# & \textbf{Name} & \textbf{Peak Flux} & \textbf{Fluence} & \textbf{Photon index} & $\mathbf{\chi^{2}_{red}}$ & \textbf{D.o.f} & \textbf{T$_{90}$} \\
& & ($15-150$ keV) & ($20-200$ keV) &  \multicolumn{3}{c}{(power law model)} & \\
& & [$10^{-8}$ erg/cm$^2$s] & [$10^{-8}$ erg/cm$^2$]& & & & [s]\\
\hline\hline
\endfirsthead
\caption[]{(continued)}\\
\hline\hline
\# & \textbf{Name} & \textbf{Peak Flux} & \textbf{Fluence} & \textbf{Photon index} & $\mathbf{\chi^{2}_{red}}$ & \textbf{D.o.f} & \textbf{T$_{90}$} \\
& & ($15-150$ keV) & ($20-200$ keV) &  \multicolumn{3}{c}{(power law model)} & \\
& & [$10^{-8}$ erg/cm$^2$s] & [$10^{-8}$ erg/cm$^2$]& & & & [s]\\
\hline\hline
\endhead
\hline
\multicolumn{8}{c}{(Continued on the next page)}\\
\endfoot
\hline
\hline
\endlastfoot

2  &  021219 &   	$>$28.7 &   $>$94.6 & 1.69 \hspace{5pt}$^{-0.08}_{+0.09}$ &  1.46 &  46 &    5.0 \\[3pt]
4  &  030227 &   	 7.0 &   74.9 & 1.75 \hspace{5pt}$^{-0.10}_{+0.11}$ &  1.0 &  46 &   15.0 \\[3pt]
5  &  030320 &   	68.0 & 1131.9 & 1.28 \hspace{5pt}$^{-0.07}_{+0.07}$ &  1.8 &  24 &   48.0 \\[3pt]
6  &  030501 &   	18.4 &  282.1 & 1.83 \hspace{5pt}$^{-0.07}_{+0.08}$ &  1.5 &  46 &   25.0 \\[3pt]
7  &  030529 &   	 1.1 &    5.2 & 3.50 \hspace{5pt}$^{-0.43}_{+0.51}$ &  1.0 &   9 &   16.0 \\[3pt]
8  &  031203 &   	12.8 &  134.8 & 1.50 \hspace{5pt}$^{-0.09}_{+0.09}$ &  1.3 &  46 &   19.0 \\[3pt]
9  &  040106 &   	 6.2 &  105.9 & 1.58 \hspace{5pt}$^{-0.14}_{+0.15}$ &  1.1 &  46 &   48.0 \\[3pt]
10 &  040223 &  	 2.1 &  147.7 & 2.11 \hspace{5pt}$^{-0.16}_{+0.17}$ &  0.8 &  46 &  198.0 \\[3pt]
11 &  040323 &         17.15 &  190.4 & 1.07 \hspace{5pt}$^{-0.06}_{+0.07}$ &  1.7 &  46 &   14.0 \\[3pt]
12 &  040403 &  	 3.3 &   28.5 & 1.84 \hspace{5pt}$^{-0.15}_{+0.16}$ &  1.0 &  46 &   15.0 \\[3pt]
13 &  040422 &  	20.7 &   44.2 & 2.01 \hspace{5pt}$^{-0.08}_{+0.09}$ &  1.7 &  46 &    4.0 \\[3pt]
14 &  040624 &  	 4.5 &   48.1 & 2.02 \hspace{5pt}$^{-0.18}_{+0.20}$ &  0.8 &  46 &   27.0 \\[3pt]
15 &  040730 &  	 3.0 &   57.8 & 1.44 \hspace{5pt}$^{-0.15}_{+0.16}$ &  0.8 &  39 &   42.0 \\[3pt]
16 &  040812 &  	4.55 &   14.0 & 2.20 \hspace{5pt}$^{-0.21}_{+0.22}$ &  0.9 &  28 &    8.0 \\[3pt]
17 &  040827 &  	6.35 &  101.0 & 1.58 \hspace{5pt}$^{-0.19}_{+0.21}$ &  1.1 &  15 &   32.0 \\[3pt]
18 &  040903 &  	 2.3 &    9.6 & 2.90 \hspace{5pt}$^{-0.39}_{+0.46}$ &  0.5 &  23 &    7.0 \\[3pt]
19 &  041015 &  	 2.3 &   51.2 & 1.13 \hspace{5pt}$^{-0.18}_{+0.18}$ &  1.0 &  35 &   30.0 \\[3pt]
20 &  041218 &  	22.7 &  491.9 & 1.57 \hspace{5pt}$^{-0.05}_{+0.05}$ &  1.3 &  46 &   38.5 \\[3pt]
21 &  041219 &       $>$ 130 &$>$2100 & 1.89 \hspace{5pt}$^{ 0.01}_{+0.01}$ &  1.1 &  46 &  239.0 (460)\footnote{If we include the precursors, see text.}\\[3pt]
22 &  050129 &  	 2.4 &   41.0 & 1.79 \hspace{5pt}$^{-0.25}_{+0.27}$ &  0.7 &  27 &   30.0 \\[3pt]
23 &  050223 &  	 4.0 &   81.5 & 1.64 \hspace{5pt}$^{-0.22}_{+0.24}$ &  1.0 &  46 &   30.0 \\[3pt]
24 &  050502 &  	12.3 &$>$108.6& 1.51 \hspace{5pt}$^{-0.05}_{+0.06}$ &  1.6 &  46 &$>$11.0 \\[3pt]
25 &  050504 &  	 3.8 &  116.0 & 1.20 \hspace{5pt}$^{-0.09}_{+0.09}$ &  1.5 &  46 &   44.0 \\[3pt]
26 &  050520 &  	 8.9 &  159.9 & 1.64 \hspace{5pt}$^{-0.06}_{+0.07}$ &  1.1 &  46 &   52.5 \\[3pt]
27 &  050522 &  	 1.5 &    6.9 & 2.65 \hspace{5pt}$^{-0.48}_{+0.59}$ &  1.0 &  21 &   10.8 \\[3pt]
28 &  050525 &      $>$314.8 &$>$1300.0 & 1.93 \hspace{5pt}$^{-0.05}_{+0.00}$ &  1.9 &  39 &  9.0 \\[3pt]
29 &  050626 &  	 2.7 &   66.5 & 2.04 \hspace{5pt}$^{-0.14}_{+0.15}$ &  0.6 &  39 &   56.0 \\[3pt]
30 &  050714 &  	 2.9 &   42.7 & 2.03 \hspace{5pt}$^{-0.19}_{+0.21}$ &  1.0 &  43 &   34.0 \\[3pt]
31 &  050918 &  	15.3 &  480.0 & 1.77 \hspace{5pt}$^{-0.09}_{+0.10}$ &  0.9 &  39 &  280.0 \\[3pt]
32 &  050922 &  	 1.1 &    5.9 & 1.85 \hspace{5pt}$^{-0.58}_{+0.69}$ &  0.7 &  11 &   10.0 \\[3pt]
33 & 051105B &  	 3.6 &   20.4 & 1.84 \hspace{5pt}$^{-0.23}_{+0.26}$ &  1.0 &  46 &   14.0 \\[3pt]
34 & 051211B &  	 7.0 &  179.0 & 1.54 \hspace{5pt}$^{-0.09}_{+0.10}$ &  1.0 &  39 &   47.0 \\[3pt]
35 &  060114 &  	 1.6 &   98.9 & 0.95 \hspace{5pt}$^{-0.18}_{+0.19}$ &  1.1 &  21 &   80.0 \\[3pt]
36 &  060130 &  	 1.9 &   22.5 & 1.59 \hspace{5pt}$^{-0.31}_{+0.34}$ &  1.4 &  22 &   19.0 \\[3pt]
37 &  060204 &  	 1.8 &   46.8 & 1.35 \hspace{5pt}$^{-0.23}_{+0.24}$ &  1.0 &  25 &   52.0 \\[3pt]
38 & 060428C &  	30.1 &  201.0 & 1.55 \hspace{5pt}$^{-0.04}_{+0.05}$ &  1.9 &  46 &   10.4 \\[3pt]
39 &  060901 &  	69.2 &  564.6 & 1.43 \hspace{5pt}$^{-0.06}_{+0.06}$ &  1.1 &  46 &   16.0 \\[3pt]
40 & 060912B &  	 1.3 &   69.4 & 1.65 \hspace{5pt}$^{-0.28}_{+0.30}$ &  1.2 &  25 &  140.0 \\[3pt]
41 &  060930 &  	 4.9 &   26.3 & 1.51 \hspace{5pt}$^{-0.27}_{+0.30}$ &  1.2 &  33 &    9.0 \\[3pt]
42 &  061025 &  	 9.7 &   97.5 & 1.34 \hspace{5pt}$^{-0.07}_{+0.08}$ &  1.6 &  46 &   11.0 \\[3pt]
43 &  061122 &         146.2 &  459.2 & 1.71 \hspace{5pt}$^{-0.04}_{+0.05}$ &  1.6 &  39 &   11.9 \\[3pt]
44 &  070309 &  	 2.6 &   35.7 & 1.36 \hspace{5pt}$^{-0.30}_{+0.31}$ &  1.3 &  16 &   22.0 \\[3pt]
45 &  070311 &           8.5 &  137.9 & 1.34 \hspace{5pt}$^{-0.12}_{+0.13}$ &  1.0 &  46 &   32.0 \\[3pt]

\end{longtable}

\begin{longtable}{lccccccc}
\caption{\label{tab:cutoff} Results of fitting spectra of some \int\ GRBs with a cut-off power law model}\\
\hline\hline
\# & \textbf{Name} & \textbf{Fluence} & \multicolumn{4}{c}{Cut-off power law model} & \textbf{F-Test} \\
& & ($20-200$ keV) & \textbf{Photon index} & \textbf{Cutoff Energy $E_0$} & $\mathbf{\chi^{2}_{red}}$ & \textbf{D.o.f} & probability  \\
& & [$10^{-8}$ erg/cm$^2$]& &[keV] & & & \\
\hline\hline
\endfirsthead
\caption{continued.}\\
\hline\hline
\# & \textbf{Name} & \textbf{Fluence} & \textbf{Photon index} & \textbf{Cutoff Energy $E_0$} & $\mathbf{\chi^{2}_{red}}$ & \textbf{D.o.f} & \textbf{F-Test} \\
& & ($20-200$ keV) &  \multicolumn{4}{c}{Cut-off power law model} & probability  \\
& & [$10^{-8}$ erg/cm$^2$]& &[keV] & & & \\
\hline\hline
\endhead
\hline\hline
\endfoot
2  &  021219 & $>$83.4 & 0.67 \hspace{5pt}$^{-0.30}_{+0.59}$ & 50  \hspace{5pt}$^{-11}_{+69}$ &  1.1 &    45 & $ 4 \times 10^{-4}$ \\[3pt]
11 &  040323 &   189.9 & 0.44 \hspace{5pt}$^{-0.29}_{+0.26}$ & 120 \hspace{5pt}$^{-39}_{+84}$ &  1.3 &    45 & $ 2 \times 10^{-4}$ \\[3pt]
13 &  040422 &    38.9 & 0.89 \hspace{5pt}$^{-0.44}_{+0.40}$ & 45  \hspace{5pt}$^{-13}_{+25}$ &  1.1 &    45 & $ 1 \times 10^{-5}$ \\[3pt]
24 &  050502 &   101.2 & 0.86 \hspace{5pt}$^{-0.24}_{+0.22}$ & 91  \hspace{5pt}$^{-25}_{+47}$ &  1.0 &    45 & $ 3 \times 10^{-6}$ \\[3pt]
25 &  050504 &   104.4 & 0.28 \hspace{5pt}$^{-0.48}_{+0.42}$ & 69  \hspace{5pt}$^{-24}_{+60}$ &  1.2 &    45 & $ 6 \times 10^{-4}$ \\[3pt]
28 &  050525 &$>$1027.7& 1.30 \hspace{5pt}$^{-0.22}_{+0.21}$ & 83  \hspace{5pt}$^{-22}_{+40}$ &  1.2 &    38 & $ 2 \times 10^{-6}$ \\[3pt]
38 & 060428C &   185.9 & 0.86 \hspace{5pt}$^{-0.20}_{+0.20}$ & 85  \hspace{5pt}$^{-19}_{+34}$ &  0.9 &    45 & $ 2 \times 10^{-8}$ \\[3pt]
42 &  061025 &    88.9 & 0.41 \hspace{5pt}$^{-0.39}_{+0.35}$ & 66  \hspace{5pt}$^{-20}_{+40}$ &  1.1 &    45 & $ 1 \times 10^{-5}$ \\[3pt]
43 &  061122 &   438.6 & 1.24 \hspace{5pt}$^{-0.16}_{+0.16}$ & 122 \hspace{5pt}$^{-31}_{+60}$ &  0.8 &    45 & $ 2 \times 10^{-7}$ \\[3pt]

\end{longtable}

\begin{longtable}{lccccc}
\caption{\label{tab:qt} Results of fitting some spectra of \int\ GRBs with a quasi-thermal model}\\
\hline\hline
\# & \textbf{Name} & \multicolumn{4}{c|}{Quasi-thermal model}  \\
& & \textbf{Photon index} & \textbf{kT} & $\mathbf{\chi^{2}_{red}}$ & \textbf{D.o.f}  \\
& & &[keV] & & \\
\hline\hline
\endfirsthead
\caption{continued.}\\
\hline\hline
\# & \textbf{Name} & \multicolumn{4}{c|}{Quasi-thermal model}  \\
& & \textbf{Photon index} & \textbf{kT} & $\mathbf{\chi^{2}_{red}}$ & \textbf{D.o.f}  \\
& & &[keV] & & \\
\hline\hline
\endhead
\hline
\endfoot
2  &  021219  &    1.47 \hspace{5pt} $^{-0.38}_{+0.29}$ & 50 \hspace{5pt} $^{-2}_{+3}$ &  1.0 &   44 \\[3pt]
13 &  040422  &    2.20 \hspace{5pt} $^{-0.32}_{+0.52}$ & 14 \hspace{5pt} $^{-3}_{+3}$ &  1.1 &       44 \\[3pt]
24 &  050502  &    1.89 \hspace{5pt} $^{-0.26}_{+0.43}$ & 22 \hspace{5pt} $^{-3}_{+2}$ &  1.0 &       44 \\[3pt]
28 &  050525  &    2.13 \hspace{5pt} $^{-0.06}_{+0.07}$ & 16 \hspace{5pt} $^{-3}_{+3}$ &  1.3 &       31 \\[3pt]
38 &  060428C &    1.70 \hspace{5pt} $^{-0.15}_{+0.20}$ & 19 \hspace{5pt} $^{-3}_{+2}$ &  0.8 &       44 \\[3pt]
39 &  060901  &    1.70 \hspace{5pt} $^{-0.21}_{+0.35}$ & 29 \hspace{5pt} $^{-5}_{+5}$ &  0.9 &       44 \\[3pt]
42 &  061025  &    1.50 \hspace{5pt} $^{-0.30}_{+0.54}$ & 21 \hspace{5pt} $^{-4}_{+4}$ &  1.1 &       44 \\[3pt]
43 &  061122  &    1.85 \hspace{5pt} $^{-0.11}_{+0.16}$ & 19 \hspace{5pt} $^{-3}_{+3}$ &  0.8 &       44 \\[3pt]

\end{longtable}

\newpage
\onecolumn
\begin{figure}[p]
\vspace{1.0cm}
\begin{tabular}{cc}
\psfig{figure=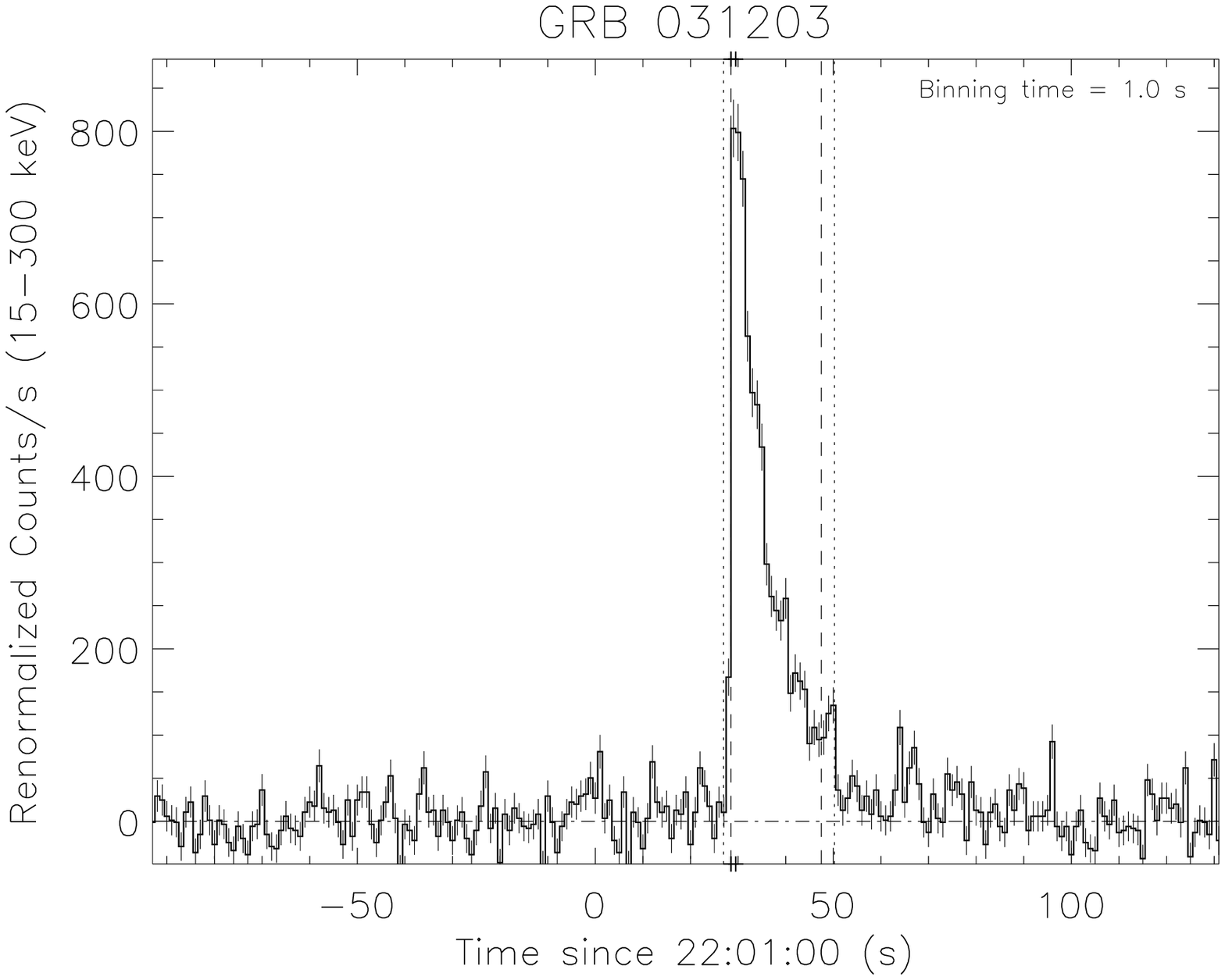,width=0.5\linewidth} &
 \psfig{figure=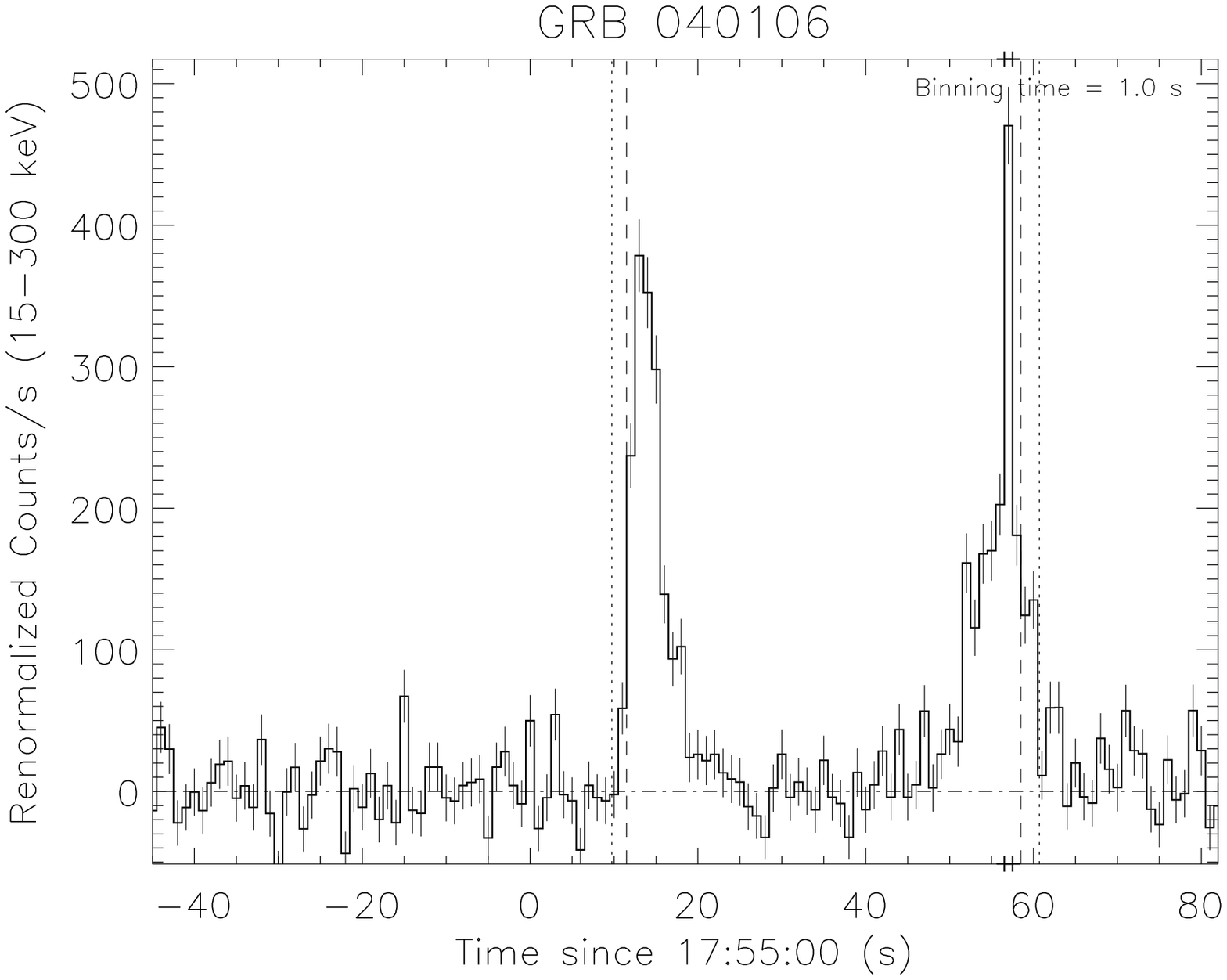,width=0.5\linewidth} \\
\psfig{figure=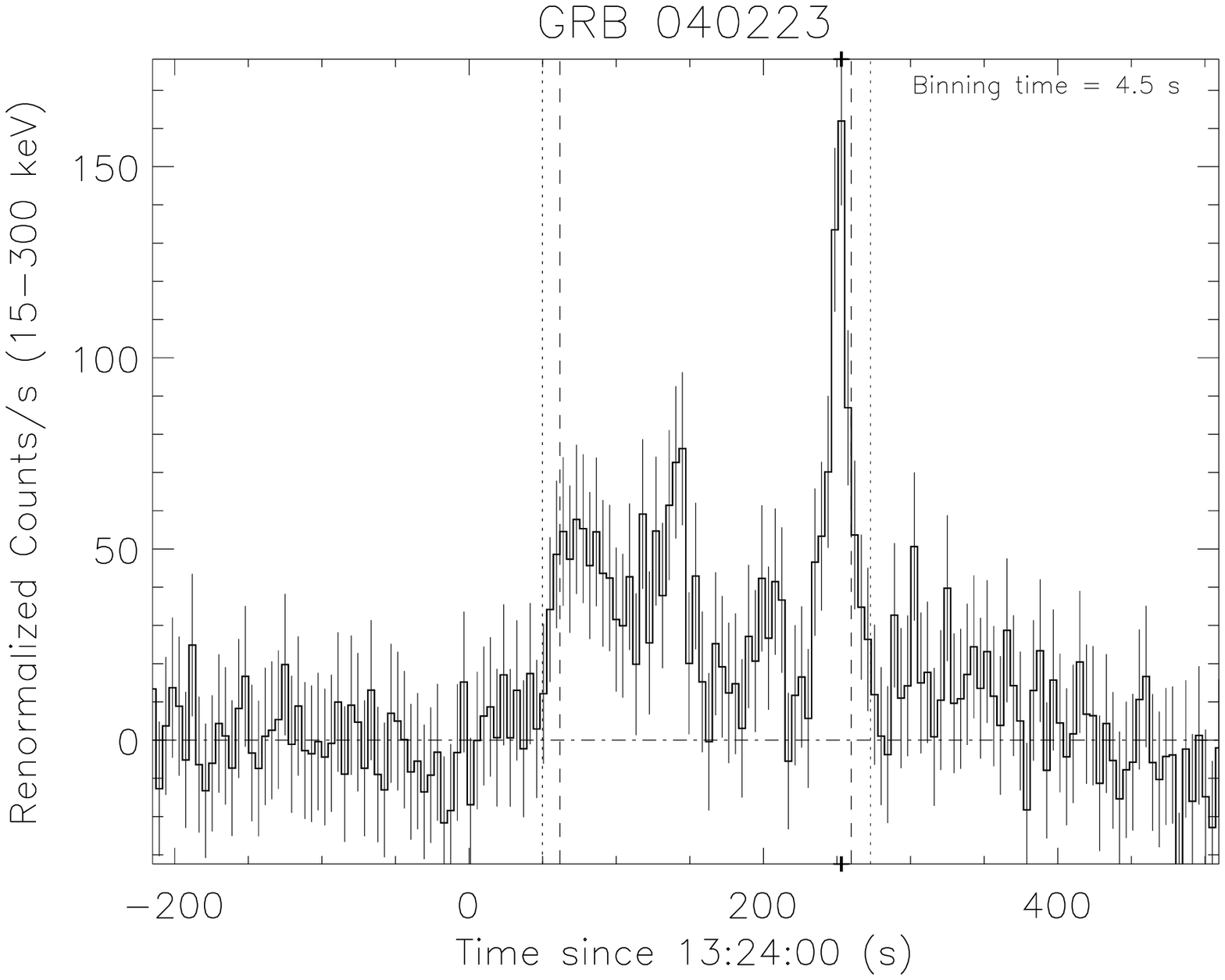,width=0.5\linewidth} &
 \psfig{figure=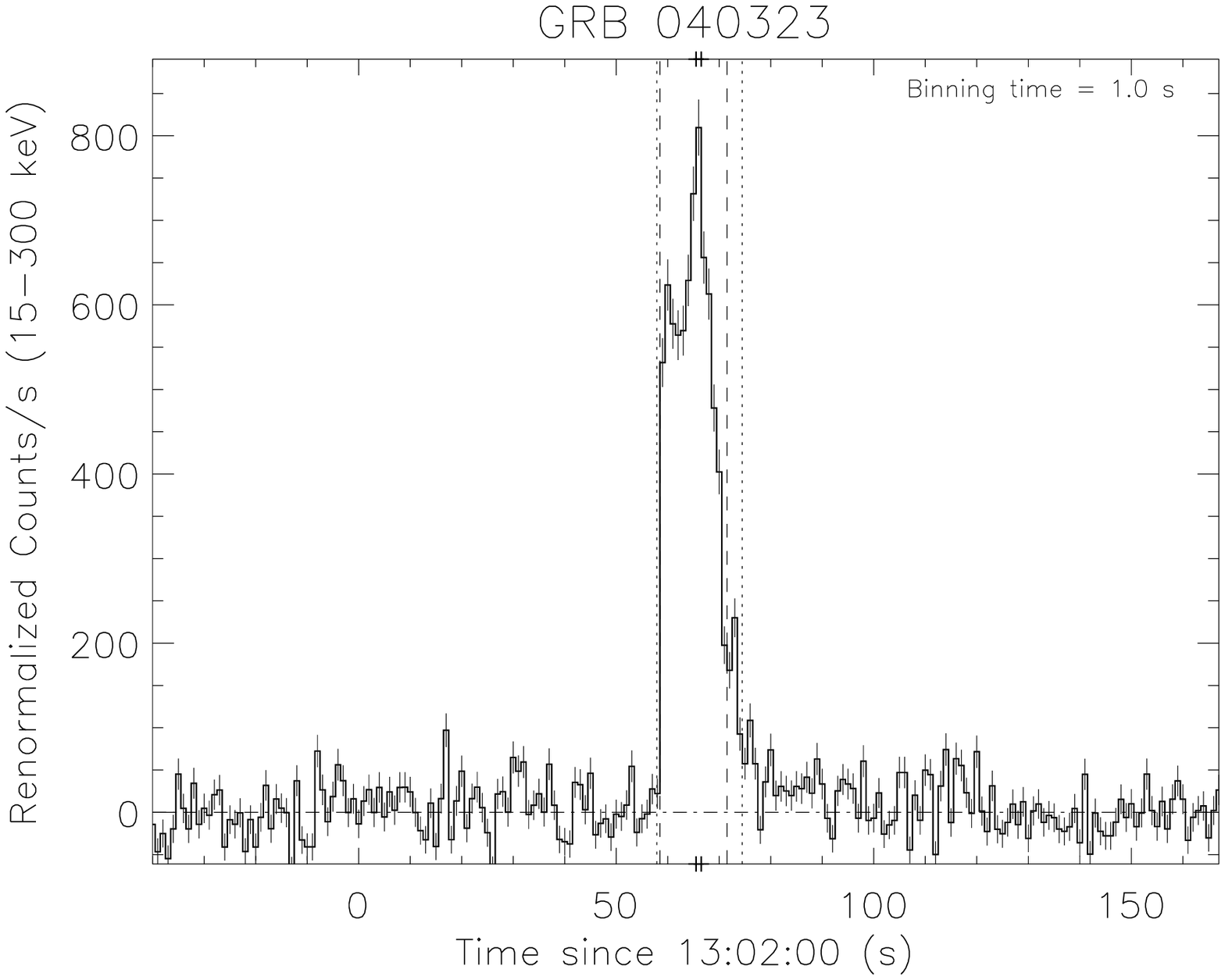,width=0.5\linewidth} \\
\psfig{figure=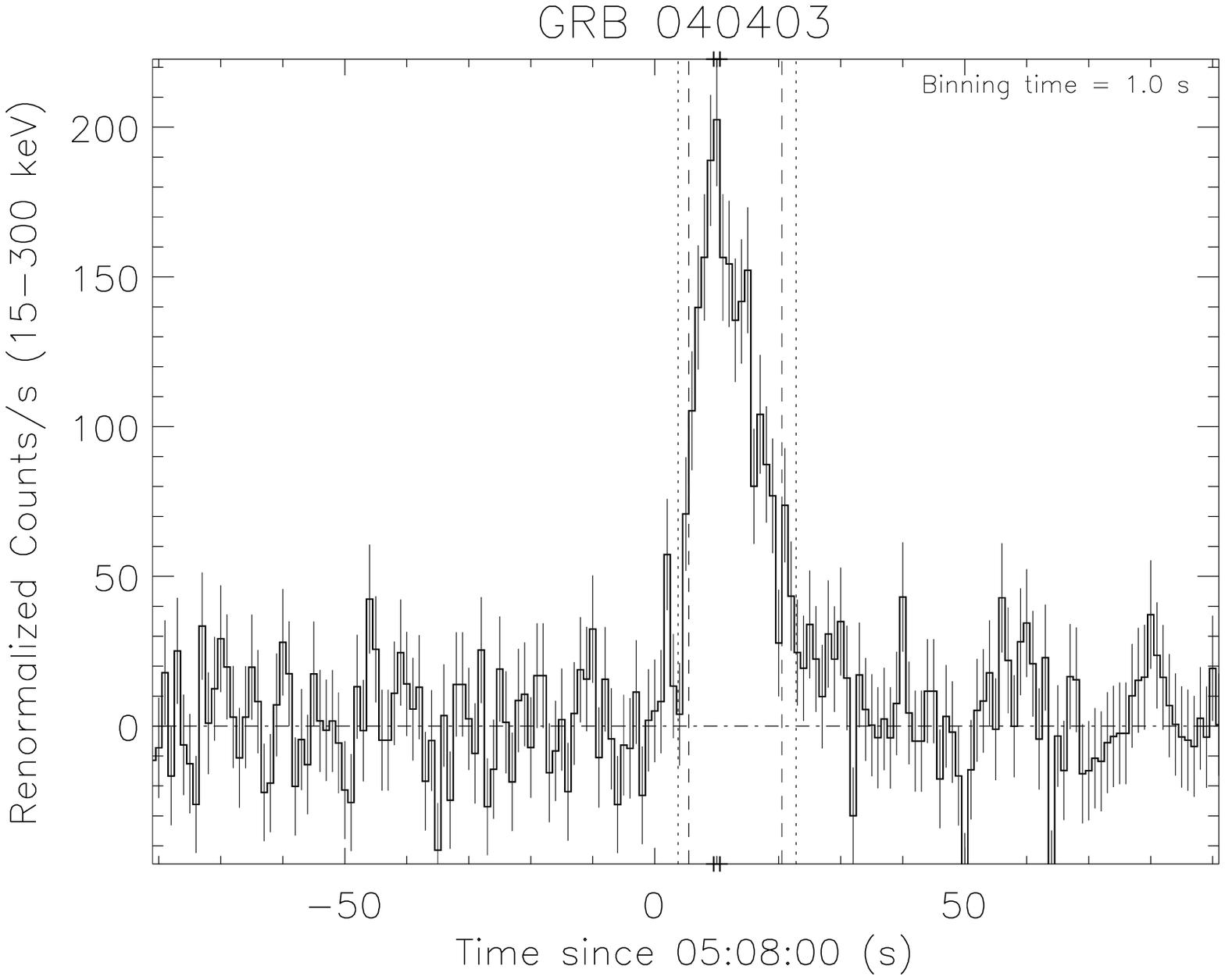,width=0.5\linewidth} &
 \psfig{figure=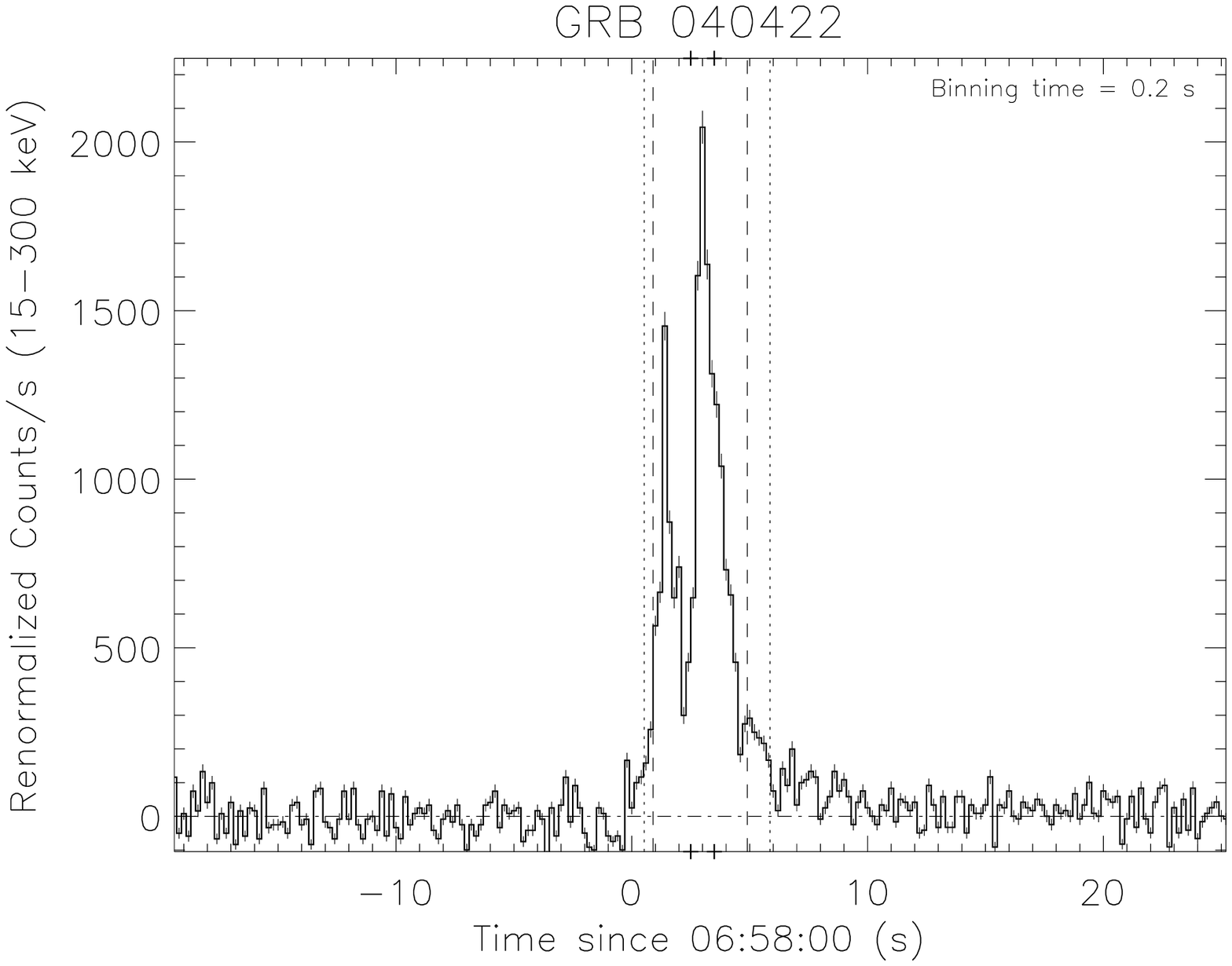,width=0.5\linewidth} \\
\end{tabular}
\textbf{Fig.~\ref{fig:light_curves}.} Light curves of \int\ GRBs (continued)
\end{figure}
\begin{figure}[p]
\vspace{1.0cm}
\begin{tabular}{cc}
\psfig{figure=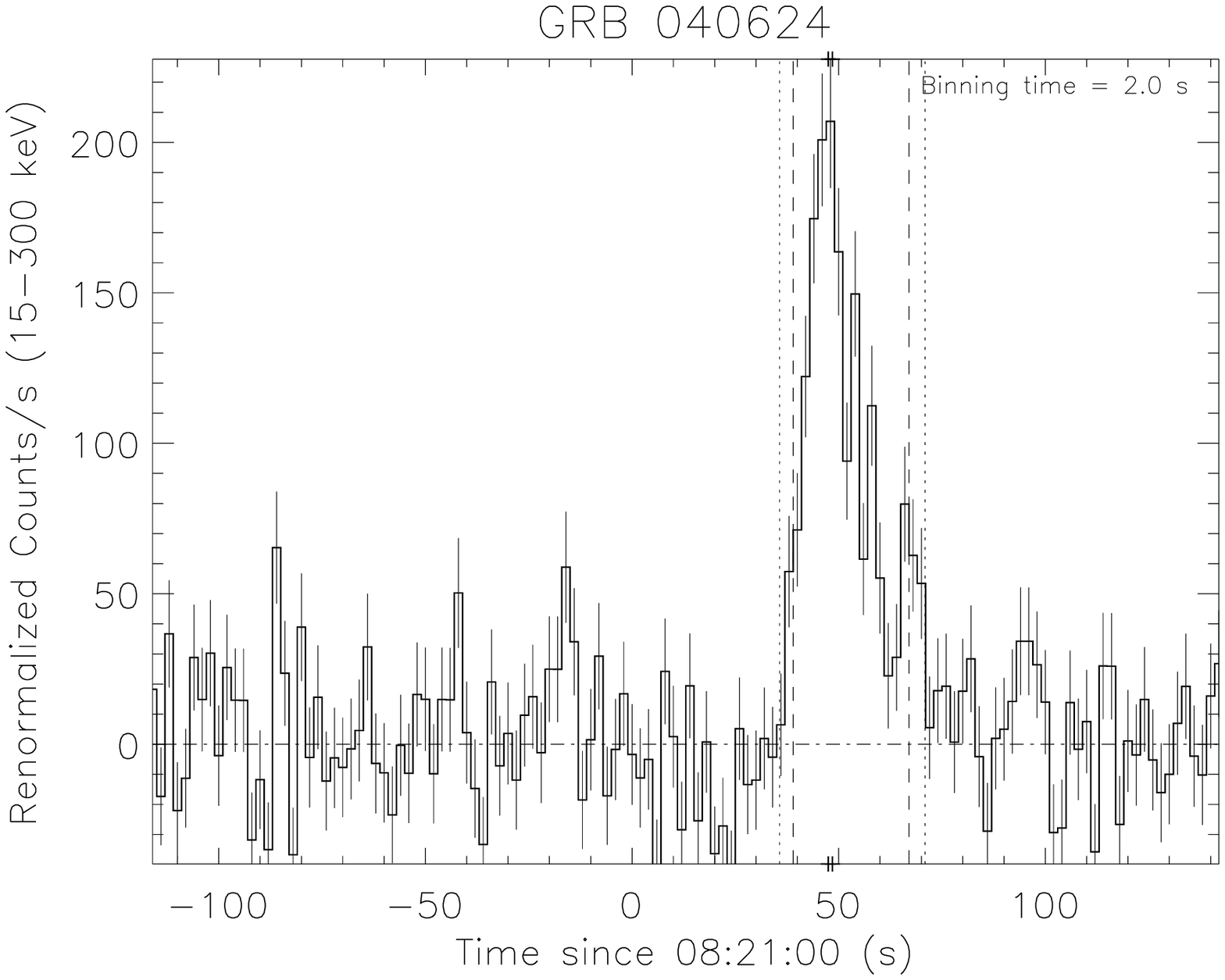,width=0.5\linewidth} &
 \psfig{figure=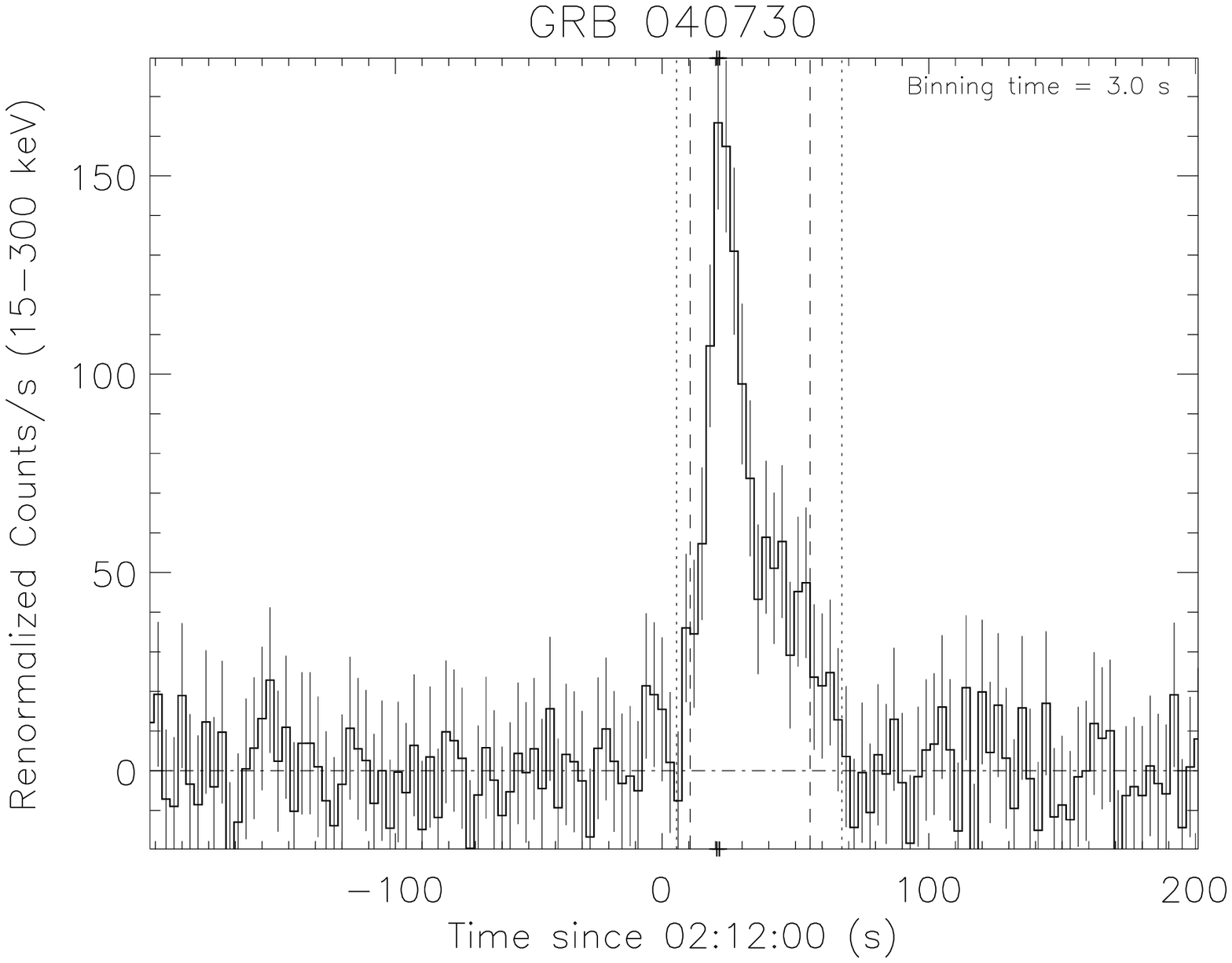,width=0.5\linewidth} \\
\psfig{figure=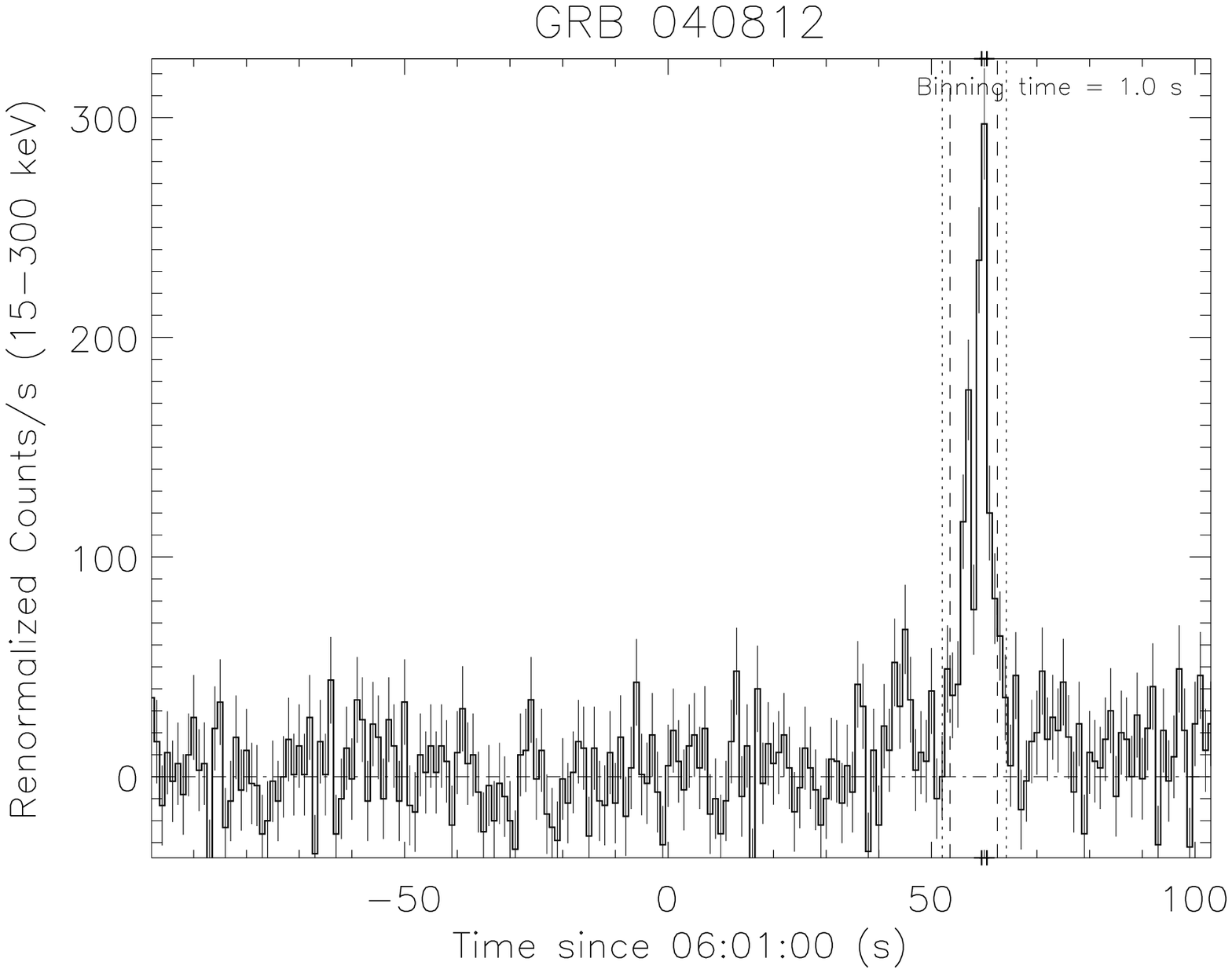,width=0.5\linewidth} &
 \psfig{figure=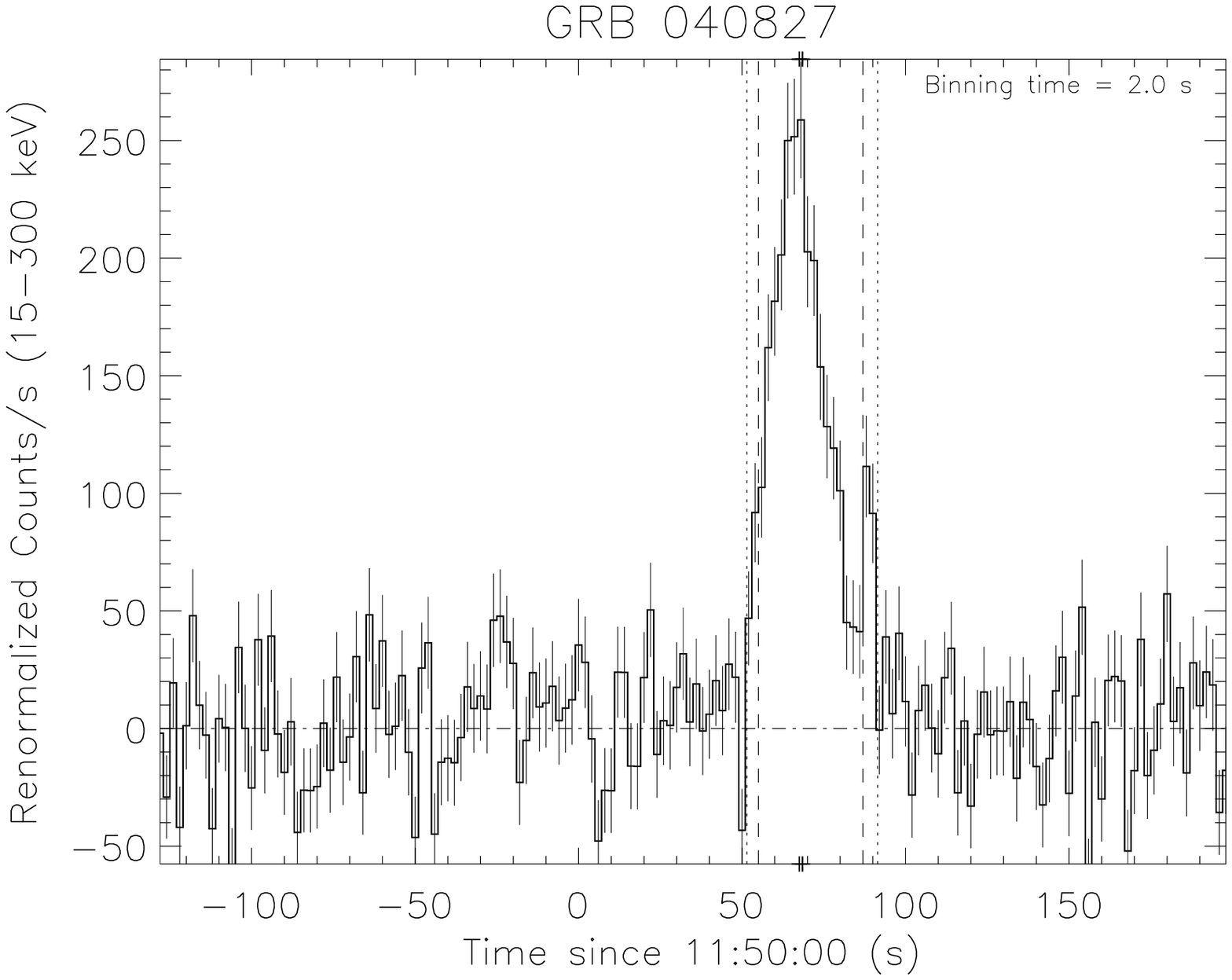,width=0.5\linewidth} \\
\psfig{figure=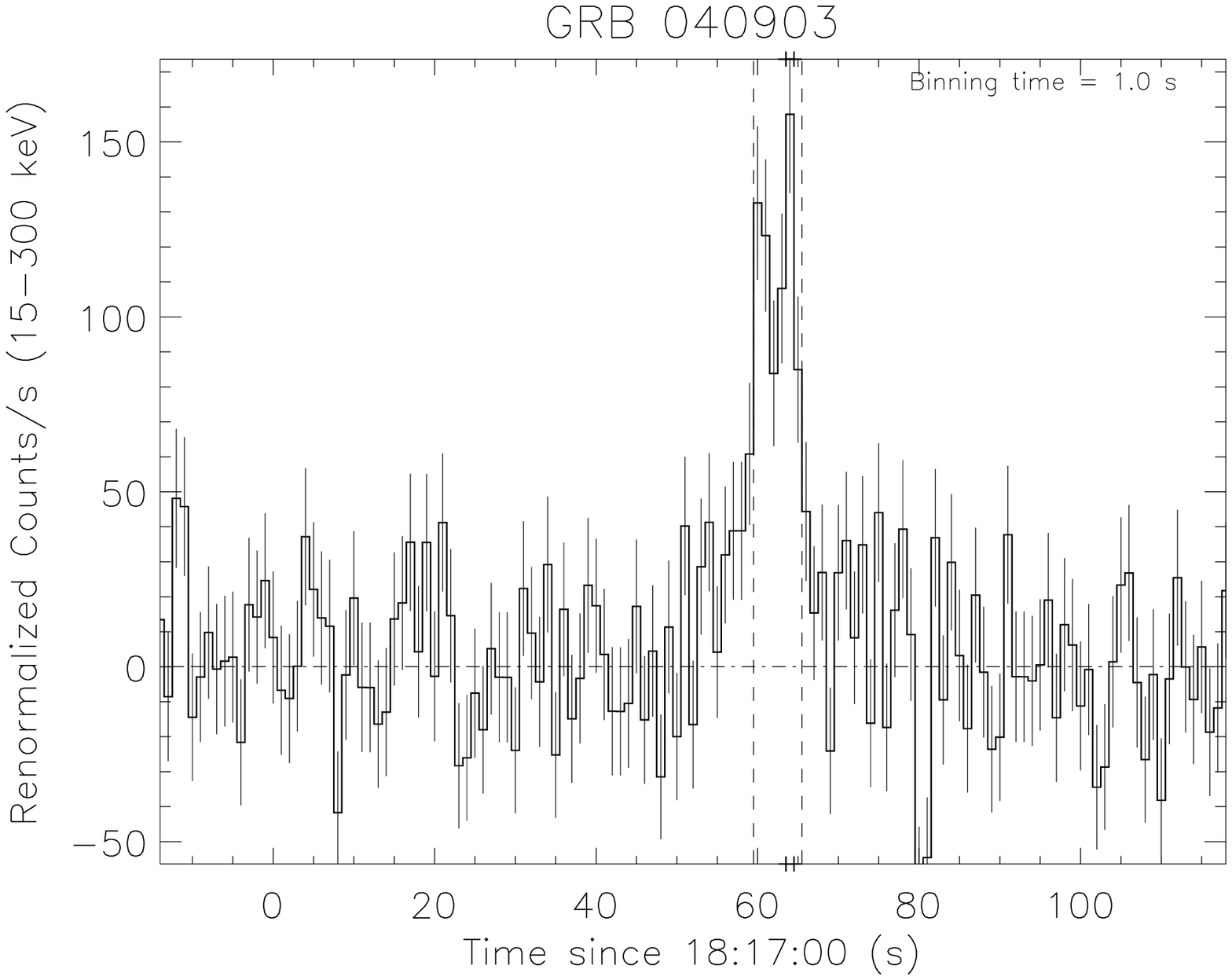,width=0.5\linewidth} &
 \psfig{figure=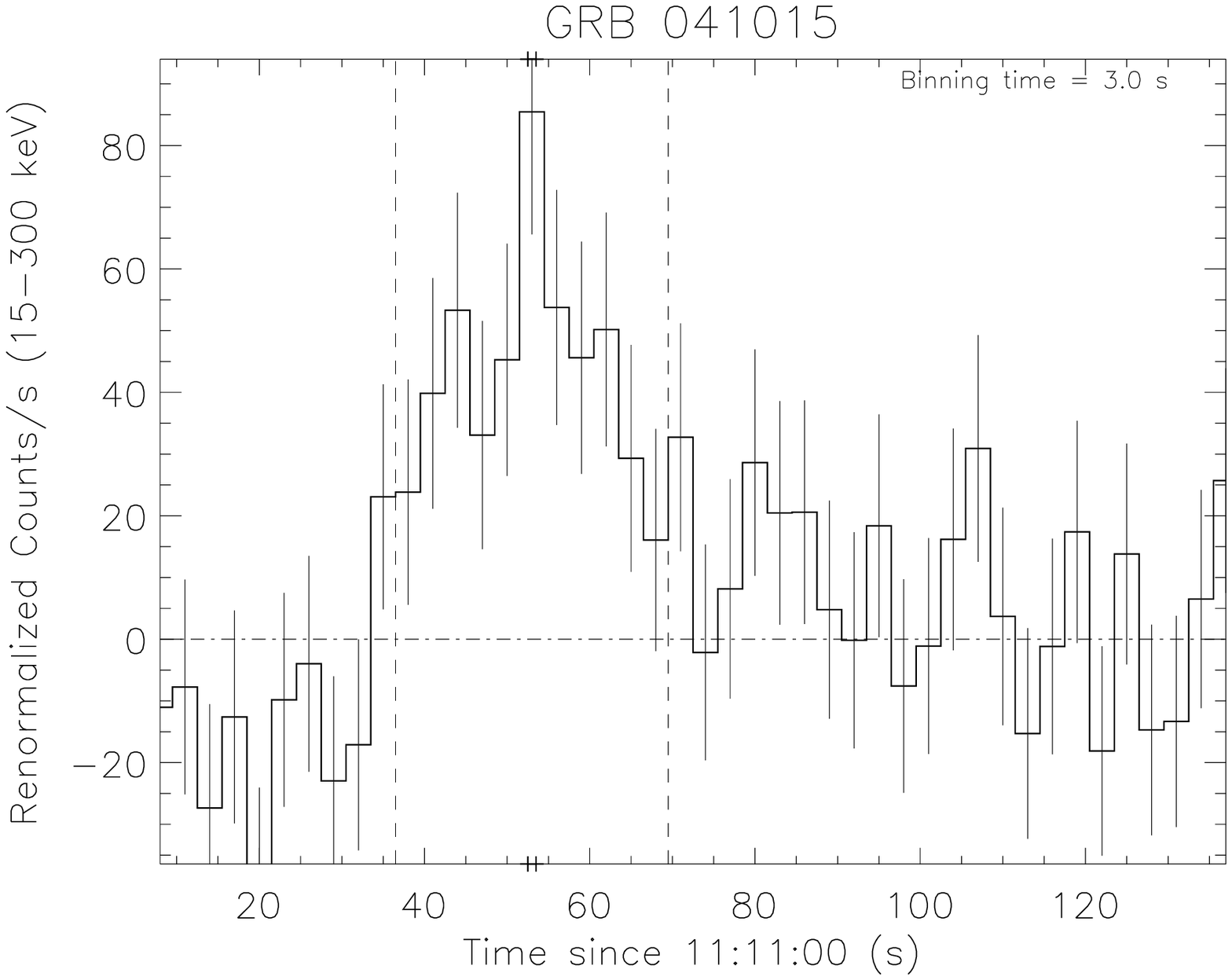,width=0.5\linewidth} \\
\end{tabular}
\textbf{Fig.~\ref{fig:light_curves}.} Light curves of \int\ GRBs (continued)
\end{figure}
\begin{figure}[p]
\vspace{1.0cm}
\begin{tabular}{cc}
\psfig{figure=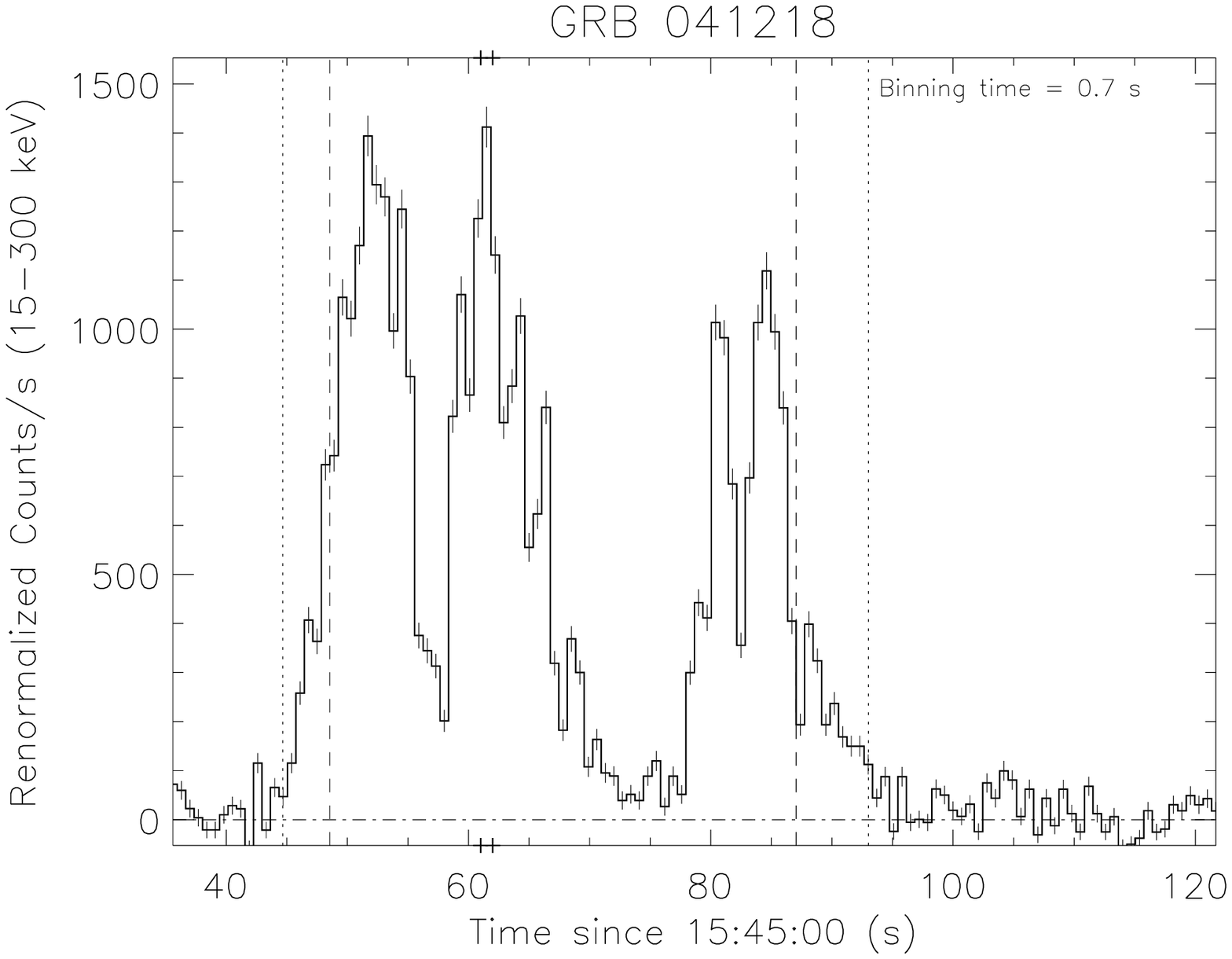,width=0.5\linewidth} &
 \psfig{figure=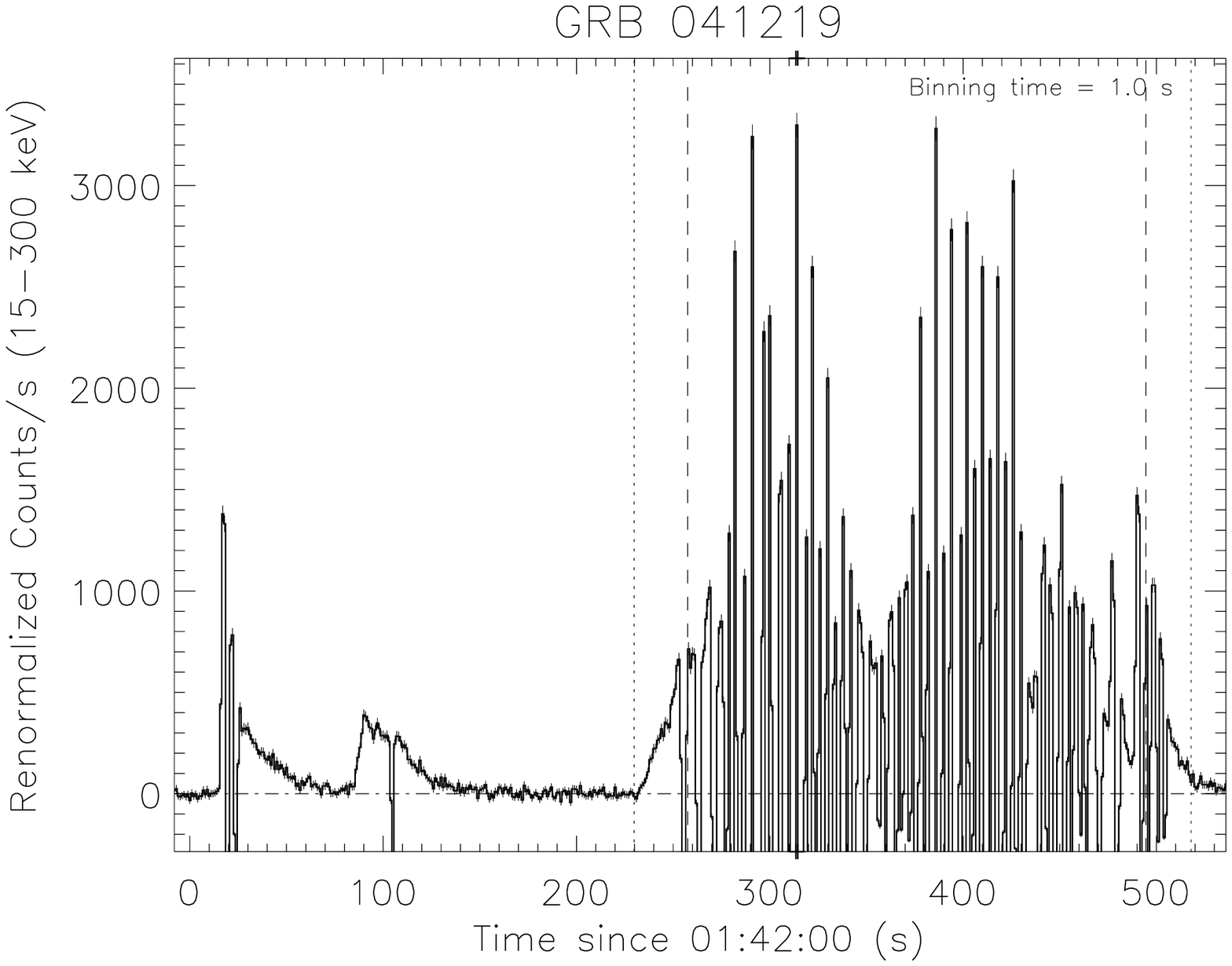,width=0.5\linewidth} \\
\psfig{figure=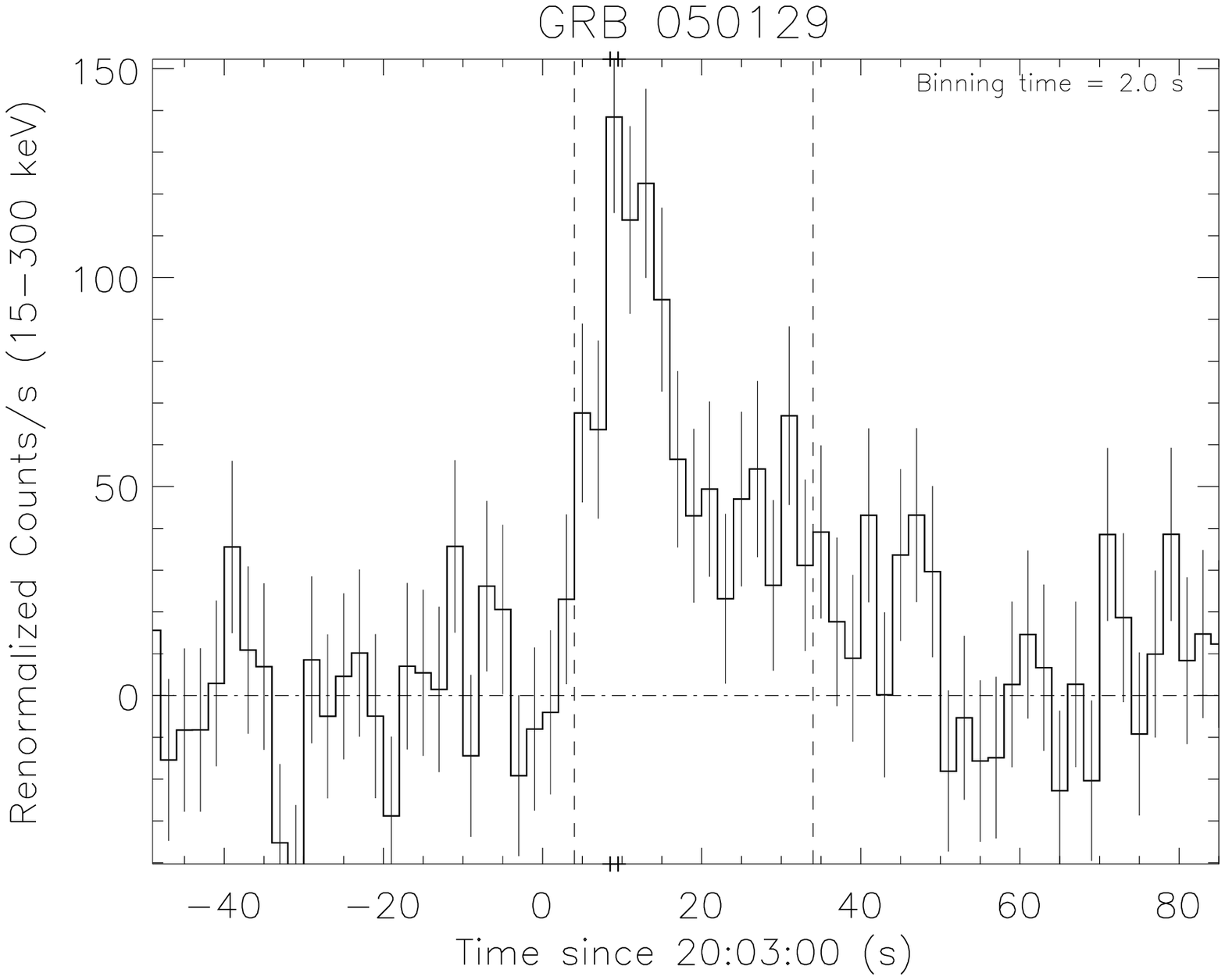,width=0.5\linewidth} &
 \psfig{figure=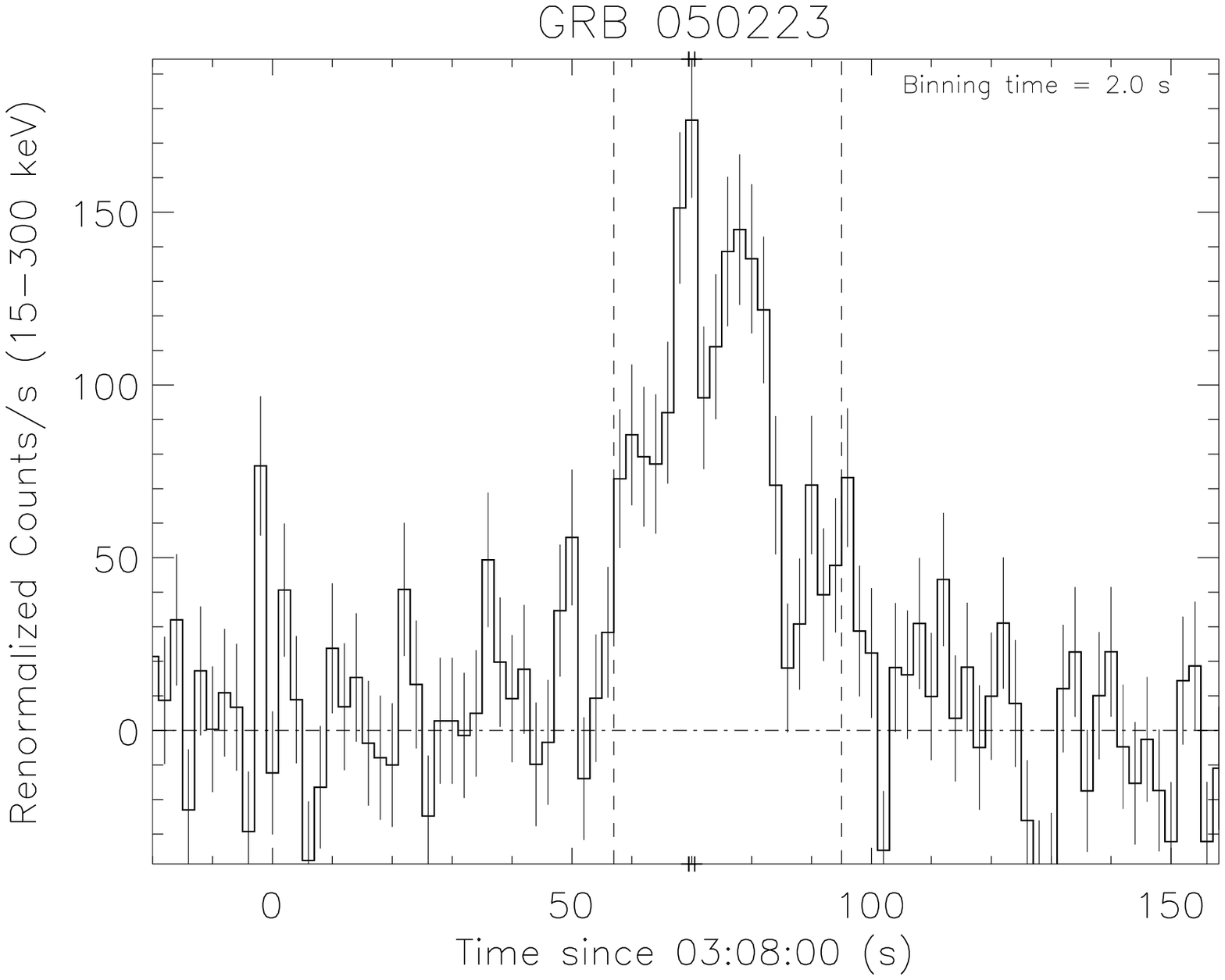,width=0.5\linewidth} \\
\psfig{figure=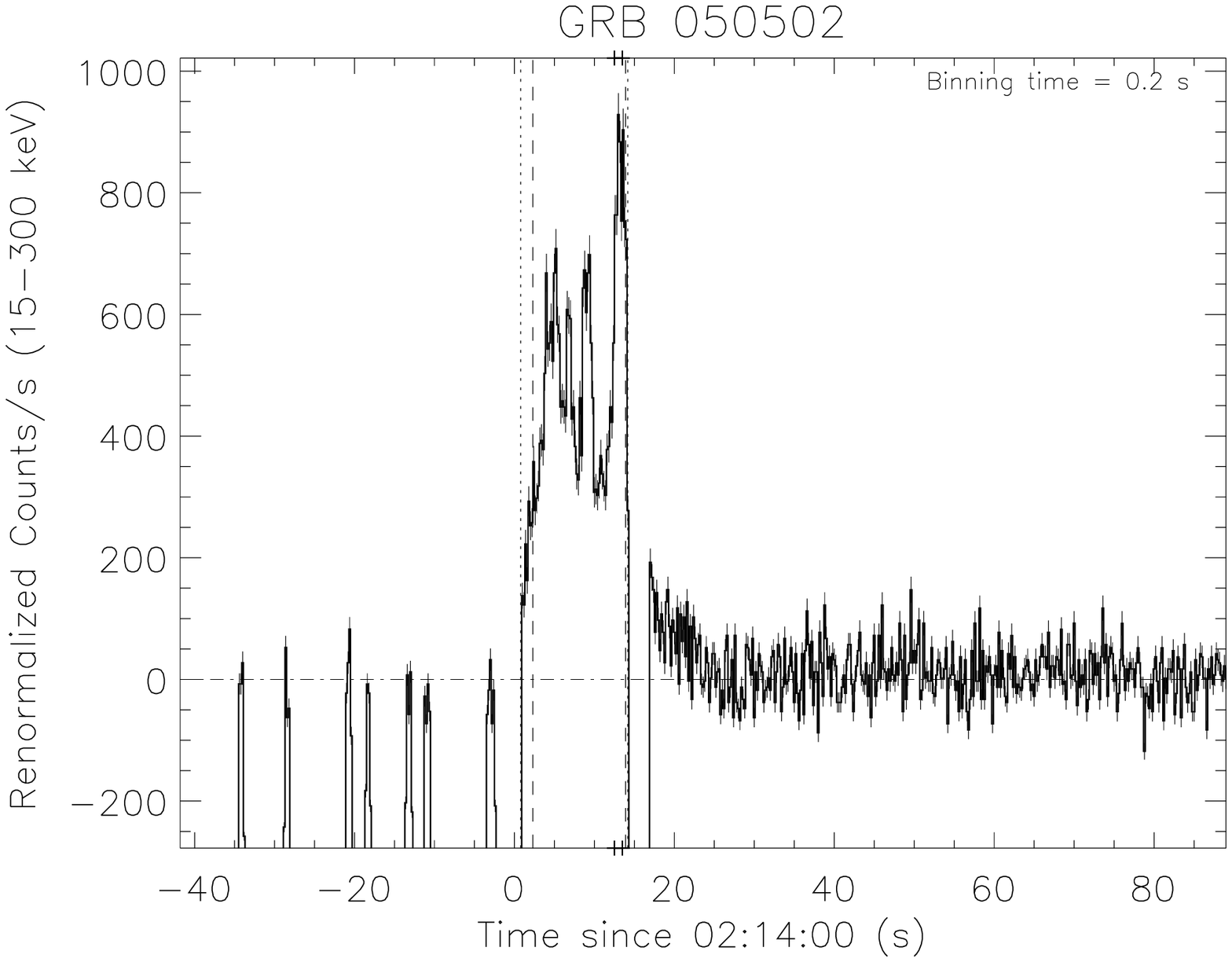,width=0.5\linewidth} &
 \psfig{figure=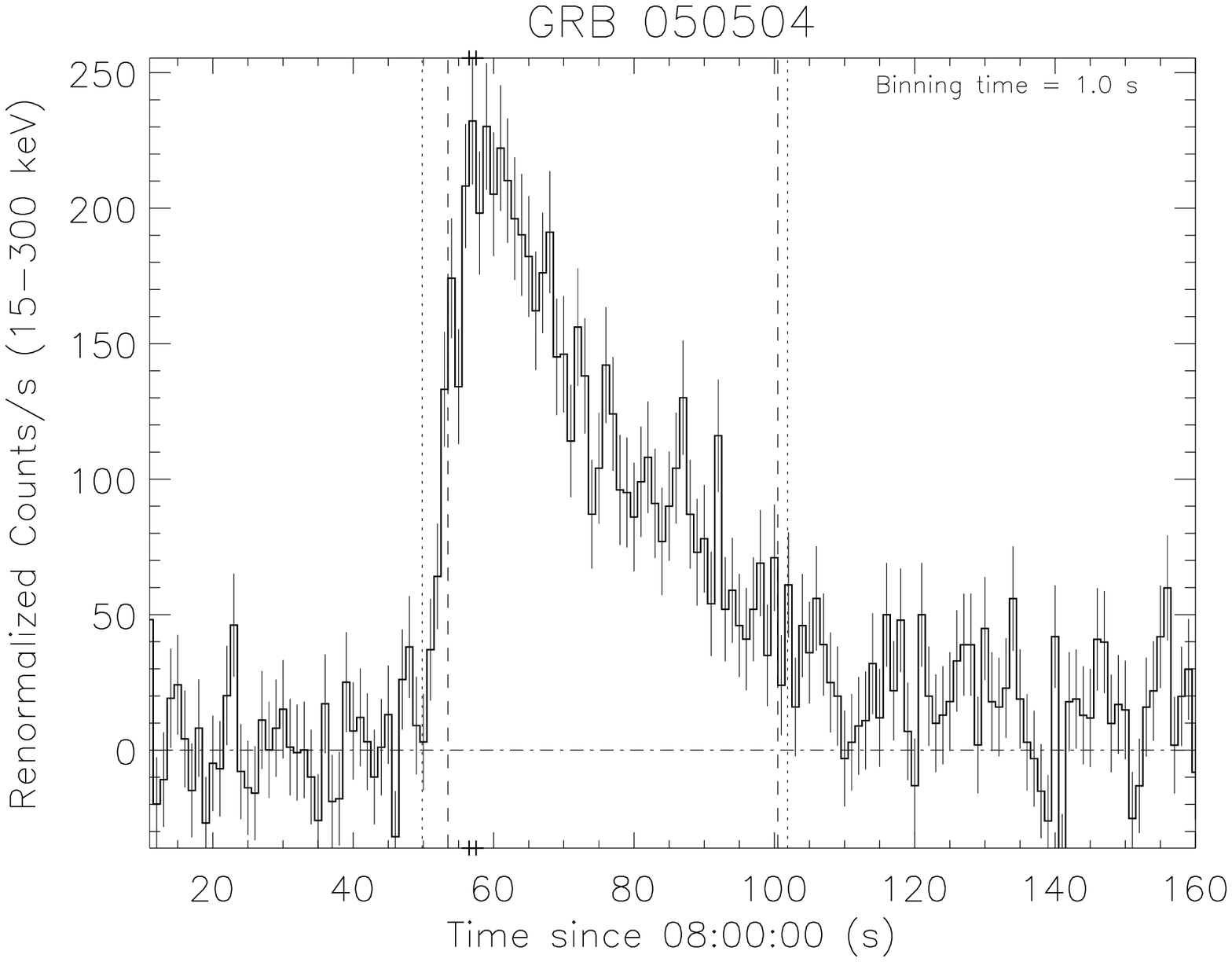,width=0.5\linewidth} \\
\end{tabular}
\textbf{Fig.~\ref{fig:light_curves}.} Light curves of \int\ GRBs (continued)
\end{figure}
\begin{figure}[p]
\vspace{1.0cm}
\begin{tabular}{cc}
\psfig{figure=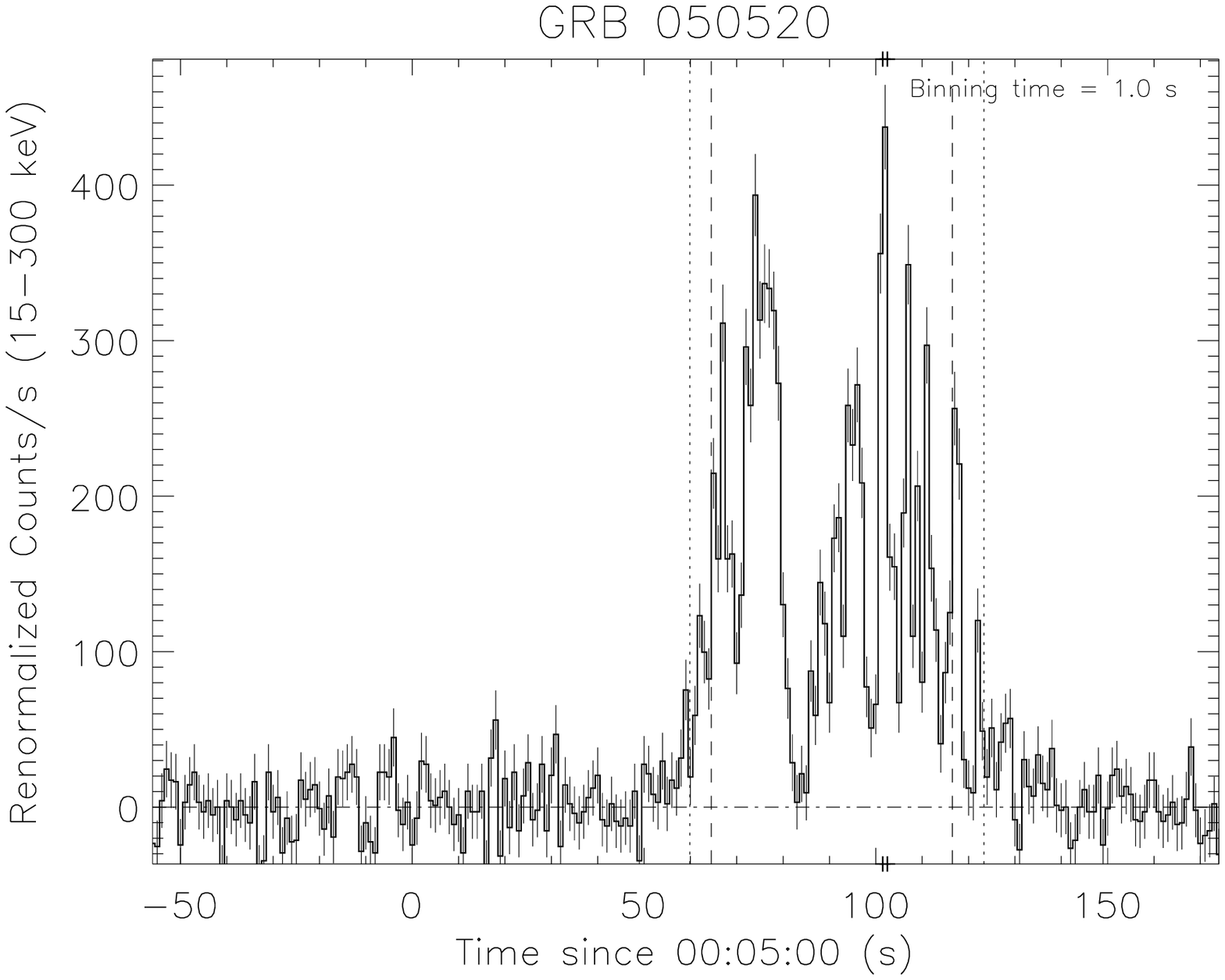,width=0.5\linewidth} &
 \psfig{figure=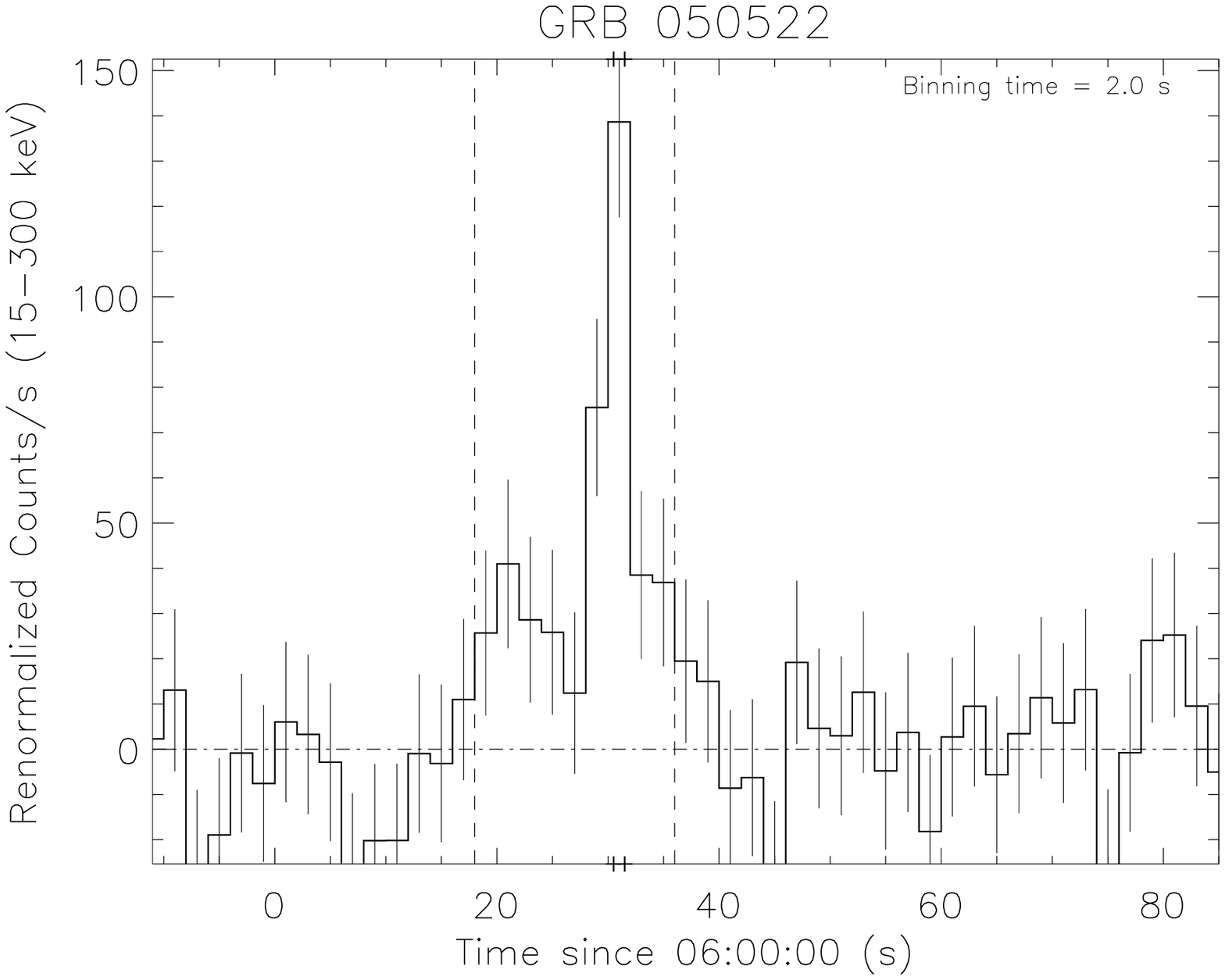,width=0.5\linewidth} \\
\psfig{figure=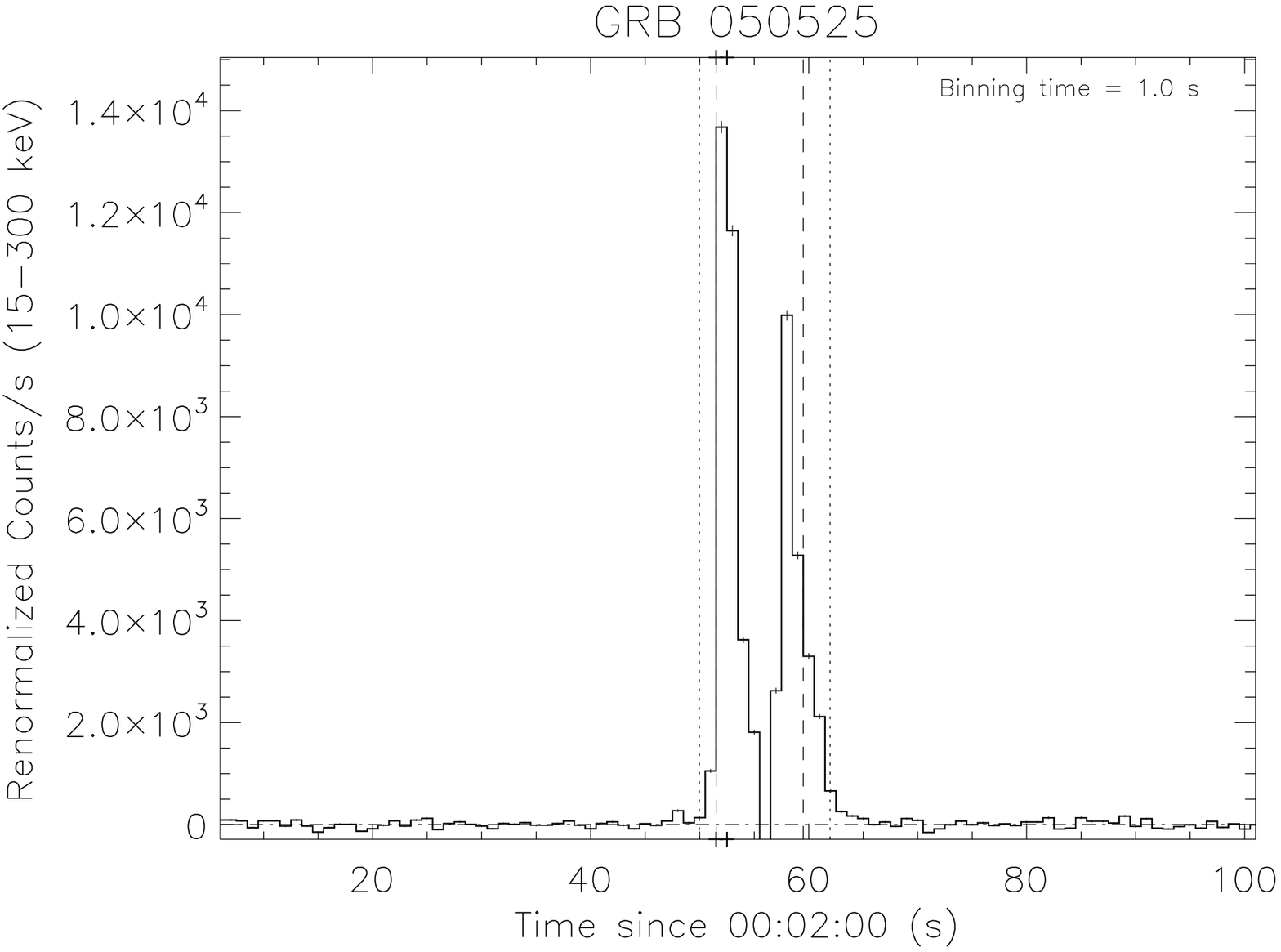,width=0.5\linewidth} &
 \psfig{figure=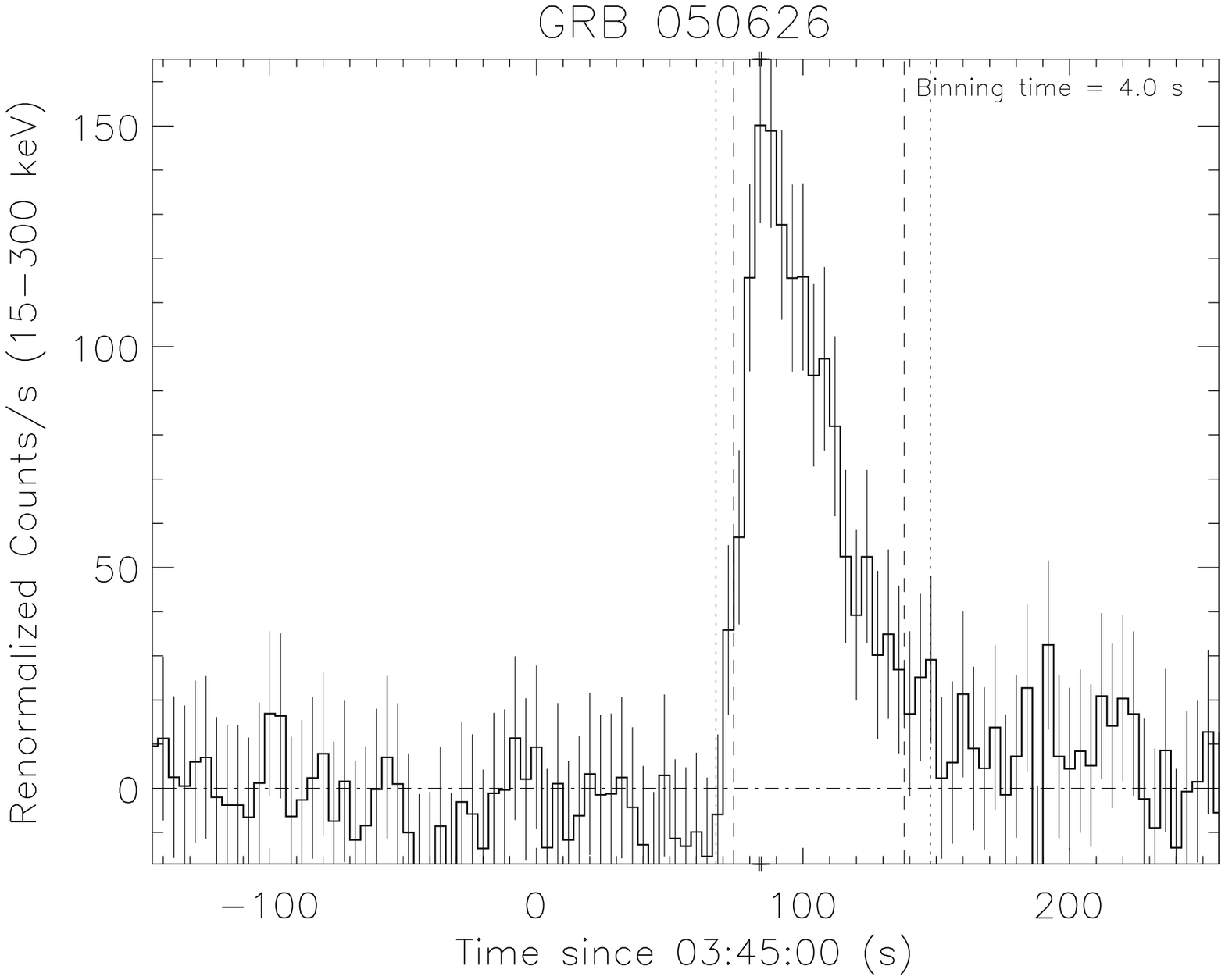,width=0.5\linewidth} \\
\psfig{figure=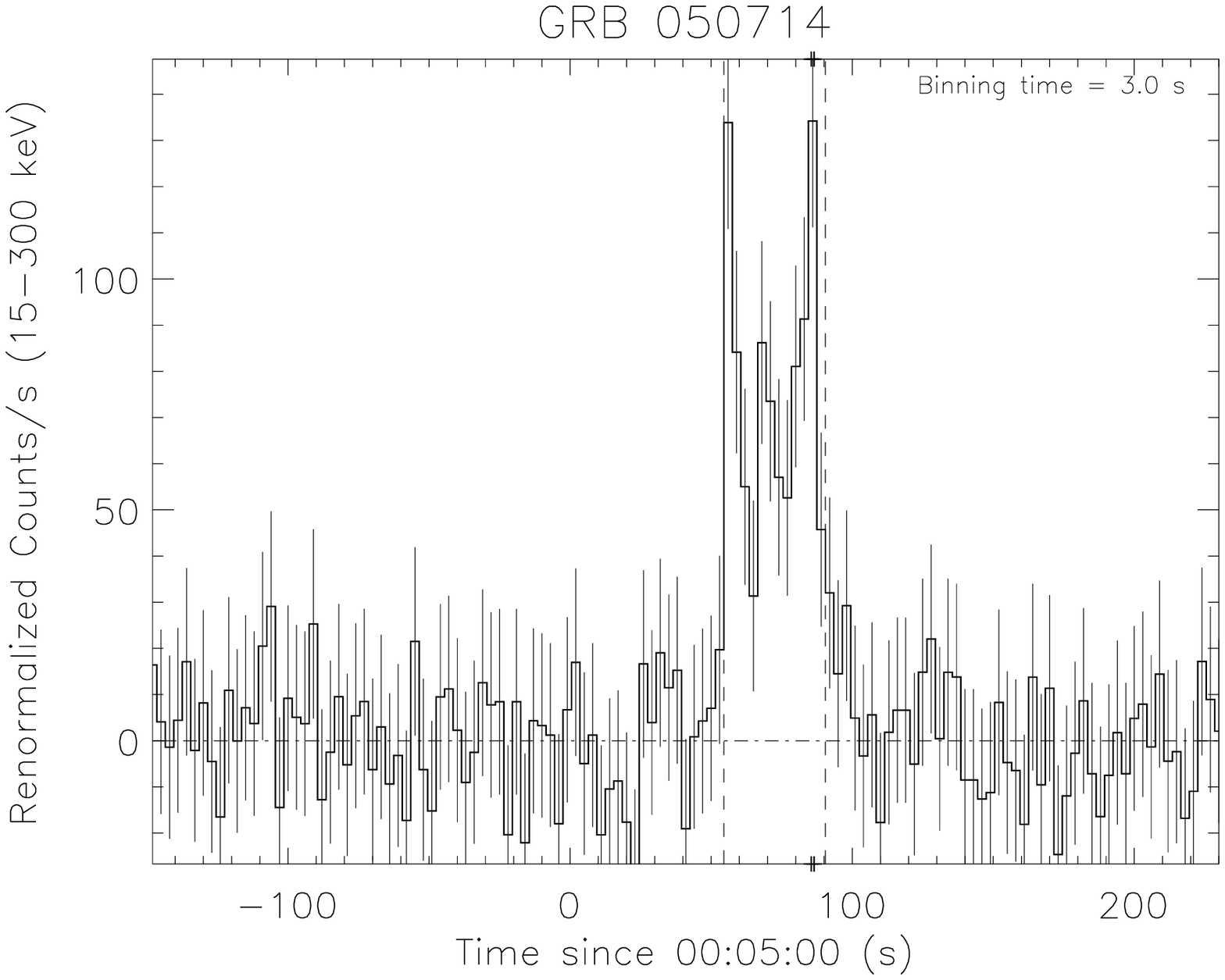,width=0.5\linewidth} &
 \psfig{figure=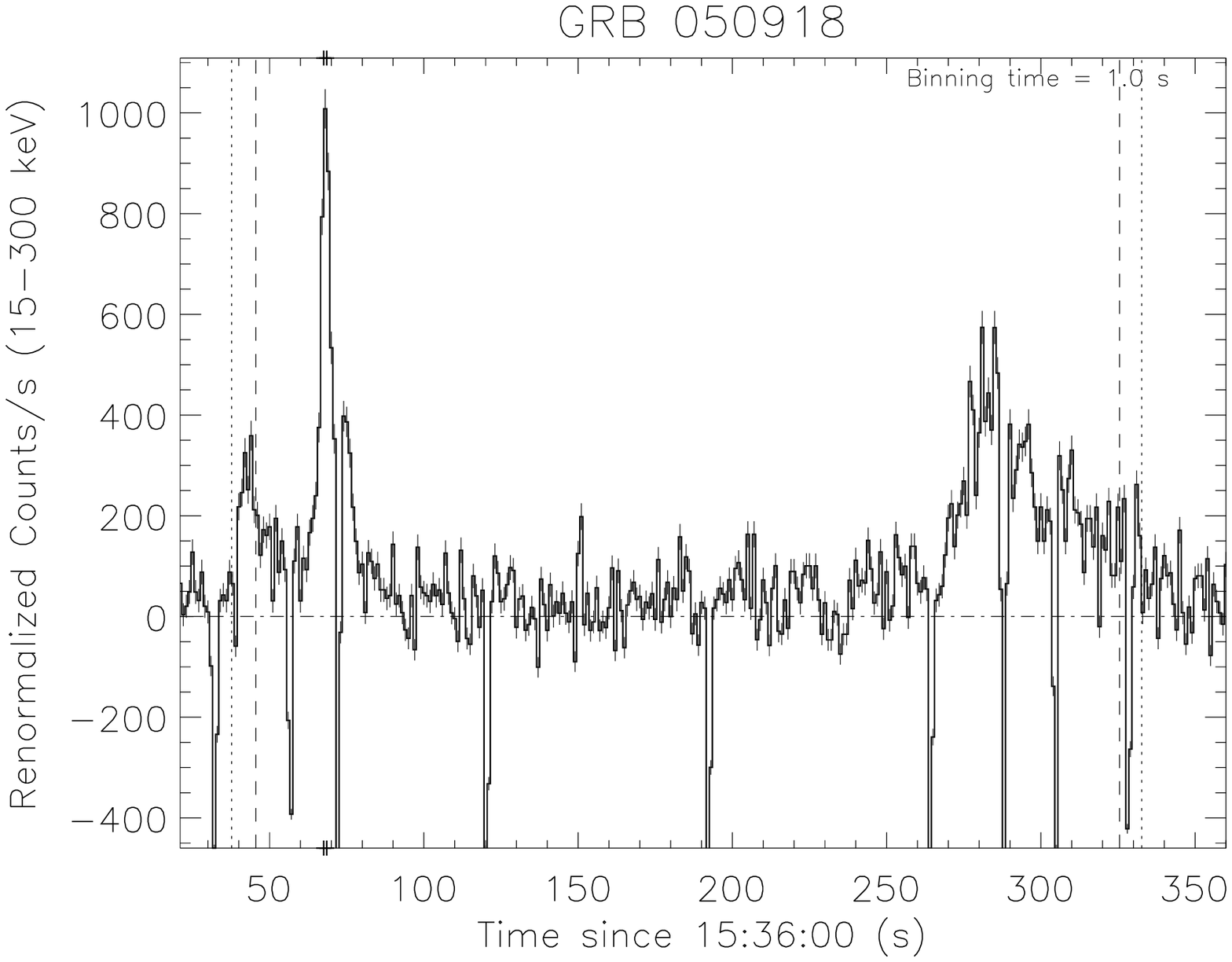,width=0.5\linewidth} \\
\end{tabular}
\textbf{Fig.~\ref{fig:light_curves}.} Light curves of \int\ GRBs (continued)
\end{figure}
\begin{figure}[p]
\vspace{1.0cm}
\begin{tabular}{cc}
\psfig{figure=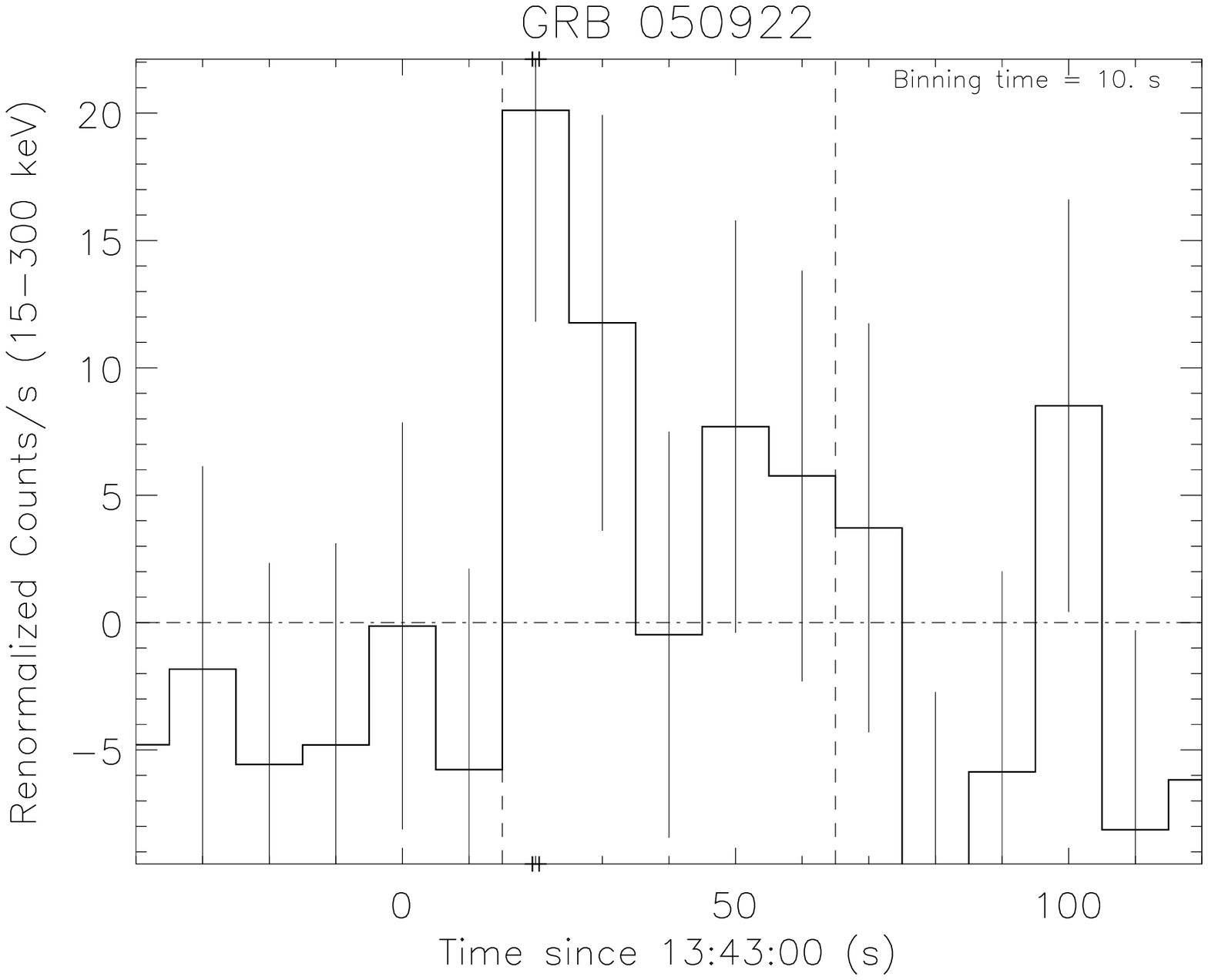,width=0.5\linewidth} &
 \psfig{figure=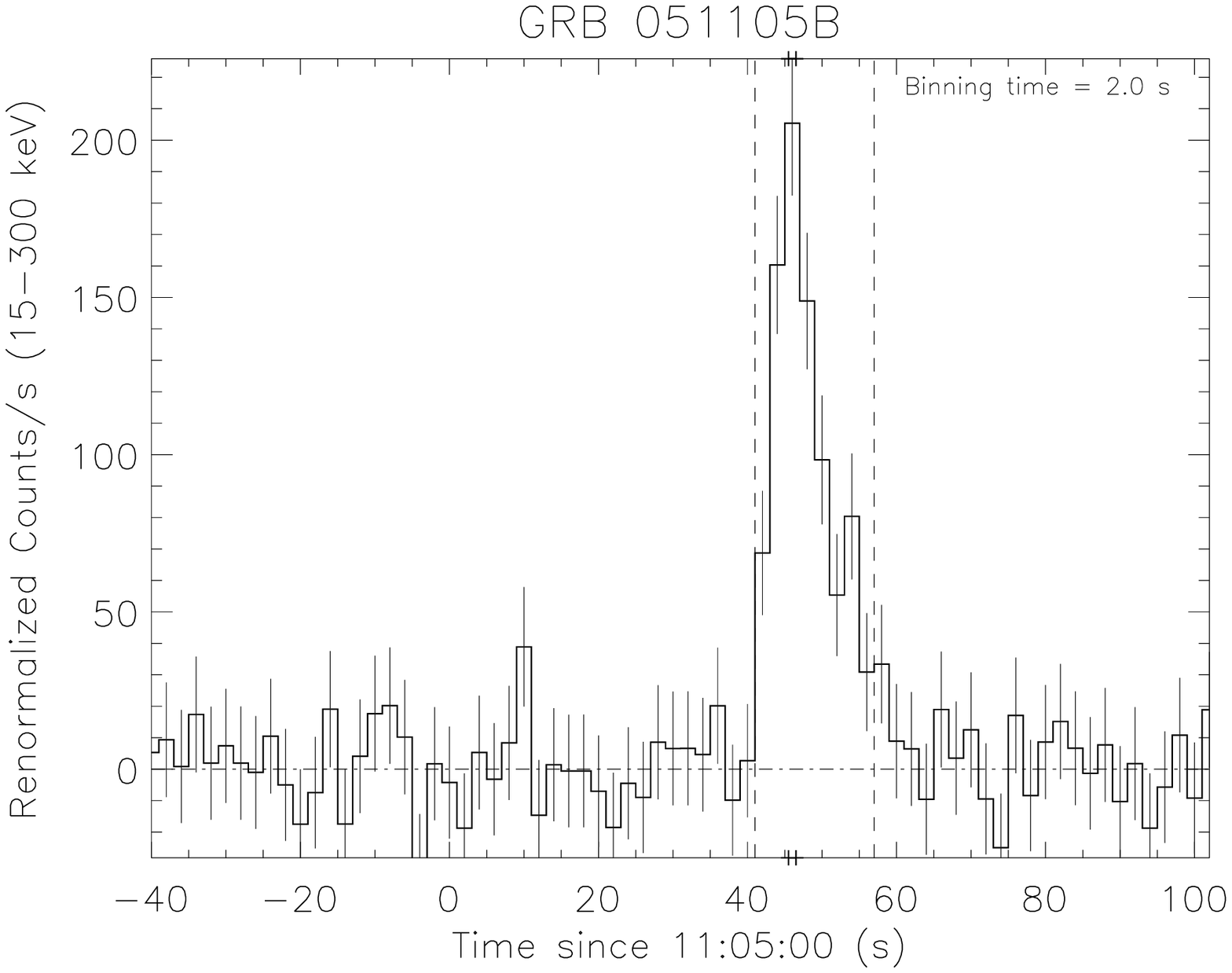,width=0.5\linewidth} \\
\psfig{figure=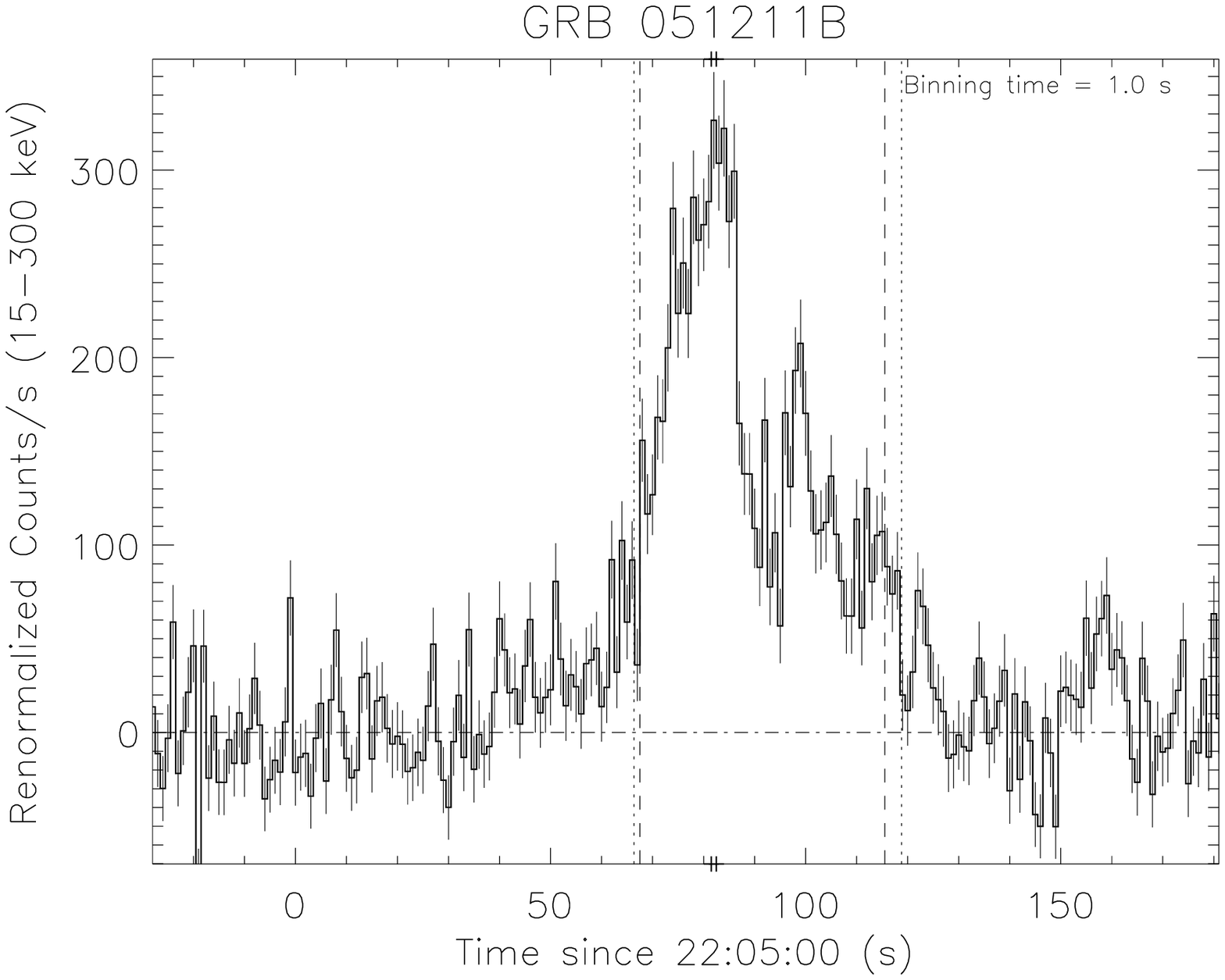,width=0.5\linewidth} &
 \psfig{figure=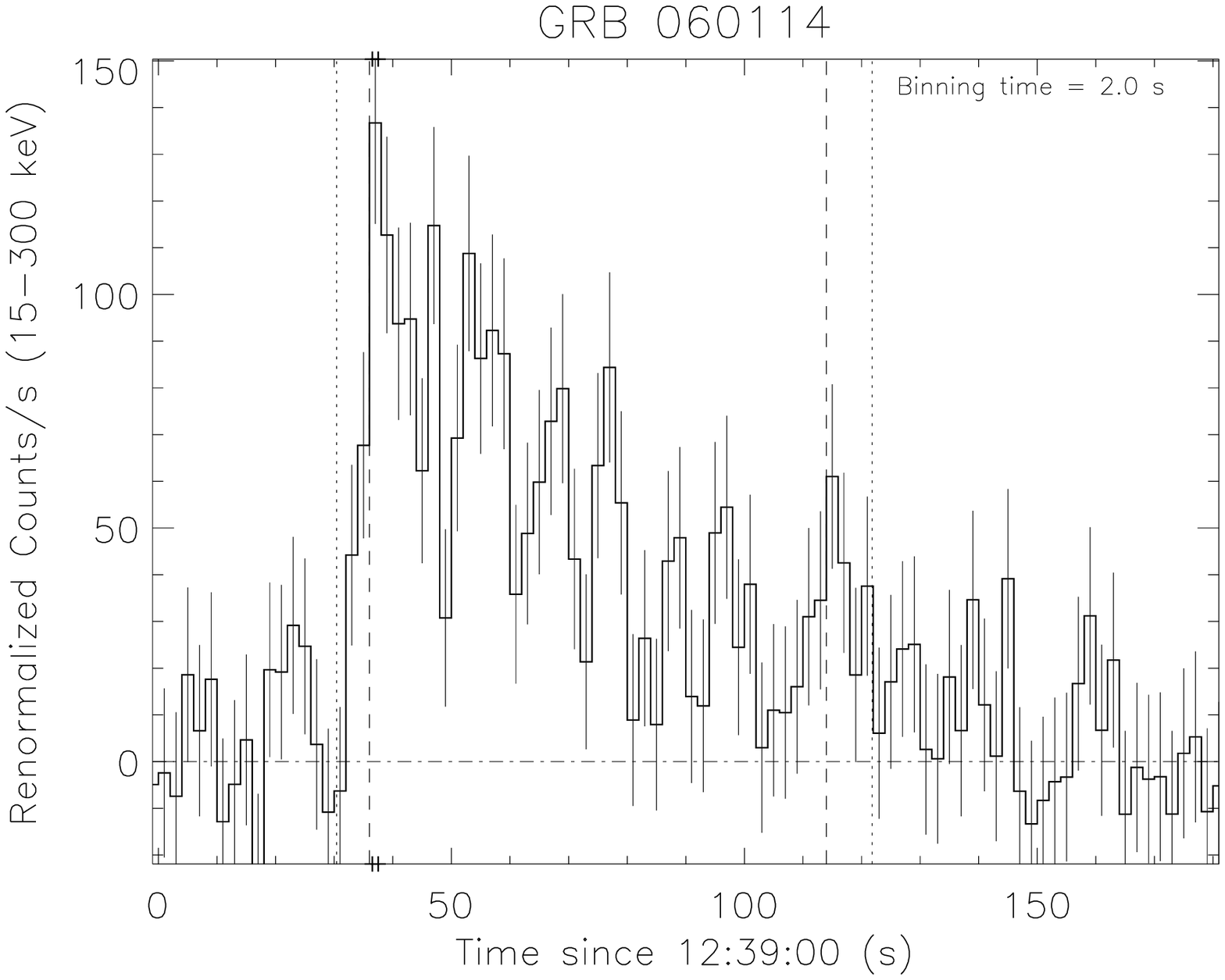,width=0.5\linewidth} \\
\psfig{figure=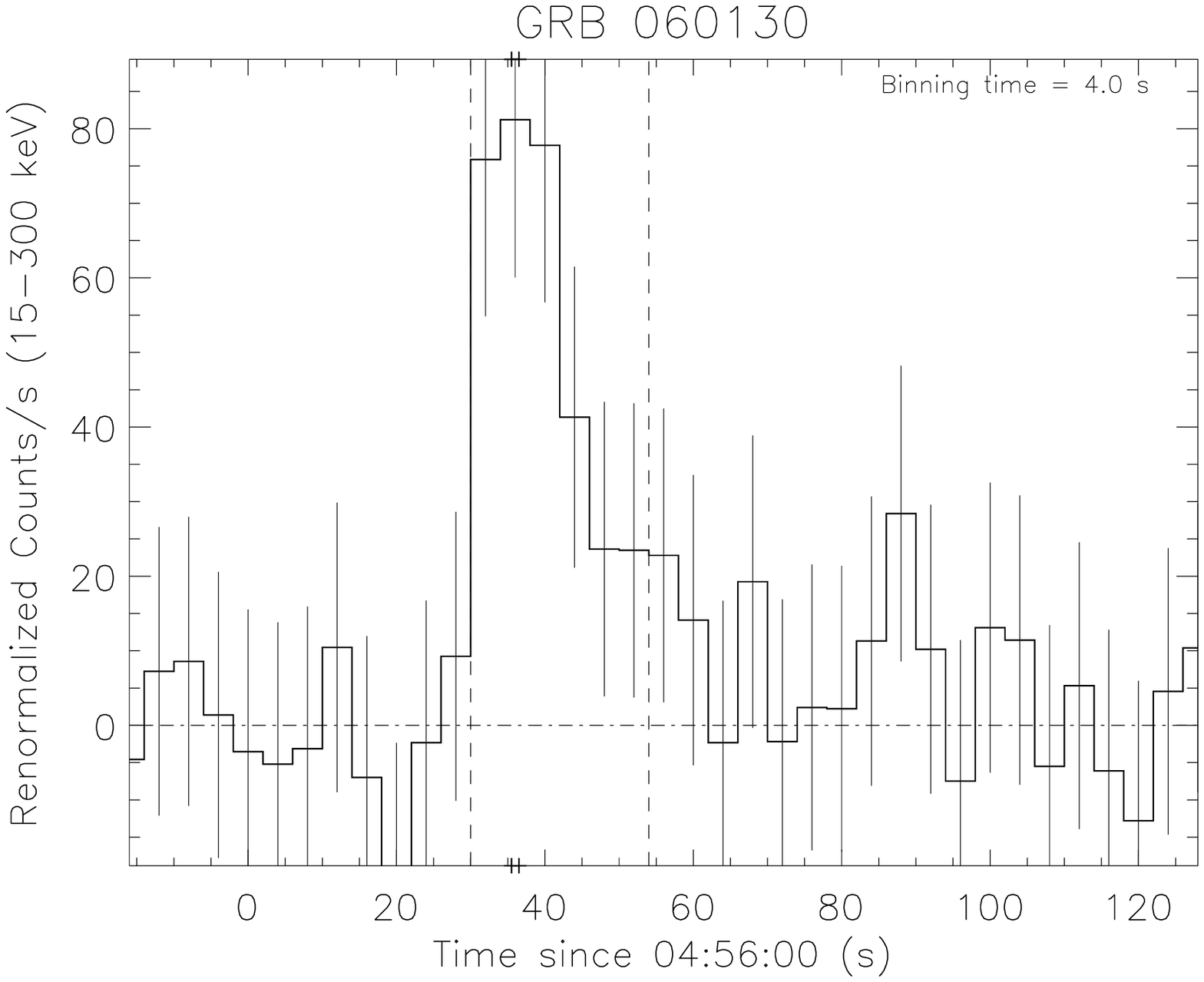,width=0.5\linewidth} &
 \psfig{figure=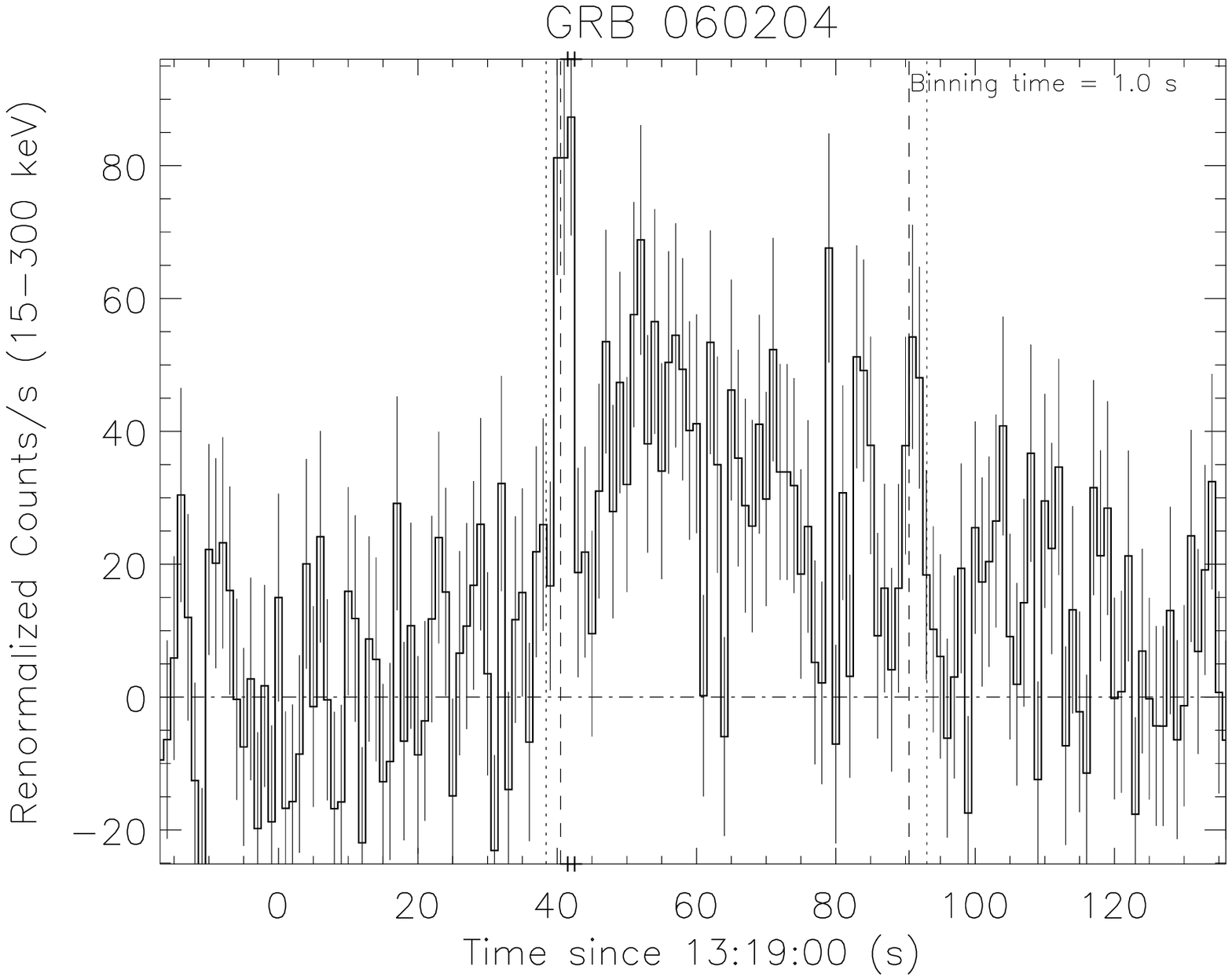,width=0.5\linewidth} \\
\end{tabular}
\textbf{Fig.~\ref{fig:light_curves}.} Light curves of \int\ GRBs (continued)
\end{figure}
\begin{figure}[p]
\vspace{1.0cm}
\begin{tabular}{cc}
\psfig{figure=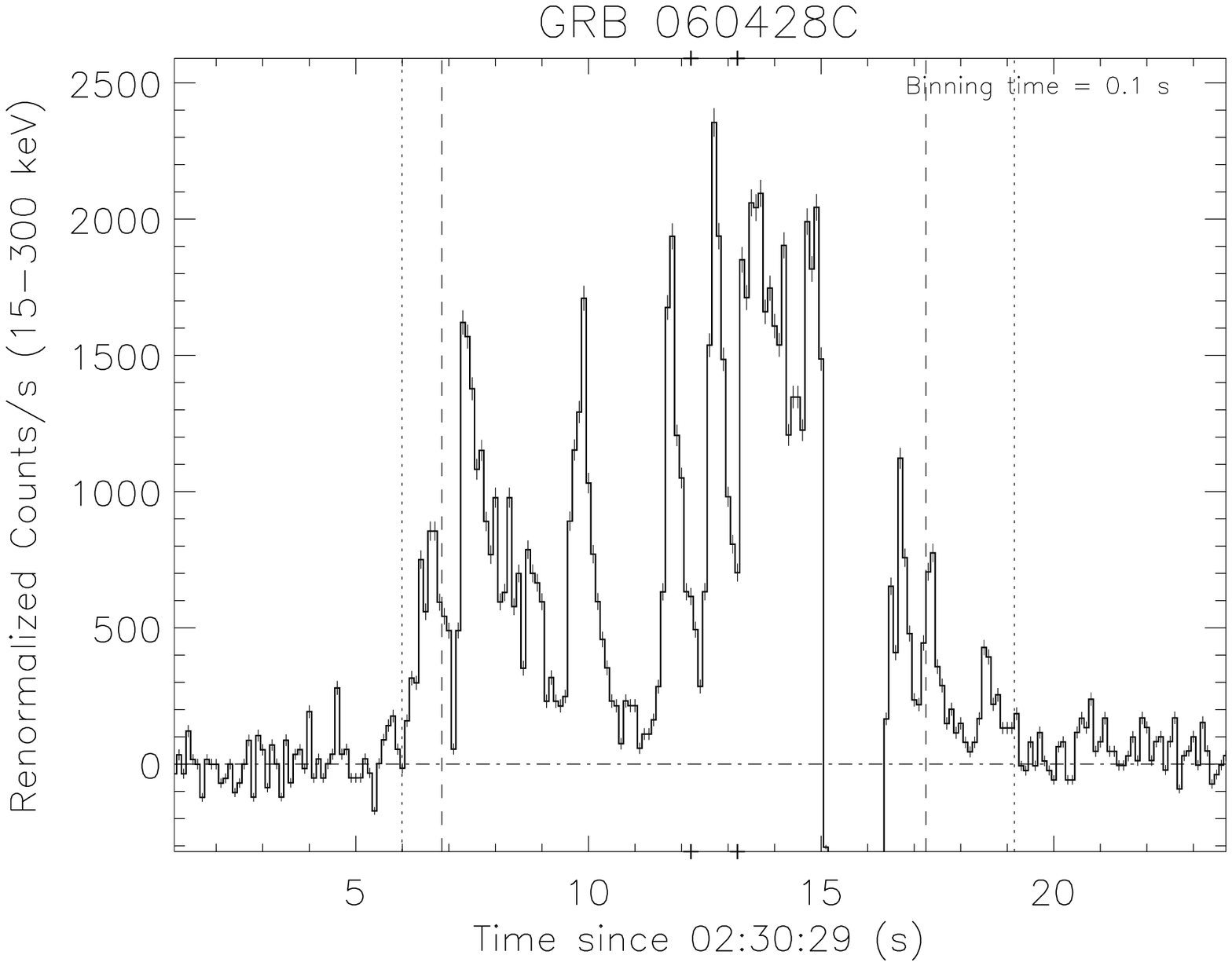,width=0.5\linewidth} &
 \psfig{figure=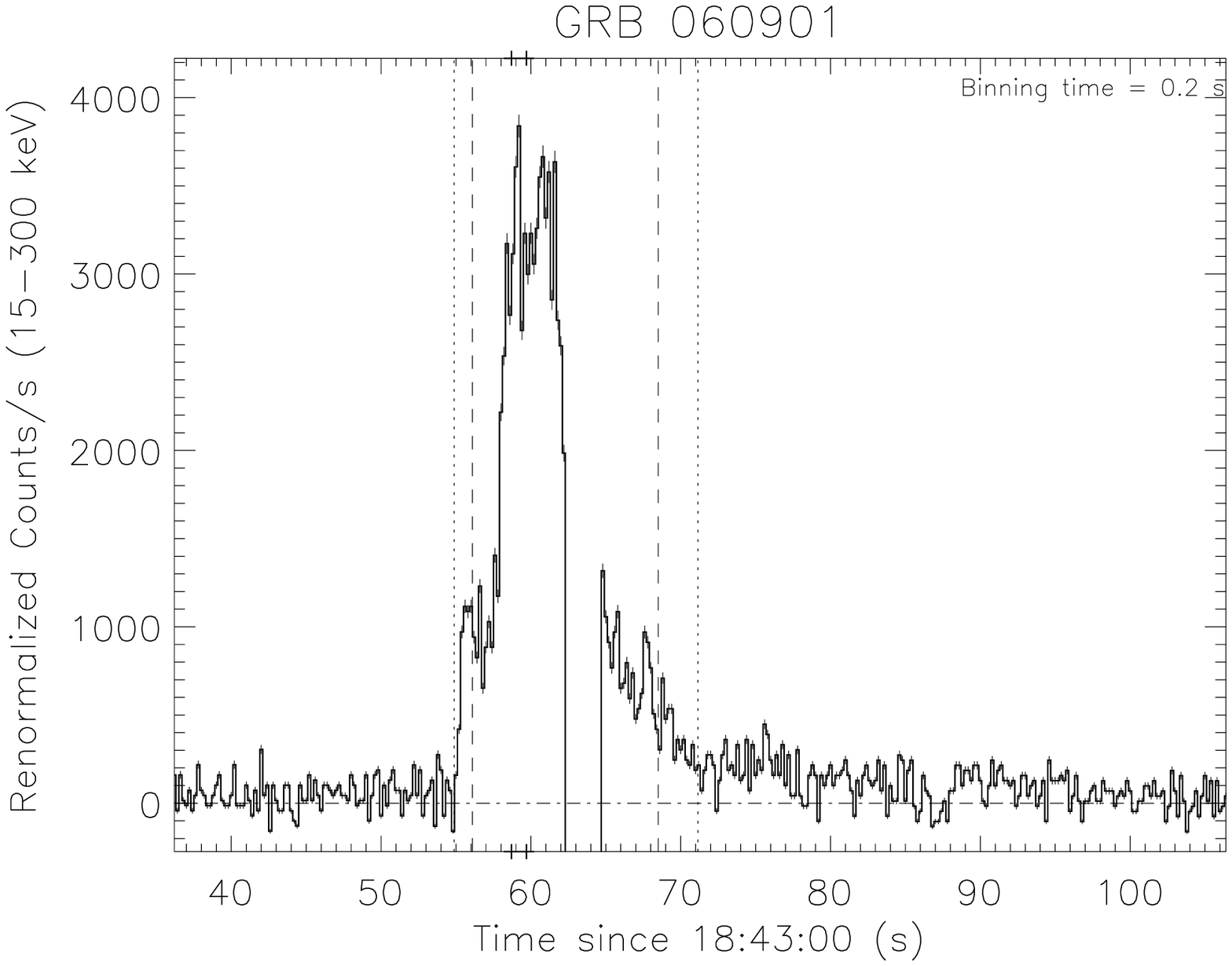,width=0.5\linewidth} \\
\psfig{figure=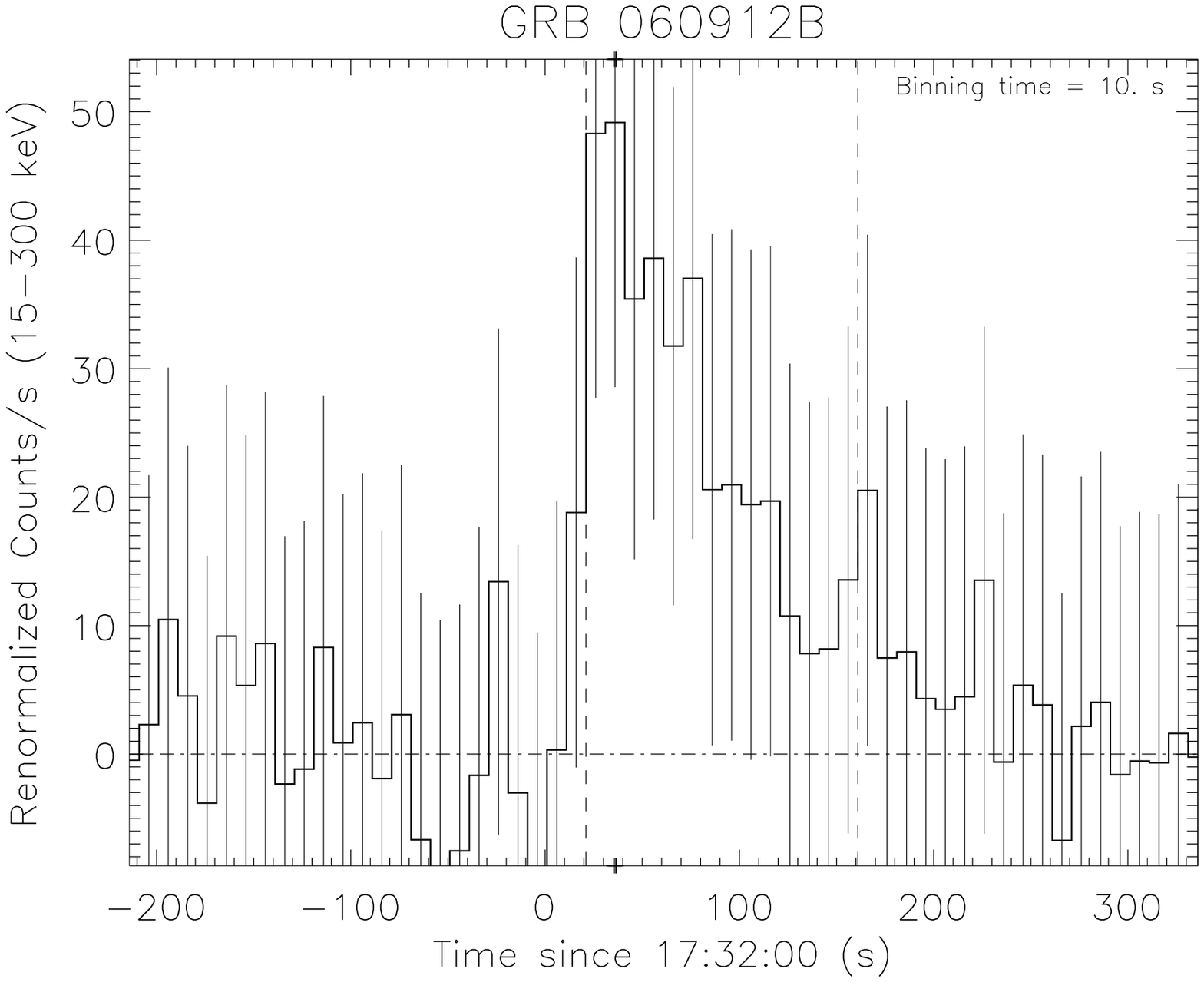,width=0.5\linewidth} &
 \psfig{figure=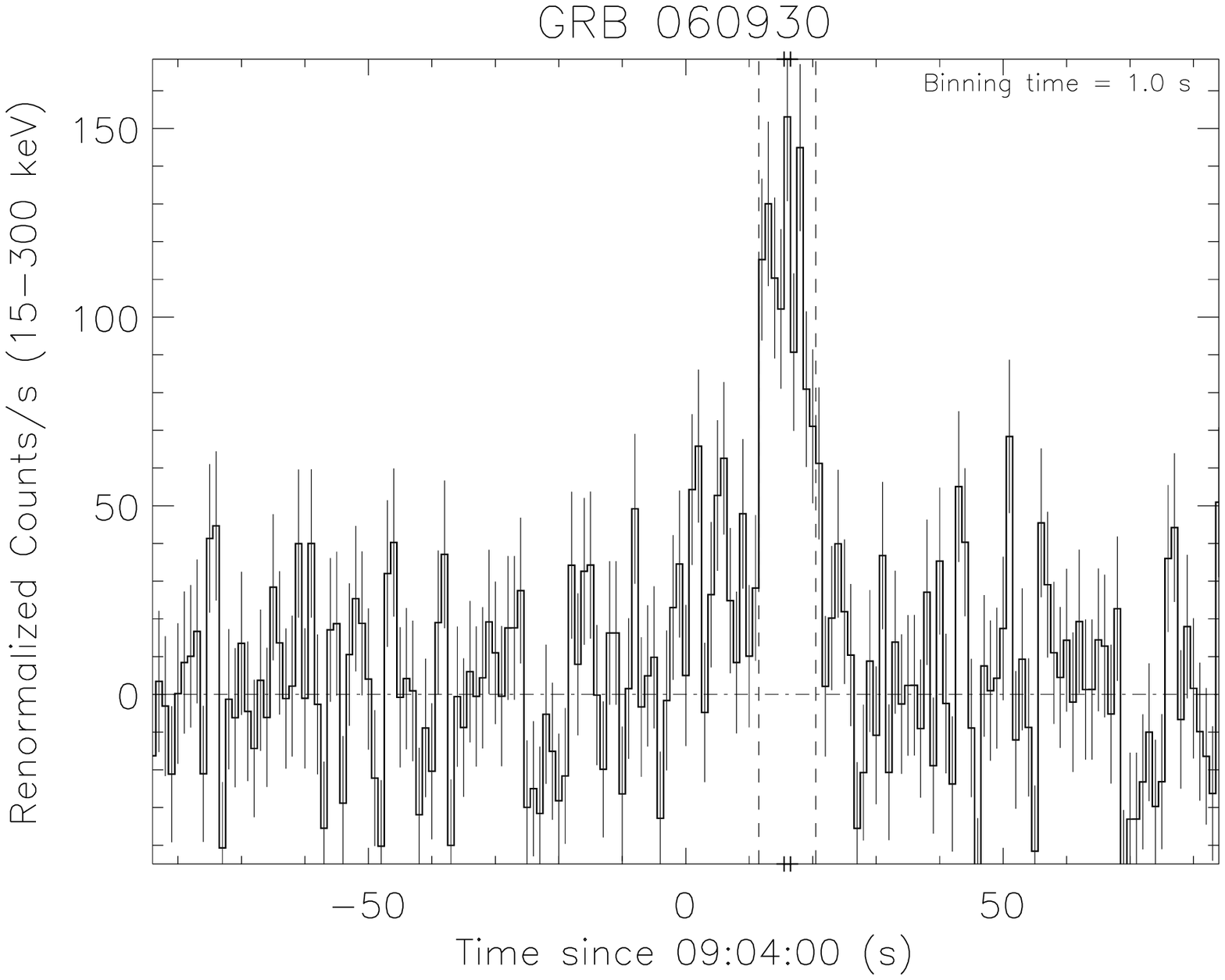,width=0.5\linewidth} \\
\psfig{figure=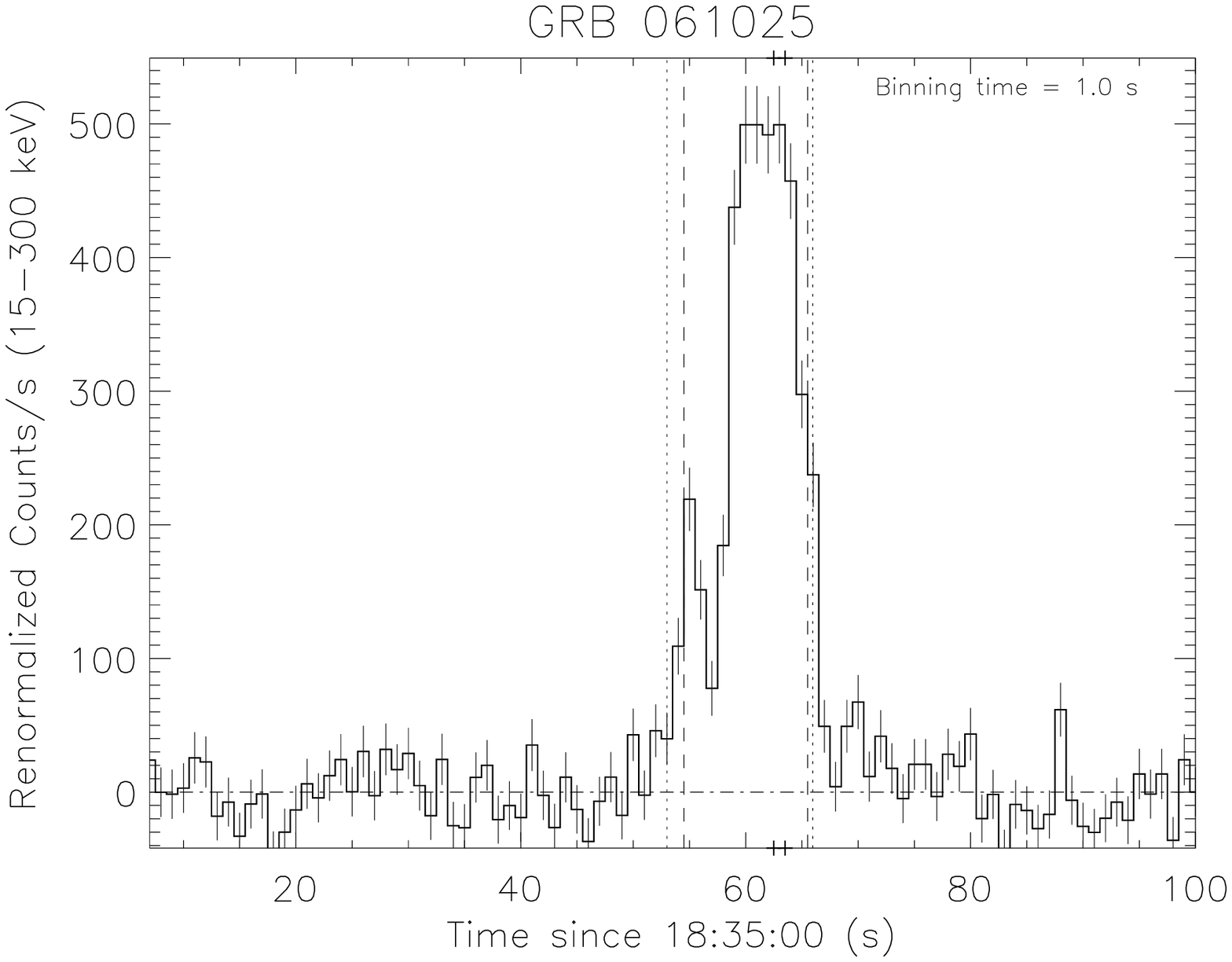,width=0.5\linewidth} &
 \psfig{figure=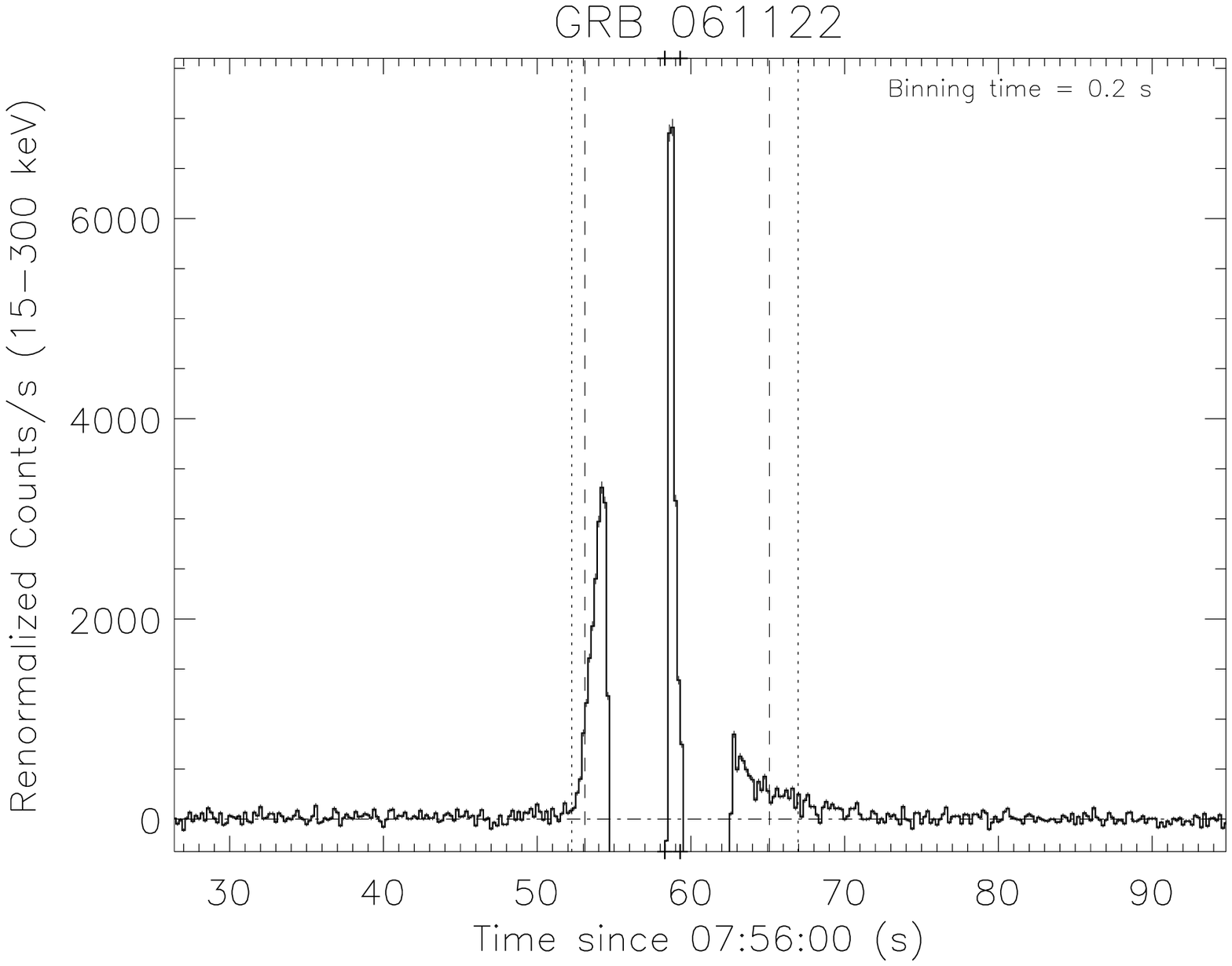,width=0.5\linewidth} \\
\end{tabular}
\textbf{Fig.~\ref{fig:light_curves}.} Light curves of \int\ GRBs (continued)
\end{figure}
\begin{figure}[p]
\vspace{1.0cm}
\begin{tabular}{cc}
\psfig{figure=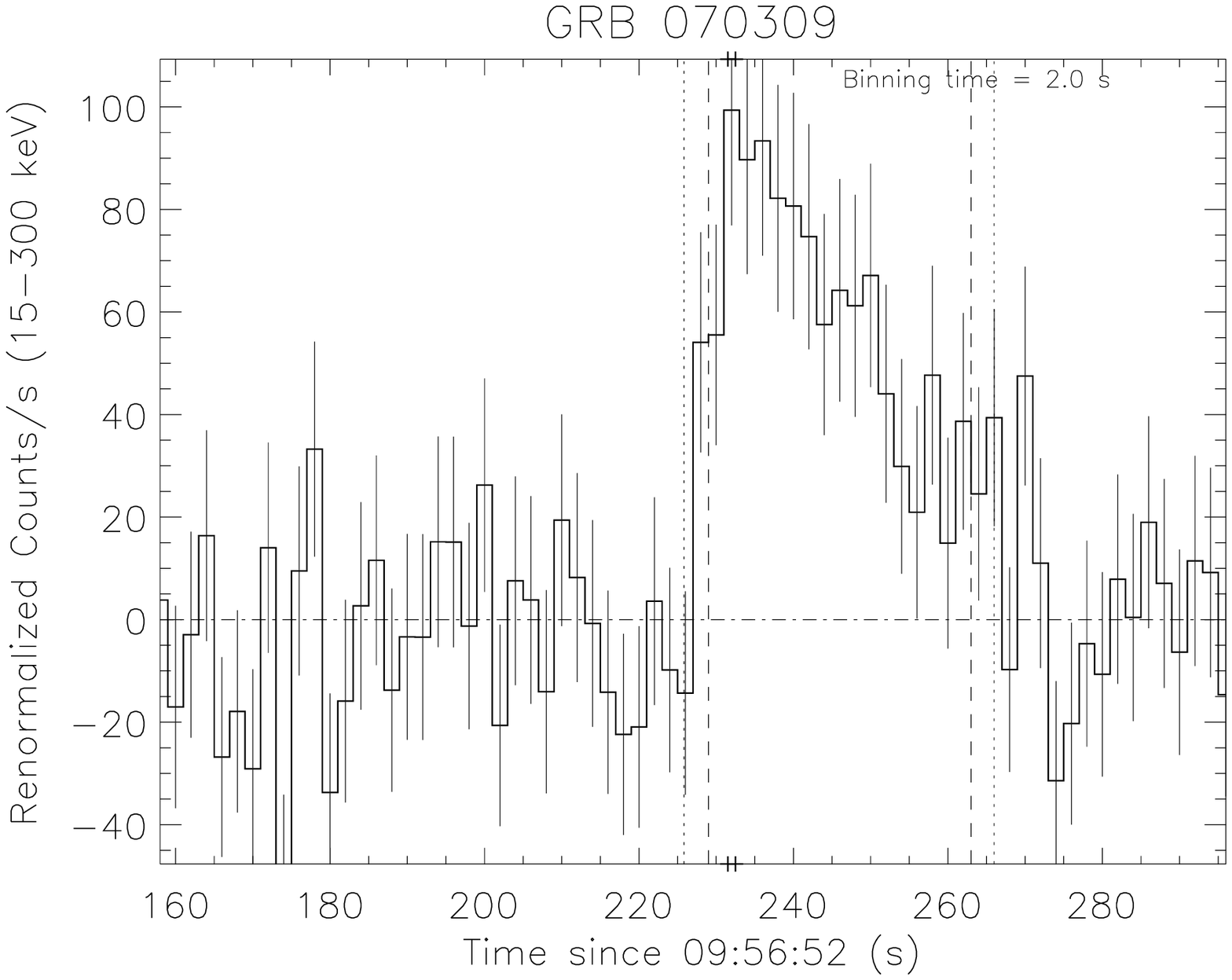,width=0.5\linewidth} &
 \psfig{figure=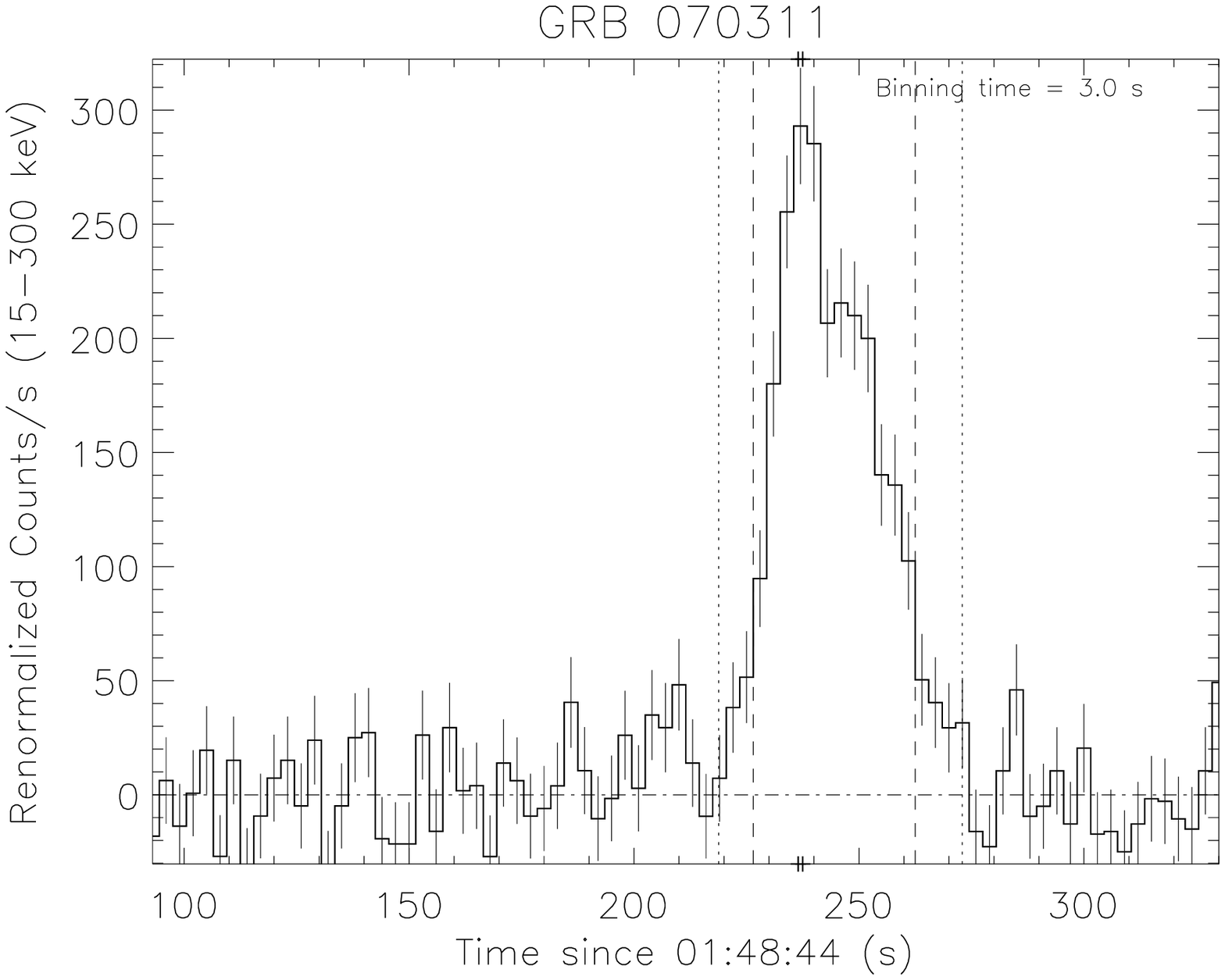,width=0.5\linewidth} \\
\psfig{figure=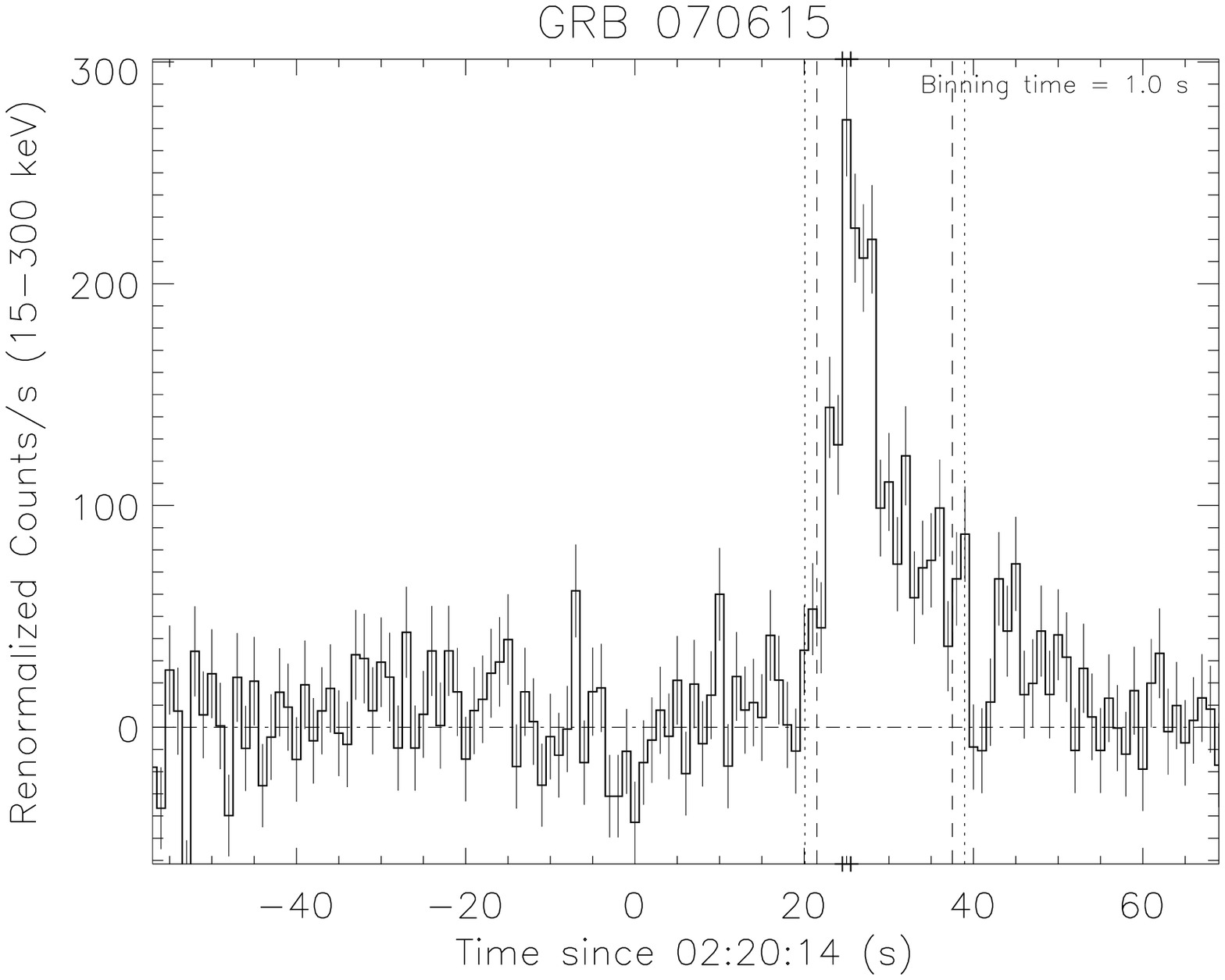,width=0.5\linewidth} &
 \psfig{figure=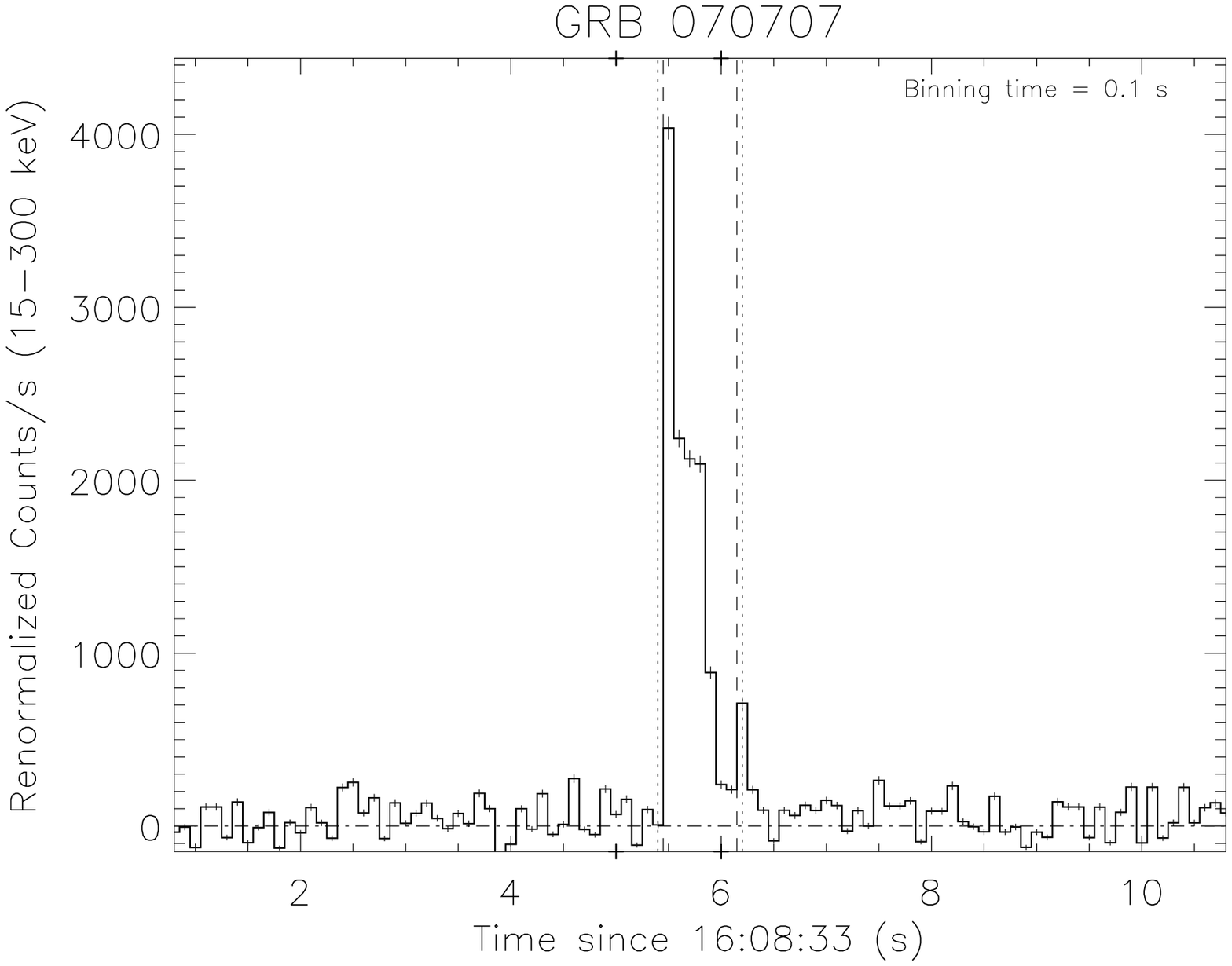,width=0.5\linewidth} \\
\psfig{figure=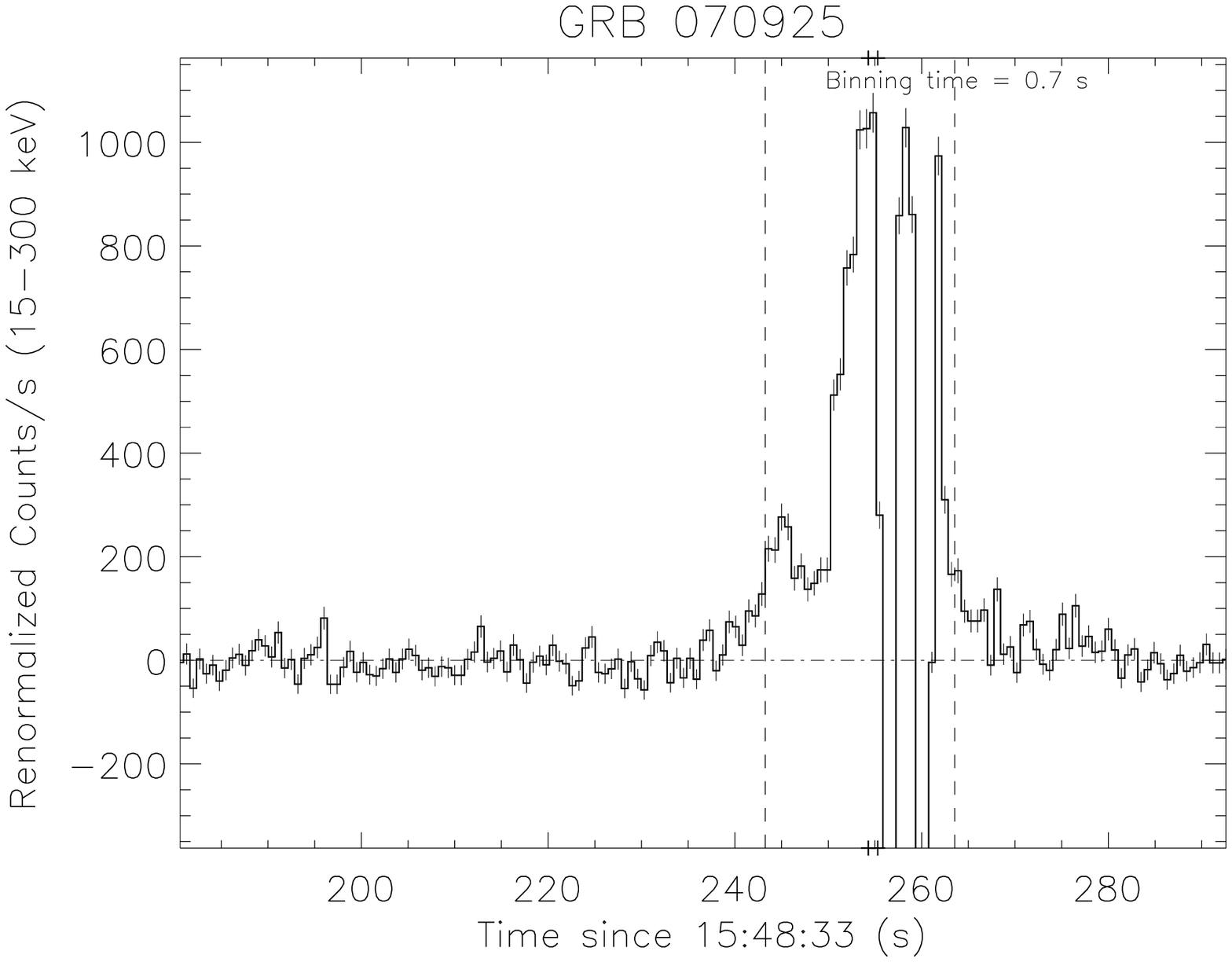,width=0.5\linewidth} &
 \psfig{figure=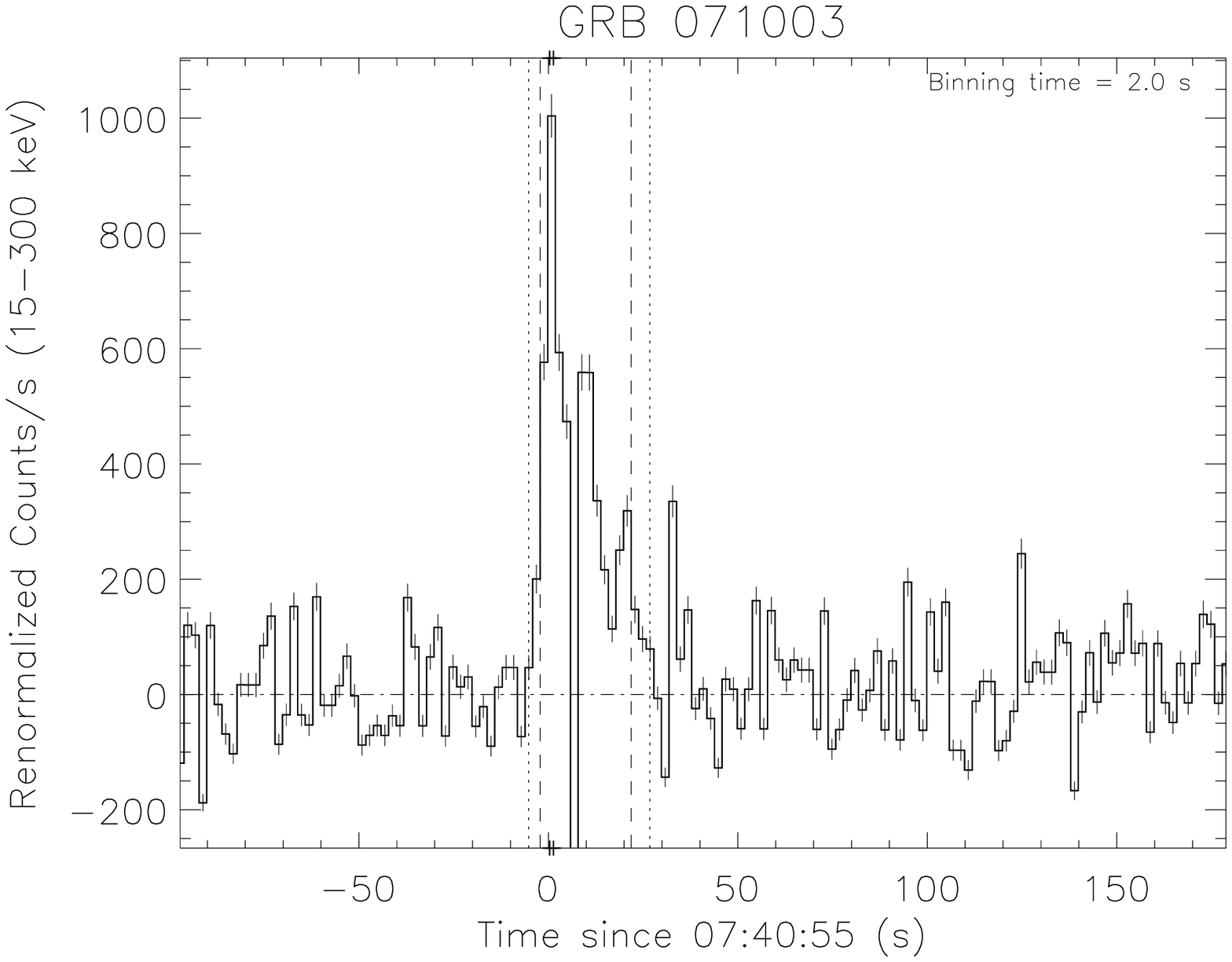,width=0.5\linewidth} \\
\end{tabular}
\textbf{Fig.~\ref{fig:light_curves}.} Light curves of \int\ GRBs (continued)
\end{figure}
\begin{figure}[p]
\vspace{1.0cm}
\psfig{figure=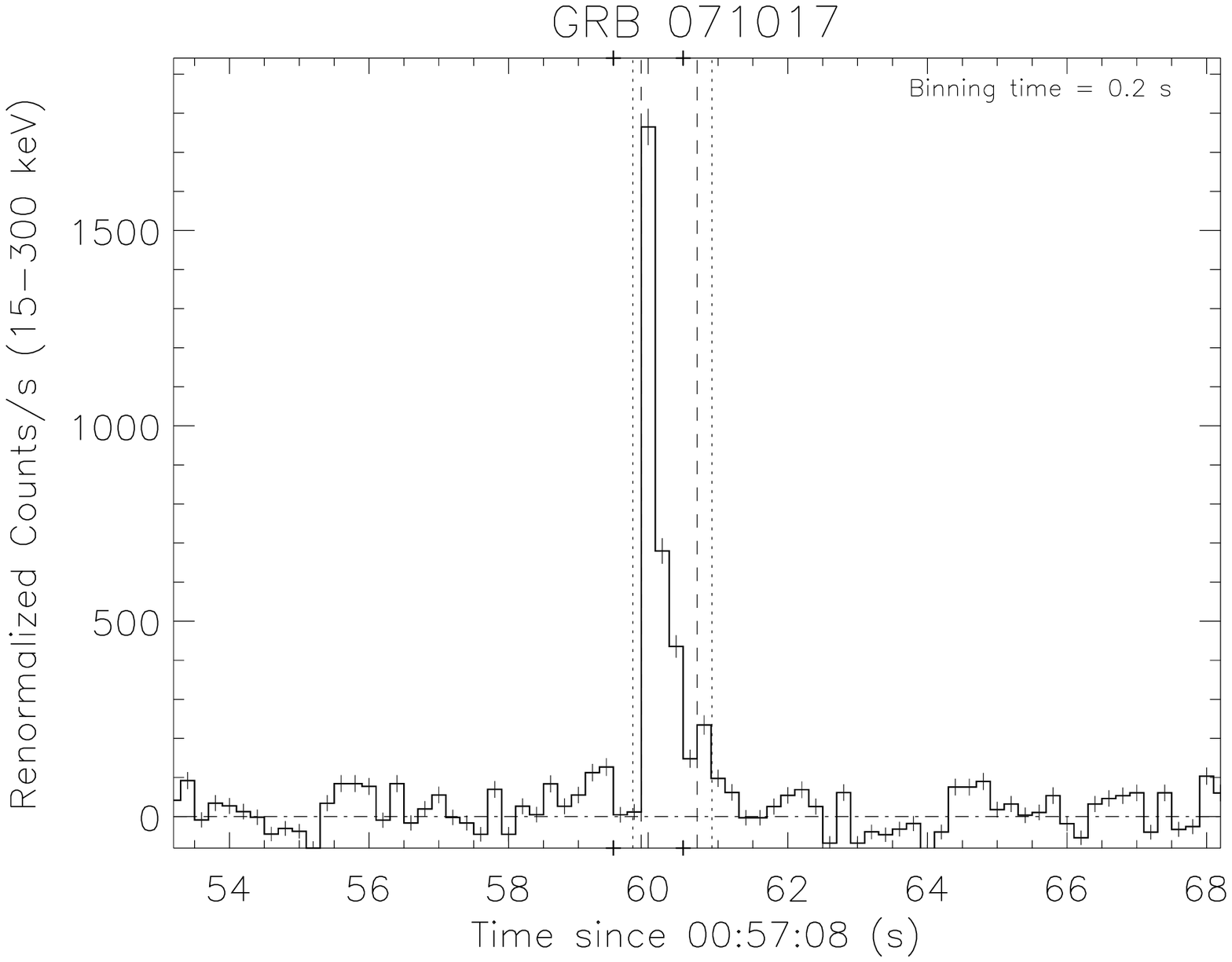,width=0.5\linewidth}\\
\textbf{Fig.~\ref{fig:light_curves}.} Light curves of \int\ GRBs (end)
\end{figure}
\end{document}